\documentclass[preprint,3p,times]{elsarticle}

\usepackage{amssymb}
\usepackage{amsthm}
\usepackage{amsmath}
\usepackage{epsfig}
\usepackage{subfigure}
\usepackage[usenames]{color}
\usepackage{algorithm}
\usepackage{algorithmic}
\usepackage{nicefrac}
\epsfverbosetrue
\graphicspath{{fig/}{../fig/}}

\newcommand{\mb}[1]{\ensuremath{\mathbf{#1}}}

\journal{Computer Methods in Applied Mechanics and Engineering}

\begin{document}
 \setlength{\parindent}{0.0ex}
 \setcounter{secnumdepth}{4}
 \setcounter{tocdepth}{4}
\begin{frontmatter}



\title{A Unified Approach for Beam-to-Beam Contact}


\author[lnm]{Christoph Meier\corref{cor1}}
\ead{meier@lnm.mw.tum.de}
\author[lnm]{Wolfgang A. Wall}
\author[lnm]{Alexander Popp}

\address[lnm]{Institute for Computational Mechanics, Technical University of Munich, Boltzmannstrasse 15, D--85748 Garching b. M\"unchen, Germany}

\cortext[cor1]{Corresponding author}

\begin{abstract}
Existing beam contact formulations can be categorized in point-to-point contact models that consider a discrete contact force at the closest point of the beams, and line-to-line contact models that assume distributed contact forces. In this work, it will be shown that line contact formulations applied to slender beams provide accurate and robust mechanical models in the range of small contact angles, whereas the computational efficiency considerably decreases with increasing contact angles. On the other hand, point contact formulations serve as sufficiently accurate and very efficient models in the regime of large contact angles, while they are not applicable for small contact angles as a consequence of non-unique closest point projections. In order to combine the advantages of these basic formulations, a novel all-angle beam contact (\textit{ABC}) formulation is developed that applies a point contact formulation in the range of large contact angles and a recently developed line contact formulation in the range of small contact angles, the two being smoothly connected by means of a variationally consistent model transition. Based on a stringent analysis, two different transition laws are investigated, optimal algorithmic parameters are suggested and conservation of linear momentum, angular momentum and total energy is shown. All configuration-dependent quantities within the point-, the line- and the transition-contact regime are consistently linearized, thus allowing for their application within implicit time integration schemes. Furthermore, a step size control of the nonlinear solution scheme is proposed that allows for displacement increments per time step that exceed the order of magnitude of the beam cross-section radius. For many standard beam-to-beam contact algorithms, this is the typical limitation concerning possible time step sizes, especially when considering high beam slenderness ratios. Finally, an efficient two-stage contact search based on dynamically adapted search segments is proposed. This algorithm yields in a tight set of potential contact pairs and enables a subdivision into potential point and potential line contact pairs, which is essential in order to fully exploit the efficiency potential of the proposed contact formulation. A series of numerical test cases is analyzed in order to verify the accuracy and consistency of the proposed contact model transition regarding contact force distributions and conservation properties, but also for quantifying the efficiency gains as compared to standard beam contact formulations.
\end{abstract}

\begin{keyword}
Beam contact \sep 
Smooth model transition \sep
Thin fibers \sep
Finite elements \sep
$C^1$-continuous Kirchhoff beams
\end{keyword}

\end{frontmatter}

%

%
\section{Introduction}
\label{sec:intro}
%
In countless fields of application, mechanical system performance is essentially determined by highly slender fiber- or rod-like components. Industrial webbing, high-tensile ropes and cables, fiber-reinforced composite materials or polymer materials, but also biological tissue or biopolymer networks (see e.g. \cite{cyron2012}) can be identified as typical examples of such fiber-dominated systems. Geometrically nonlinear beam finite elements are an efficient and accurate tool for modeling and solving these systems numerically. In \cite{romero2008}, different types of nonlinear beam element formulations have been evaluated and compared, and the so-called geometrically exact beam formulations (see e.g \cite{crisfield1999, eugster2013, jelenic1999, romero2004, romero2002, simo1985, simo1986, sonneville2014, zupan2003}) have been recommended in terms of model accuracy and computational efficiency. While all the mentioned finite element formulations are based on the Simo-Reissner beam theory, an alternative geometrically exact element formulation based on the Kirchhoff theory of thin rods and incorporating the modes of axial tension, torsion and anisotropic bending has been proposed in the authors' recent contributions \cite{meier2014} and \cite{meier2015}. This formulation is tailored for high beam slenderness ratios as considered in this work. Furthermore, it consists of a $C^1$-continuous beam centerline representation which enables smooth beam-to-beam contact kinematics.\\

The applications mentioned above are characterized by an intensive mechanical contact interaction between individual fibers and by geometrically complex contact configurations. Some recent contributions focusing on the analytical modeling of contact interaction between thin fibers are, e.g., the investigation of ropes with single- and bi-helical fiber substructures~\cite{xiang2015}, the theoretical treatment of knot-mechanics \cite{jawed2015} or the analysis of optimal topologies and packing densities in filamentous materials based on an implicit consideration of contact \cite{grason2015}. The arguably most popular numerical contact formulation for slender continua \cite{wriggers1997} models mechanical beam-to-beam contact interaction by means of a discrete contact force acting at the closest point of the beam centerlines. This model, in the following denoted as point-to-point contact formulation, results in a rather compact and efficient numerical formulation, which subsequently has been extended to frictional problems considering friction forces \cite{zavarise2000} and friction torques \cite{konjukhov2010}, rectangular beam cross-sections \cite{litewka2002,litewka2002b}, smoothed centerline geometries \cite{litewka2007}, constraint enforcement via Lagrange multipliers instead of penalty methods \cite{litewka2005} and adhesion effects \cite{kulachenko2012}. Quite recently, it has also been applied to self-contact problems \cite{neto2015}. In the recent works \cite{litewka2013} and \cite{litewka2015}, additional contact points located in the neighborhood of the actual closest point have been proposed in order to improve the accuracy of this purely point-based approach when applied in the regime of small contact angles. Nevertheless, this formulation still relies on the existence of a unique closest point projection between the beams. First investigations concerning existence and uniqueness of closest point projections as well as possible shortcomings of purely point-based procedures have been made in \cite{konjukhov2008} and \cite{konjukhov2010}.\\

In \cite{meier2015b}, it has been shown analytically that the existence of the closest point solution can not be guaranteed and consequently that point-to-point contact approaches cannot be applied in a considerable range of small contact angles. Since such configurations are very likely in complex fibrous systems, alternative beam contact models are required. One of the few existing alternatives is the formulation developed by Durville \cite{durville2004,durville2007,durville2010,durville2012,vu_and_durville2015}. It is based on a collocation-point-to-segment formulation and the definition of proximity zones on an intermediate geometry. A second alternative proposed by Chamekh et al. \cite{chamekh2009}, \cite{chamekh2014} is based on a 
Gauss-point-to-segment type approach and has mostly been applied to self-contact problems. In our earlier work \cite{meier2015b}, different beam contact formulations have been investigated and eventually a Gauss-point-to-segment formulation in combination with consistently linearized integration interval segmentation, a smooth contact force law and a $C^1$-continuous beam element formulation has been suggested as model of choice for the contact interaction of slender beams. Since these alternative approaches consider contact forces that are \textit{distributed} along the beams, they will be denoted as line-to-line contact formulations in the following.\\

Even though line-to-line approaches yield accurate and robust contact models in the entire range of possible contact angles, their computational efficiency decreases considerably with increasing beam slenderness ratio. In Section~\ref{sec:allangleformulation}, it will be shown that especially in the range of large contact angles the number of Gauss or collocation points required by these approaches and the resulting computational effort is prohibitively high as compared to point-to-point contact formulations. Thus, on the one hand, the point-to-point contact formulation serves as sensible mechanical model and very efficient numerical algorithm in the range of intermediate and large contact angles while it is not applicable for small contact angles. On the other hand, the line-to-line contact formulation provides a very accurate and robust mechanical model in the small-angle regime whereas the computational efficiency dramatically decreases with increasing contact angles.  These properties motivate the development of a novel all-angle beam contact (\textit{ABC}) formulation that combines the advantages of two worlds: The formulation is based on a standard point-to-point contact formulation applied in the range of large contact angles while the scope of small contact angles is covered by the line-to-line contact formulation proposed in~\cite{meier2015b}. Two different variants of a smooth model-transition procedure between the regimes of point and line contact are investigated, a variationally consistent transition on penalty potential level and a simpler variant on contact force level. Both variants lead to exact conservation of linear and angular momentum, while only the variationally consistent variant enables exact energy conservation. Based on analytical investigations, recommendations are made concerning the optimal ratio between the two penalty parameters of the point and the line contact, the required number of line contact Gauss points and the choice of the model transition interval. All configuration-dependent quantities are consistently linearized allowing for an application within implicit time integration schemes. The resulting \textit{ABC} formulation is supplemented by the contact contributions of the beam endpoints (see~\cite{meier2015b}).\\

The novel formulation successfully addresses one of the most essential challenges in the contact modeling
of highly slender structures: To achieve a sufficiently fine spatial contact resolution at manageable computational costs. A second typical limitation for many standard beam contact algorithms in the range of high slenderness ratios is the requirement of an adequately small time step size. In that regard, we propose a step size control of the nonlinear solution scheme, which allows for displacement increments per time step that exceed the order of magnitude of the beam cross-section radius. Additionally, we propose a very efficient two-stage contact search algorithm based on dynamically adapted search segments for each finite element. This does not only result in a very tight set of potential contact pairs, but it also enables a subdivision into potential point-to-point and potential line-to-line contact pairs. The latter property is essential in order to fully exploit the efficiency potential of the proposed \textit{ABC} formulation. All the presented algorithmic components are tailored for the most challenging case of arbitrary discretization orders and lengths that typically lead to high element slenderness ratios and deformations. The interplay of these individual constituents yields the first beam-to-beam contact formulation that combines a significant degree of robustness and universality in the treatment of complex contact scenarios and arbitrary beam-to-beam orientations with high computational efficiency, especially in the limit of extreme beam and element slenderness ratios.\\

The remainder of this paper is organized as follows. In Section~\ref{sec:beamformulation}, we sketch the main constituents of the beam element formulation proposed in \cite{meier2014, meier2015}. In Sections~\ref{sec:point} and \ref{sec:line}, the basics of standard point contact models and of the line contact formulation derived in \cite{meier2015b} are outlined. The novel all-angle beam contact formulation, the related theoretical investigations and the recommendations concerning optimal parameter choice are presented in Section~\ref{sec:allangleformulation}, while Section~\ref{sec:algorithmicaspects} contains algorithmic aspects such as contact search, step size control, penalty force laws and treatment of endpoint contacts. Finally, detailed numerical verifications are presented in Section~\ref{sec:numerical_examples}. While the first two examples in Sections~\ref{sec:examples_example1} and ~\ref{sec:examples_example2} aim to investigate the accuracy and consistency of the all-angle beam contact formulation regarding contact force distributions and conservation properties, the remaining two examples in Sections~\ref{sec:examples_biopolymer} and~\ref{sec:examples_rope} are intended to bridge the gap towards challenging real-world applications. Therein, also the attainable efficiency gains of the proposed algorithm as compared to standard line-to-line contact formulations are quantified.

%
\section{Beam formulation}
\label{sec:beamformulation}
%

While the proposed ABC formulation can be combined with arbitrary beam formulations with a proper centerline representation, we will exclusively consider the finite element formulation of Kirchhoff type applied to beam contact problems in the recent contribution \cite{meier2015b} within this paper. The original derivation of the 
quasi-static beam formulation can be found in \cite{meier2014} and \cite{meier2015}, while the extension to dynamic problems is presented in \cite{meier2015b}. The element residual and stiffness contributions can also 
be found in \cite{meier2015b} and are additionally summarized in \ref{anhang:reslin_beamelement}. The most relevant constituent for contact interaction is the interpolation of the 
beam centerline, which will briefly be summarized here. Concretely, we follow a Bubnov-Galerkin approach that leads to the following discretized beam centerline:
\begin{align}
\label{interpolation}
\mb{r}(\xi) \! \approx \! \mb{r}_h(\xi)  \! = \! \sum_{i=1}^{2} N^i_{d}(\xi) \mb{\hat{d}}^i + \frac{l_0}{2} \sum_{i=1}^{2} N^i_{t}(\xi) \mb{\hat{t}}^i =: \mb{N}(\xi) \mb{d}, \,\,\,\, 
\delta \mb{r}(\xi) \! \approx \! \delta \mb{r}_{h}(\xi) \! = \! \sum_{i=1}^{2} N^i_{d}(\xi) \delta \mb{\hat{d}}^i + \frac{l_0}{2} \sum_{i=1}^{2} N^i_{t}(\xi) \delta \mb{\hat{t}}^i  =: \mb{N}(\xi) \delta \mb{d}\, ,
\end{align}
It is specified by a proper finite-dimensional discrete trial space $\mb{r} \! \approx \! \mb{r}_h \!\in\!  \mathcal{S}_h \!\subset\! \mathcal{S}$ and a discrete test space $\delta \mb{r} \!\approx\! \delta \mb{r}_h \!\in\! \mathcal{V}_h \!\subset\! \mathcal{V}$. Moreover, $\mb{\hat{d}}^i, \mb{\hat{t}}^i \! \in \! \Re^3$ are positions and tangent vectors at the two element nodes $i=1,2$, $\delta \mb{\hat{d}}^i, \delta \mb{\hat{t}}^i \! \in \! \Re^3$ represent their variations, $l_0$ is the initial 
length of the initially straight beam element and $\xi \in [-1;1]$ is an element parameter coordinate that can explicitly be related to the arc-length coordinate $s \in [0,l_0] \subset \Re$ on the beam centerline according 
to $s(\xi)=s_0 + (\xi+1)l_0/2$ and $(.)_{,s}=(.)_{,\xi}\cdot J_{ele}(\xi)$ with the Jacobian $J_{ele}(\xi)=l_0/2$ if considering only initially straight beams. Here, $s_0$ represents the arc-length coordinate of the first node of the resulting two-noded finite element. Similar to the abbreviation $(.)^{\prime}\!=\!(.)_{,s}$ for the arc-length derivative, we will use the notation $(.)^{\shortmid}\!=\!(.)_{,\xi}$ for the derivative with respect to the parameter coordinate. Here and in the following, the index $h$ denotes the spatially discretized version of a quantity, but this index will often be omitted in the following when there is no danger of confusion. The third-order Hermite shape functions $N^i_{d}(\xi)$ and $N^i_{t}(\xi)$ (see \cite{meier2014} for the properties of these polynomials) defined as
\begin{align}
\label{shapefunctions}
N^1_{d}(\xi) = \frac{1}{4}(2+\xi)(1-\xi)^2, \,\,\, N^2_{d}(\xi) = \frac{1}{4}(2-\xi)(1+\xi)^2, \,\,\,
N^1_{t}(\xi) = \frac{1}{4}(1+\xi)(1-\xi)^2, \,\,\, N^2_{t}(\xi) = -\frac{1}{4}(1-\xi)(1+\xi)^2
\end{align}
provide a $C^1$-continuous beam centerline representation, thus enabling smooth contact kinematics. This property will be beneficial for the derivation of the contact 
formulation in the following sections. The abbreviations $\mb{d}$, $\delta \mb{d}$ and $\mb{N}(\xi)$ appearing in \eqref{interpolation} represent proper element-wise vector- and matrix-valued 
assemblies of the nodal variables and shape functions. The resulting element residual contributions $\mb{r}_{int}, \mb{r}_{kin}$ and $\mb{r}_{ext}$
of the internal, kinetic and external forces and their linearizations are summarized in \ref{anhang:reslin_beamelement}. An assembly of these quantities and the corresponding element-wise 
contact contributions $\mb{r}_{con}$ presented in the next sections leads to the following global system of equations:
\begin{align}
\label{global_system}
\mb{R}_{tot}=\mb{M} \ddot{\mb{D}}+\mb{R}_{int}(\mb{D})+\mb{R}_{con}(\mb{D})-\mb{R}_{ext}(\mb{D})=\mb{0}.
\end{align}
Equation \eqref{global_system} represents the spatially discretized weak form of mechanical equilibrium, where $\mb{D}$ is the assembled global vector of primary variables containing the nodal degrees of freedom $\mb{\hat{d}}^k,\mb{\hat{t}}^k$ of all $n_{node}$ nodes with $k = 1,...,n_{node}$. It is again emphasized that the employed two-noded, initially straight elements have been chosen for simplicity. The proposed ABC formulation is of a very general nature. It is in no way restricted to this class of beam elements.\\

\hspace{0.5 cm}
\begin{minipage}{15.0 cm}
\textbf{Remark:} The $C^1$-continuous Hermite shape functions guarantee for the existence of a
 unique tangent field. Nevertheless, the contact formulations presented in the following are general enough to be combined with any type of beam element formulation and shape function set. However, in case such alternative shape functions do not satisfy the smoothness requirement, additional means are necessary in order to detect and evaluate mechanical contact at positions with non-unique tangent vectors.
\end{minipage}

%
\section{Point-to-point contact formulation}
\label{sec:point}
%

\begin{figure}[t]
 \centering
  \includegraphics[width=0.5\textwidth]{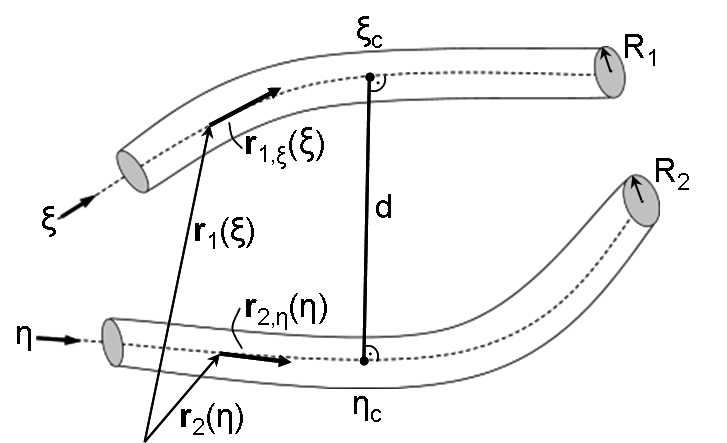}
  \caption{Kinematic quantities defining the point-to-point contact problem of two beams.}
  \label{fig:point_problemdescription1}
\end{figure}

Within this section, we briefly review the main constituents of a standard point-to-point beam contact formulation as introduced in \cite{wriggers1997}. Thereto, we consider two arbitrarily 
curved beams with cross-section radii $R_1$ and $R_2$, respectively. The beam centerlines are represented by two parametrized curves~$\mb{r}_{1}(\xi)$ and $\mb{r}_{2}(\eta)$ with curve 
parameters~$\xi$ and $\eta$. Furthermore, $\mb{r}_{1,\xi}(\xi)= \mb{r}_1^{\shortmid}(\xi)$ and $\mb{r}_{2,\eta}(\eta)=\mb{r}_2^{\shortmid}(\eta)$ denote the tangents to these curves at positions 
$\xi$ and $\eta$, respectively. In what follows, we assume that the considered space curves are at least $C^1-$continuous, thus providing a unique tangent 
vector at every position $\xi$ and $\eta$. The kinematic quantities introduced above are illustrated in Figure~\ref{fig:point_problemdescription1}.\\

The point-to-point beam contact formulation enforces the contact constraint by prohibiting penetration of the two beams at the closest point positions $\xi_c$ and $\eta_c$. Here and in the following, 
the subscript $c$ indicates, that a quantity is evaluated at the closest point coordinate $\xi_c$ or $\eta_c$, respectively. These closest point 
coordinates are determined as solution of the bilateral (''bl``) minimal distance problem, also denoted as bilateral closest point projection, i.e.
\begin{align}
\label{point_mindist}
  d_{bl}:=\min_{\xi, \eta} d(\xi, \eta) = d(\xi_c, \eta_c) \quad \text{with} \quad d(\xi, \eta)=||\mb{r}_{1}(\xi)-\mb{r}_{2}(\eta)||.
\end{align}
This leads to two orthogonality conditions that must be solved for the unknown closest point coordinates $\xi_c$ and $\eta_c$:
\begin{align}
\label{point_orthocond}
\begin{split}
  p_{1}(\xi,\eta)&=\mb{r}^T_{1,\xi}(\xi)\left( \mb{r}_{1}(\xi)-\mb{r}_{2}(\eta) \right) \quad \rightarrow \, p_{1}(\xi_c,\eta_c) \dot{=} 0, \\
  p_{2}(\xi,\eta)&=\mb{r}^T_{2,\eta}(\eta)\left( \mb{r}_{1}(\xi)-\mb{r}_{2}(\eta) \right) \quad \rightarrow \, p_{2}(\xi_c,\eta_c) \dot{=} 0.
\end{split}
\end{align}
The non-penetration condition at the closest point is formulated by means of the inequality constraint
\begin{align}
  g \geq 0 \quad \text{with} \quad g:=d_{bl}-R_1-R_2,
\end{align}
where $g$ is the gap function. This constraint will be included into a variational problem setting via a penalty potential
\begin{align}
\label{point_penaltypotential}
     \quad \Pi_{c\varepsilon}=\frac{1}{2} \varepsilon \langle g\rangle^2 \quad \text{with} \quad 
           \langle x\rangle=\left\{\begin{array}{ll}
                                 x, & x \leq 0 \\
                                 0, & x > 0
                    \end{array}\right. .
\end{align}
Variation of \eqref{point_penaltypotential} yields the point contact contribution to the weak form and the definition of the contact force $\mb{f}_{c\varepsilon}$:
\begin{align}
\label{point_weakform}
  \delta \Pi_{c\varepsilon} =
      \varepsilon \langle g\rangle \delta g =
      \varepsilon \langle g\rangle \left(\delta \mb{r}_{1c} - \delta \mb{r}_{2c} \right)^T \! \mb{n}, \,\,\,\,\,\,
      \mb{f}_{c\varepsilon}= \underbrace{-\varepsilon \langle g\rangle}_{=:f_{c\varepsilon}} \mb{n}, \,\,\,\,\,\,  \mb{n}:=\frac{\mb{r}_{1}(\xi_c)-\mb{r}_{2}(\eta_c)}{||\mb{r}_{1}(\xi_c)-\mb{r}_{2}(\eta_c)||} \,\,.
\end{align}
According to \eqref{point_weakform}, the point-to-point beam contact formulation models the contact force $\mb{f}_{c\varepsilon}$ that is transferred between the two beams 
as a discrete point force acting at the closest points of the beam centerlines in normal direction $\mb{n}$. For later use, we also define the so-called contact angle as the angle between the tangent vectors at the contact point:
\begin{align}
\label{point_contactangle}
    \alpha = \arccos{ \left(z \right) }
  \quad \text{with} \quad z=\frac{ ||\mb{r}_1^{\shortmid T}(\xi_c) \mb{r}_2^{\shortmid}(\eta_{c})|| }{ ||\mb{r}_1^{\shortmid}(\xi_c)|| \cdot ||\mb{r}_2^{\shortmid}(\eta_{c})|| },
  \quad \alpha \in [0;90^{\circ}].
\end{align}
In a next step, spatial discretization has to be performed. Since, for simplicity, we only consider the contact contribution of one contact point, the indices $1$ and $2$ are directly transferred 
to the two finite elements where the contact actually takes place. Inserting the spatial discretization \eqref{interpolation} into the orthogonality conditions \eqref{point_orthocond} allows to solve the latter for the unknown closest point parameter coordinates $\xi_c$ and $\eta_c$. Since, in general, the system of equations provided by \eqref{point_orthocond} is nonlinear in $\xi$ and $\eta$, a local Newton-Raphson scheme is applied for 
its solution. The corresponding linearizations of \eqref{point_orthocond} as well as all residual and stiffness contributions of the point-to-point contact formulation can for example be found in \cite{wriggers1997} and are additionally summarized in~\ref{anhang:linearizationendpoint}. Inserting equations \eqref{interpolation} into equation \eqref{point_weakform} leads to the following contact residual contributions $\mb{r}_{con,1}$ and $\mb{r}_{con,2}$ of the two considered elements:
\begin{align}
\label{point_discreteweakform}
  \delta \Pi_{c\varepsilon}=\delta \mb{d}_1^T \underbrace{\varepsilon \langle g\rangle \mb{N}_{1}^T(\xi_c) \mb{n}}_{=:\mb{r}_{con,1}} - 
  \delta \mb{d}_2^T \underbrace{\varepsilon \langle g\rangle \mb{N}_{2}^T(\eta_c) \mb{n}}_{=:\mb{r}_{con,2}}.
\end{align}

%
\section{Line-to-line contact formulation}
\label{sec:line}
\begin{figure}[ht]
 \centering
  \subfigure[Space continuous problem setting.]
   {
    \includegraphics[width=0.4\textwidth]{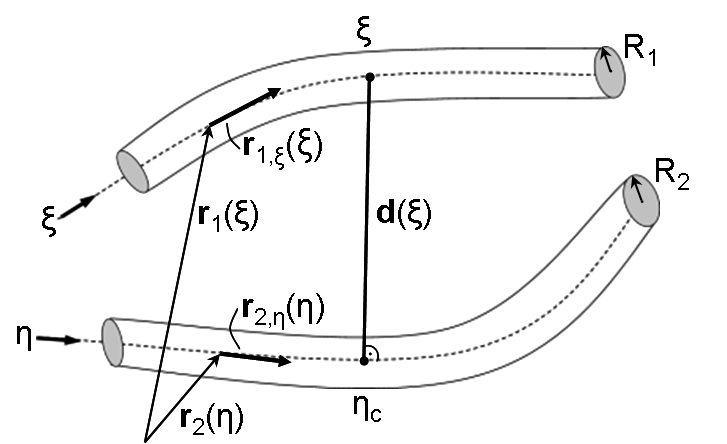}
    \label{fig:line_problemdescription_contiuous}
   }
   \hspace{0.05 \textwidth}
   \subfigure[Discretized problem setting.]
   {
    \includegraphics[width=0.4\textwidth]{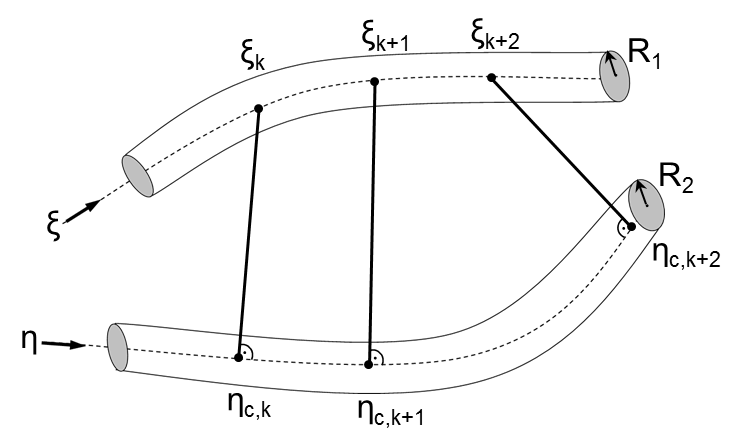}
    \label{fig:line_problemdescription_discrete}
   }
  \caption{Kinematic quantities defining the line-to-line contact problem of two beams.}
  \label{fig:line_problemdescription}
\end{figure}

%
Here, we will briefly repeat the most important aspects of the line-to-line contact formulation developed in \cite{meier2015b}. In contrary to the point-to-point contact model, this formulation is based on 
a line constraint enforced along the entire beam length. The relevant kinematic quantities of this approach are illustrated in Figure~\ref{fig:line_problemdescription_contiuous}.\\

Here, a distinction has to be made between a master beam (beam $1$) and a slave beam (beam $2$). The closest master point~$\eta_c$ to a given slave point~$\xi$ is determined as solution of the following unilateral (``ul'') minimal distance problem:
\begin{align}
\label{line_mindist}
  d_{ul}(\xi):=\min_{\eta} d(\xi,\eta)= d(\xi,\eta_c) \quad \text{with} \quad d(\xi,\eta) = ||\mb{r}_{1}(\xi)-\mb{r}_{2}(\eta)||.
\end{align}
Condition~\eqref{line_mindist} leads to one orthogonality condition that has to be solved for the unknown parameter coordinate $\eta_c$:
\begin{align}
\label{line_orthocond}
\begin{split}
  p_2(\xi,\eta)&=\mb{r}^T_{2,\eta}(\eta)\left( \mb{r}_{1}(\xi)-\mb{r}_{2}(\eta) \right) \quad \rightarrow \, p_{2}(\xi,\eta_c) \dot{=} 0
\end{split}
\end{align}
Thus, in contrary to the procedure of the last section, the normal vector is still perpendicular to the master beam $2$ but not to the slave beam $1$ anymore. Furthermore, in the context of line contact, 
the subscript $c$ indicates that a quantity is evaluated at the closest master point $\eta_c$ of a given slave point $\xi$. The non-penetration condition becomes
\begin{align}
\label{line_constraint}
  g(\xi) \geq 0 \, \forall \, \xi \quad \text{with} \quad g(\xi):=d_{ul}(\xi)-R_1-R_2,
\end{align}
and is integrated into the variational formulation by means of an inequality-constraint enforced via a penalty potential:
\begin{align}
\label{line_pen_totalpotential}
     \Pi_{c\varepsilon}=\frac{1}{2} \varepsilon \int \limits_0^{l_1} \langle g(\xi) \rangle^2 ds_1.
\end{align}
The space-continuous penalty potential in~\eqref{line_pen_totalpotential} does not only serve as a purely mathematical tool for constraint enforcement, but can rather be regarded as a mechanical model for the cross-section flexibility of the contacting beams. Variation of the penalty potential defined in~\eqref{line_pen_totalpotential} leads to the contact contribution to the weak form:
\begin{align}
\label{line_pen_weakform}
  \delta \Pi_{c\varepsilon} = \varepsilon \int \limits_0^{l_1} \langle g(\xi)\rangle \delta g (\xi)  ds_1
  \quad \text{and} \quad \delta g(\xi) = ( \delta \mb{r}_1(\xi) - \delta \mb{r}_2(\xi) )^T \mb{n}(\xi).
\end{align}
In the virtual work expression~\eqref{line_pen_weakform}, we can identify the contact force vector $\mb{f}_{c\varepsilon}(\xi)$ and the normal vector $\mb{n}(\xi)$:
\begin{align}
\label{line_pen_contactforce}
  \mb{f}_{c\varepsilon}(\xi)= \underbrace{- \varepsilon \langle g(\xi)\rangle }_{=:f_{c\varepsilon}(\xi)} \mb{n}(\xi), \quad  \mb{n}(\xi):=\frac{\mb{r}_{1}(\xi)-\mb{r}_{2}(\eta_c)}{||\mb{r}_{1}(\xi)-\mb{r}_{2}(\eta_c)||}.
\end{align}
According to \eqref{line_pen_contactforce}, a line-to-line beam contact formulation models the contact force $\mb{f}_{c\varepsilon}(\xi)$ that is transferred between the beams as a distributed 
line force. For later use in Section~\ref{sec:allangleformulation}, we can again define the contact angle field as:
\begin{align}
\label{line_contactangle}
  \alpha(\xi) = \arccos{ \left(z(\xi) \right) }
  \quad \text{with} \quad z(\xi)=\frac{ ||\mb{r}_1^{\shortmid T}(\xi) \mb{r}_2^{\shortmid}(\eta_{c})|| }{ ||\mb{r}_1^{\shortmid}(\xi)|| \cdot ||\mb{r}_2^{\shortmid}(\eta_{c})|| },
  \quad \alpha \in [0;90^{\circ}].
\end{align}
Next, spatial discretization has to be carried out. For simplicity, we only consider the contact contribution stemming from one pair of finite elements on the slave beam and on the master beam, that are assigned to each other via the projection in~\eqref{line_orthocond}. Therefore, in the following, the indices $1$ of the slave beam and $2$ of the master beam will also be used in order to denote the two considered finite elements lying on these beams. Inserting the discretization \eqref{interpolation} into equation \eqref{line_pen_weakform} and replacing the analytical integral by a Gauss quadrature leads to the contributions of element $1$ and $2$ to the discretized weak form. Similar to \cite{meier2015b}, we allow for $n_{II} \geq 1$ contact integration intervals per slave beam element with $n_{GR}$ integration points defining a Gauss rule of order $p=2 n_{GR}-1$ on each of these integration intervals, thus leading to
$n_{GP}=n_{II} \cdot n_{GR}$ integration points per slave element. In order to realize such a procedure, one has to introduce a mapping between the element parameter space $\xi$ and the $n_{II}$ contact parameter spaces $\bar{\xi}$ per slave element: 
\begin{align}
\label{line_integrationparamvalues}
\begin{split}
  \xi_{ij}=\frac{1.0-\bar{\xi}_j}{2}\xi_{1,i} + \frac{1.0+\bar{\xi}_j}{2}\xi_{2,i} \quad \text{for} \quad i=1,...,n_{II}, \,\,\, j=1,...,n_{GR}.
\end{split}
\end{align}
Here,  $\xi_{ij}$ denotes the element parameter coordinate of the Gauss point $j$ with contact parameter coordinate $\bar{\xi}_j \in [-1;1]$ lying within the contact parameter interval $i$ confined by the element parameter coordinate values $\xi_{1,i}~\in~[-1;1]$ and $\xi_{2,i}~\in~[-1;1]$. The coordinates $\bar{\xi}_j \in [-1;1]$ represent constant, i.e. deformation independent, parameters defined by the respective Gauss rule. In the simplest case, the parameter coordinates $\xi_{1,i}$ and $\xi_{2,i}$ confining the $i^{th}$ integration interval are chosen equidistantly within the slave element. For illustration, we want to give the following example: When considering a three-point Gauss rule $n_{GR}=3$ and four integration intervals $n_{II}=4$ per slave element, we have $\xi_{1,i}~\in~\{-1,-0.5,0,0.5\}$, $\xi_{2,i}~\in~\{-0.5,0,0.5,1\}$ for $i=1,2,3,4$ and $\bar{\xi}_j~\in~\{-1/\sqrt{3},0,1/\sqrt{3}\}$ for $j=1,2,3$. \\

Using~\eqref{interpolation} and \eqref{line_integrationparamvalues}, the discretized version of the contact contributions of element $1$ and $2$ according to \eqref{line_pen_weakform} read:
\begin{align}
\label{line_pen_discreteweakform_multipleIS}
\begin{split}
  \mb{r}_{con,1} \! = \!\! \sum \limits_{i=1}^{n_{II}}\! \sum \limits_{j=1}^{n_{GR}}  \underbrace{w_j J(\xi_{ij},\xi_{1,i},\xi_{2,i}) \varepsilon \langle g(\xi_{ij}) \rangle \mb{N}_{1}^T(\xi_{ij}) \mb{n}(\xi_{ij})}_{\mb{r}_{con,1}^{ij}}, \,\,
  \mb{r}_{con,2} \! = \! \!\sum \limits_{i=1}^{n_{II}} \! \sum \limits_{j=1}^{n_{GR}} \underbrace{-w_j J(\xi_{ij},\xi_{1,i},\xi_{2,i}) \varepsilon \langle g(\xi_{ij}) \rangle \mb{N}_{2}^T(\eta_{c}(\xi_{ij})) \mb{n}(\xi_{ij})}_{\mb{r}_{con,2}^{ij}}
\end{split}
\end{align}
Here, the terms $\mb{r}_{con,1}^{ij}$ and $\mb{r}_{con,2}^{ij}$ denote the residual contributions of one individual Gauss point $j$ in the integration interval $i$. The element parameter 
coordinate $\xi_{ij}$ is evaluated according to \eqref{line_integrationparamvalues}. Furthermore, $w_j$ is the corresponding Gauss quadrature weight and $\eta_c(\xi_{ij})$ is the closest master point coordinate assigned to the Gauss point coordinate $\xi_{ij}$ on the slave beam. Inserting~\eqref{interpolation} into the orthogonality condition \eqref{line_orthocond} allows to solve the latter for the unknown $\eta_c(\xi_{ij})$ 
for any given $\xi_{ij}$. The linearizations of \eqref{line_orthocond} required for an iterative solution can be found in~\ref{anhang:linearizationlinecontact}. Finally,
\begin{align}
\label{line_totaljacobian}
  J(\xi_{ij},\xi_{1,i},\xi_{2,i})=J_{ele}(\xi_{ij}) \cdot \frac{\xi_{2,i}-\xi_{1,i}}{2} \quad \text{with} \quad i=1,...,n_{II},
\end{align}
represents the total Jacobian, where the mapping $J_{ele}(\xi_{ij})$ results from the applied beam element formulation (see Section~\ref{sec:beamformulation}). In \cite{meier2015b}, it has 
been shown that the overall integration error can be reduced drastically by applying an additional integration interval segmentation at the master beam endpoints $\eta_{EP}$ in order to avoid 
an integration across strong discontinuities in the integrand occurring at these points. Accordingly, the constant integration interval bound $\xi_{1,i}$ or $\xi_{2,i}$ at such a point has 
to be replaced by a deformation dependent and consistently linearized projection point $\xi_{B1}(\eta_{EP}, \mb{d}_{12})$ or $\xi_{B2}(\eta_{EP}, \mb{d}_{12})$ with 
$\mb{d}_{12}=(\mb{d}_{1}^T,\mb{d}_{2}^T)^T$ (see \cite{meier2015b} for details). The proposed formulation according to \eqref{line_pen_discreteweakform_multipleIS} consists of a Gauss-point-to-segment type contact 
discretization and a penalty regularization of the contact constraint. In \cite{meier2015b}, theoretical considerations concerning alternative constraint enforcement strategies by means of Lagrange multipliers and alternative 
contact discretizations based on mortar methods have been made. However, detailed theoretical and numerical investigations of these different approaches suggest the penalty-based Gauss-point-to-segment formulation 
according to  \eqref{line_pen_discreteweakform_multipleIS} as the variant that is most suitable for beam-to-beam contact. 

%
\section{All-angle Beam Contact (\textit{ABC}) formulation}
\label{sec:allangleformulation}
%

%
\subsection{Limitations of existing beam-to-beam contact formulations}
\label{sec:limitations}
%
In the last two sections, we have presented two basic contact formulations that enable the mechanical modeling of beam-to-beam contact in the sense of a point-to-point and a line-to-line contact interaction. 
In the next two Subsections \ref{sec:limitationspoint} and \ref{sec:limitationsline}, practically 
relevant limitations of these basic formulations will be investigated. The results of this study will serve as foundation for the development of a new general beam-to-beam contact formulation 
that combines the advantages of the point-to-point and the line-to-line contact formulation without exhibiting their limitations.

%
\subsubsection{Limitations of point-to-point contact formulation}
\label{sec:limitationspoint}
%
Compared to the line-to-line contact model, the point-to-point contact model has advantages in terms of implementation effort and computational efficiency. However, its limitation 
lies in the requirement of a unique closest point solution~\eqref{point_orthocond}, which can not be guaranteed for arbitrary geometrical configurations. In \cite{meier2015b}, it has 
been shown that no unique closest point solution can be guaranteed in the range of small contact angles $\alpha$, thus leading to the requirement:
\begin{align}
\label{limitationspoint_requirementsecondderiv8}
     \alpha > \alpha_{min}=\arccos \left(1 - 2 \mu_{max} \right).
\end{align}
Here, $\mu_{max}$ represents the maximal ratio of cross section radius $R$ to bending curvature radius $\bar{r}$ according to
\begin{align}
\label{limitationspoint_kappa2}
  \mu_{max}=\frac{R}{min \, (\bar{r})} \ll 1 \quad \text{with} \quad \bar{r}=\frac{1}{\bar{\kappa}}.
\end{align}
The bending curvature $\bar{\kappa}$ is represented by the geometrical curvature $\kappa$ of the beam centerline according to
\begin{align}
\label{limitationspoint_kappa}
  \bar{\kappa}:= \frac{\kappa}{||\mb{r}^{\prime}||}=\frac{||\mb{r}^{\prime} \times \mb{r}^{\prime \prime}||}{||\mb{r}^{\prime}||^3}.
\end{align}
The implication of requirement \eqref{limitationspoint_requirementsecondderiv8} is clear: As soon as we can provide an upper bound $\mu_{max}$ for the admissible ratio 
of cross section to curvature radius, \eqref{limitationspoint_requirementsecondderiv8} yields a lower bound for the admissible contact angles above which the closest 
point solution is unique. Since $\mu_{max}\ll 1$ is typically limited by the applied beam theory, condition~\eqref{limitationspoint_requirementsecondderiv8} possesses the desirable feature that in general 
the lower bound $\alpha_{min}$ is an a priori known quantity that does not depend on the actual deformation state. This a priori knowledge will allow for the development of contact algorithms 
that apply the point-to-point contact formulation only within a fixed range of contact angles $\alpha \in ]\alpha_{min};90^{\circ}]$ while the remaining range $\alpha \in [0^{\circ};\alpha_{min}]$ is treated 
by a different formulation. As explained in \cite{meier2015b}, the contact model of point-to-point type seems to be rather suitable for the range of large contact angles also from a purely mechanical point of view.

%
\subsubsection{Limitations of line-to-line contact formulation}
\label{sec:limitationsline}
\begin{figure}[t]
    \centering
    \includegraphics[width=0.95\textwidth]{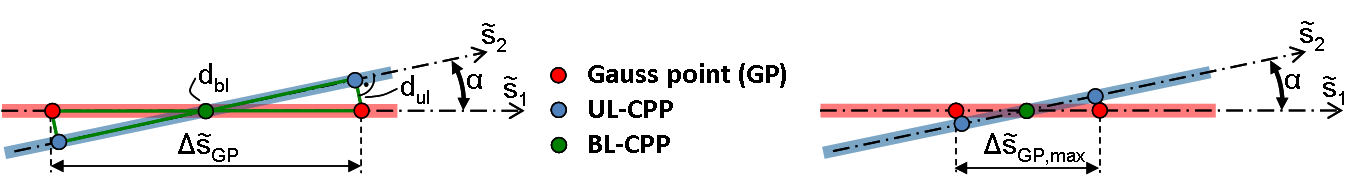}
    \caption{Contact interaction of two straight beams enclosing a given contact angle $\alpha$: Top view.}
    \label{fig:limitations_numGP}
\end{figure}

%
In~\cite{meier2015b}, it has been shown that in the range of admissible beam curvatures and for a sufficiently small distance between the contacting beams the unilateral closest point solution \eqref{line_orthocond} required for the line-to-line contact formulation will be unique. However, the practical limitation of the line-to-line contact formulation is of a different nature and appears in terms of computational effort. In order to explain this statement, we have visualized the top-view of two contacting straight beams with given contact angle $\alpha$ (see Figure~\ref{fig:limitations_numGP}, left). The crucial question is which distance $\Delta \tilde{s}_{GP}$ between two successive Gauss points (visualized by red circles) is admissible such that the normalized gap
\begin{align}
\label{limitationsline_GPdistance_normalizedgap}
g_{n}:=g/R=(d_{bl}-2R)/R
\end{align}
at the bilateral closest points (visualized by green circles) does not exceed a prescribed minimal value. From simple geometrical considerations, we can derive the following 
relation between the unilateral distance function $d_{ul}$ associated with the Gauss point and its projection onto the master beam (visualized by blue circles), the bilateral distance function 
$d_{bl}$ at the closest point pair $\tilde{s}_{1c}, \tilde{s}_{2c}$ (perpendicular to the slide plane) and the contact angle $\alpha$:
\begin{align}
\label{limitationsline_GPdistance1a}
d_{ul}^2=d_{bl}^2+\left( \frac{\Delta \tilde{s}_{GP} \sin{(\alpha)}}{2} \right)^2.
\end{align}
We assume that the penalty parameter $\varepsilon$ of the line contact formulation is high 
enough such that $d_{ul} \approx 2R$ in case of an active/contacting Gauss point. Using this estimation together with~\eqref{limitationsline_GPdistance_normalizedgap}, 
we can exploit~\eqref{limitationsline_GPdistance1a} in order to derive the maximal admissible distance $\Delta \tilde{s}_{GP,max}$ such that a prescribed minimal normalized gap $g_{n,min}$ will not 
be exceeded:
\begin{align}
\label{limitationsline_GPdistance1b}
\Delta \tilde{s}_{GP,max}=\sqrt{1-\left(\frac{g_{n,min}}{2}+1\right)^2}\frac{4R}{\sin{(\alpha)}}.
\end{align}
From~\eqref{limitationsline_GPdistance1b}, we can deduce two extreme cases: Requiring (theoretically) a vanishing gap at the bilateral closest point $g_{n,min}=0$ means that the distance between the Gauss 
points also would have to vanish. On the other hand, in order to simply prevent the beams from an undetected crossing ($g_{n,min}=2$ represents the minimal possible gap which occurs in the case 
of crossing centerlines) a maximal distance of $4R/\sin{(\alpha)}$ (see Figure~\ref{fig:limitations_numGP}, right) must not be exceeded. From~\eqref{limitationsline_GPdistance1b}, the minimal number of 
(evenly distributed) Gauss points $n_{min}$ required for a slave beam of length $l_1$ and an expected maximal contact angle $\alpha_{max}$ in order to limit 
the normalized gap to a minimal value of $g_{n,min}$ yields
\begin{align}
\label{limitationsline_GPdistance2}
n_{min}= k_{GP} \left[1-\left(\frac{g_{n,min}}{2}+1\right)^2 \right]^{-0.5}\frac{\sin{( \alpha_{max})}}{4}\rho_1 \quad \text{with} \quad \rho_1=\frac{l_1}{R},
\end{align}
where $\rho_1$ denotes the slenderness ratio of the slave beam. Here, $k_{GP}$ represents a safety factor, which takes into account that the Gauss points are actually not evenly distributed, that in fact we will have 
some small penetration at the Gauss points (and not exactly $d_{ul} = 2R$) and finally that in practical simulations one often requires a certain number of intermediate Gauss points for reasons of 
integration accuracy. Nevertheless, the required minimal number of Gauss points increases linearly with the slenderness ratio of the beams. Practical simulations of slender filaments confirm the prediction in~\eqref{limitationsline_GPdistance2}, i.e. that the computational effort increases with the slenderness ratio of the beams and that the overall computational cost is dominated by the numerical evaluation of Gauss point quantities. The second interesting information provided by \eqref{limitationsline_GPdistance2} is that 
$n_{min}$ increases with the expected contact angle. Thus, the computational effort could be reduced dramatically, if the expensive line-to-line contact formulation were only applied in 
the range of small contact angles. Again, it is not only the numerical point of view that suggests a confinement of the line-to-line contact formulation to the range of small contact angles: With increasing contact angle and increasing penalty parameter, the force evolution resulting from the line contact model degenerates more and more to a Dirac-delta distribution, thus advocating the point 
contact model as mechanical model of choice to be applied in the range of large contact angles.

%
\subsection{Derivation of \textit{ABC} formulation}
\label{sec:derivationabs}
%

The quintessence of the last two sections is rather straightforward: The point-to-point contact formulation serves as a sensible mechanical model and very efficient numerical algorithm in the range 
of intermediate and large contact angles, while it represents an insufficient mechanical model in the range of small contact angles and is even inapplicable for contact angles below the lower bound provided by~\eqref{limitationspoint_requirementsecondderiv8}. On the other hand, the line-to-line contact formulation provides a very accurate mechanical model and a robust and rather efficient numerical algorithm in the range of small contact angles, whereas the model quality and especially the computational efficiency dramatically decrease with increasing contact angles. According to~\eqref{limitationsline_GPdistance2}, this situation aggravates with increasing beam slenderness ratio. The novel approach presented in the following is based on the simple idea of combining the advantages of these two types of formulations, while abstaining from their 
disadvantages. Thus, we apply a standard point-to-point contact formulation in the range of large contact angles, while the range of small contact angles is covered by a line-to-line contact 
formulation. The smooth model-transition between these two regimes within a prescribed angle interval
\begin{align}
\label{ABC_shiftangles}
[\alpha_1;\alpha_2] \quad \text{with} \quad \alpha_1,\alpha_2 \in [0^{\circ};90^{\circ}], \quad \alpha_1 < \alpha_2,
\end{align}
is realized in a variationally consistent manner without loosing essential properties such as conservation of linear momentum, angular momentum and energy. 
Furthermore, all configuration-dependent quantities describing the point, line and transition contact 
range are consistently linearized thus allowing for their application within an implicit time integration scheme. Next, two different possibilities of carrying out the model transition will be investigated.

%
\subsubsection{Force-based model transition}
\label{sec:ABC_forcebased}
%

The first variant proposes a beam-to-beam contact model transition that is performed on the contact force level. Thereto, the overall contact contribution to the weak form 
will be defined as follows:
\begin{align}
\label{ABC_forcebased_weakform}
\delta \Pi_{c\varepsilon} = \underbrace{\left[1-k(z_c)\right] \varepsilon_{\perp} \langle g \rangle}_{=:-f_{c\varepsilon  \perp}} \delta g + 
\int \limits_0^{l_1} \underbrace{k(z(\xi)) \varepsilon_{\parallel} \langle g(\xi)\rangle}_{=:-f_{c \varepsilon  \parallel}(\xi)} \delta g (\xi) ds_1.
\end{align}
Here and in the following, the indices $\perp$ and $\parallel$ of a quantity refer to the point-contact or the line-contact formulation, respectively. Additionally, we have 
applied the common notations $z_c=z(\xi_c,\eta_c)$ as well as $z(\xi)=z(\xi,\eta_c(\xi))$. The angle-dependent transition factor $k(z)$ occurring in \eqref{ABC_forcebased_weakform} is represented by the 
following analytical expression:
\begin{align}
\label{ABC_forcebased_transition}
     k(z)= \left\{\begin{array}{lll}
                                    1, & \alpha < \alpha_1 \\
                                    0.5 \left(1-\cos\left(\pi\frac{z-z_2}{z_1-z_2}\right)\right), & \alpha_2 \geq \alpha \geq \alpha_1 \\
                                    0, & \alpha > \alpha_2
        \end{array}\right.
        \quad \text{with} \quad z=\cos(\alpha)=\frac{ ||\mb{r}_{1,\xi}^{T} \mb{r}_{2,\eta}|| }{ ||\mb{r}_{1,\xi}|| \cdot ||\mb{r}_{2,\eta}|| }.
\end{align}
Thus, \eqref{ABC_forcebased_weakform} represents a pure point-contact formulation for large angles $\alpha > \alpha_2$, a pure line-contact formulation for small angles $\alpha < \alpha_1$ and 
a weighted sum of these two basic formulations for angles within the transition interval $\alpha \in [\alpha_1;\alpha_2]$. Furthermore, the transition factor $k(z)$ according to 
\eqref{ABC_forcebased_transition} provides a $C^1$-continuous transition law for the corresponding contact forces contributions. It would be an obvious choice to take the contact angle 
$\alpha$ as argument of the transition function $k$. However, we rather take $z=\cos(\alpha)$, which represents the scalar product of the two unit tangent vectors (see also \eqref{point_contactangle} 
or \eqref{line_contactangle}) at the contact point instead of the contact angle $\alpha$ itself as argument of the transition function. Firstly, $\cos(\alpha)$ is a smooth and monotonie function 
for $\alpha \in [0;90^{\circ}]$ and $k(\cos(\alpha))$ can therefore model the transition in a similar manner as $k(\alpha)$. Secondly, this way, additional nonlinearities and singularities 
resulting from the $\arccos$-function, which is necessary in order to express $\alpha$ in terms of primary variables, can be avoided.
Mechanically, the products $(1-k(z_c(\alpha_c))\varepsilon_{\perp} \langle g \rangle$ and $k(z(\alpha(\xi))\varepsilon_{\parallel} \langle g(\xi) \rangle$ of the scaling factor, 
the penalty parameter and the gap occurring in \eqref{ABC_forcebased_weakform} can be interpreted as angle-dependent penalty-force laws $f(g,\alpha)$. 
Unfortunately, it can easily be shown that no potential $\Pi(g,\alpha)$ exists for such a force law: If a potential would exist, the integrability condition
\begin{align}
\label{ABC_forcebased_integrability}
  f_{g,\alpha}=-f_{\alpha,g} \quad \text{with} \quad f_g=\frac{\partial \Pi(g,\alpha)}{\partial g}, \quad f_\alpha=\frac{\partial \Pi(g,\alpha)}{\partial \alpha}
\end{align}
would have to be fulfilled by the force law. Since \eqref{ABC_forcebased_weakform} only provides a force component $f_g=f_{c\varepsilon}$, but no force component that is work-conjugated with $\delta \alpha$ 
($f_\alpha \equiv 0$), the integrability condition \eqref{ABC_forcebased_integrability} can not be fulfilled leading to a non-conservative force law. 
In contrary to mechanically motivated non-conservative force laws (e.g. friction forces), the non-conservative nature of \eqref{ABC_forcebased_weakform} has a pure algorithmic reason. Concretely, this means 
that even for a conservative mechanical system, exact energy conservation can not be reached if the contact interaction is modeled via \eqref{ABC_forcebased_weakform}. For that reason, 
we want to propose an alternative, potential-based contact model transition in the next section.

\subsubsection{Potential-based model transition}
\label{sec:ABC_potentialbased}
%

In order to preserve energy conservation, we apply a transition similar to \eqref{ABC_forcebased_transition}, but now on the penalty potential level:
\begin{align}
\label{ABC_potentialbased_potential}
\Pi_{c\varepsilon} = \frac{1}{2}\varepsilon_{\perp} (1-k(z_c)^2) \langle g \rangle ^2 + \frac{1}{2}\varepsilon_{\parallel}\int \limits_0^{l_1} k^2(z(\xi)) \langle g(\xi)\rangle^2 ds_1.
\end{align}
Variation of the combined penalty potential according to~\eqref{ABC_potentialbased_potential} leads to the contact contribution to the weak form:
\begin{align}
\begin{split}
\label{ABC_potentialbased_weakform}
     \delta \Pi_{c\varepsilon} & = \underbrace{\varepsilon_{\perp} \left(1-k(z_c)^2\right) \langle g}_{=-f_{c\varepsilon \perp}} \rangle \,\,\,\, \delta g \quad 
 - \underbrace{\varepsilon_{\perp} \langle g \rangle^2 k(z_c) \dfrac{\partial k(z_c)}{\partial \alpha}}_{=-m_{c \varepsilon \perp}} \,\, \delta  \alpha \\
     & + \int \limits_0^{l_1} 
     \big[ \underbrace{\varepsilon_{\parallel} k^2(z(\xi)) \langle g(\xi)\rangle}_{=-f_{c \varepsilon \parallel}} \,\, \delta g(\xi) 
     + \underbrace{\varepsilon_{\parallel} \langle g(\xi)\rangle^2 k(z(\xi)) \dfrac{\partial k(z(\xi))}{\partial \alpha}}_{=-m_{c \varepsilon \parallel}} \,\,\delta \alpha(\xi)\big]    
ds_1 \,\,.
\end{split}
\end{align}
While the terms on the left represent contact force contributions similar to the ones occurring in \eqref{ABC_forcebased_weakform}, the terms on the right can be identified
as contact moment contributions that are work-conjugated to the variation of the contact angle $\alpha$. These contact moments play the role of the additional contact 
contributions $f_\alpha = -m_{c \varepsilon}$ necessary in order to fulfill the integrability condition \eqref{ABC_forcebased_integrability} and eventually make the force law conservative. 
The only reason why a quadratic transition factor $k(z)^2$ has been applied in \eqref{ABC_potentialbased_potential} instead of a linear one is the derivative $\partial k/\partial \alpha$ 
occurring in \eqref{ABC_potentialbased_weakform}. By this choice, also the transition in the contact moment contributions becomes $C^1$-continuous.
For conservative problems, where exact energy conservation is important, the weak form \eqref{ABC_potentialbased_weakform} has to be preferred. However, for non-conservative problems and/or problems 
where exact energy conservation is only of secondary interest, also the simpler variant according to \eqref{ABC_forcebased_transition} can be applied. This statement can be supported by the following three arguments: Firstly, for a sensible choice of the penalty parameter, the total energy contribution of the penalty forces is often small as compared to the internal elastic or kinetic energy. 
Secondly, the ratio of the contact moments $m_{c \varepsilon}$ to the moment contribution of the contact forces, which are typically in the order of magnitude of $f_{c \varepsilon} \cdot l$, can 
be written as
\begin{align}
\frac{m_{c \varepsilon}}{f_{c \varepsilon} \cdot l} \sim \frac{g}{l}
\end{align}
and is therefore expected to be small, since $g \ll l$ holds for a sensible choice of the penalty parameter. Thirdly, for reasonably balanced penalty parameters $\varepsilon_{\perp}$ and $\varepsilon_{\parallel}$ (see Section \ref{sec:ABC_penaltyadjustment}), the contact moments stemming from a decreasing/increasing 
point-contact potential and the ones stemming from an increasing/decreasing line-contact potential will mutually erase each other up to a certain degree, such that the total moment contribution of point- and line-contact is smaller than the individual contributions. For all these reasons, we will usually apply the force-based formulation~\eqref{ABC_forcebased_weakform} for the applications in Section~\ref{sec:numerical_examples}. For comparison purposes, also the potential-based model will be consulted.\\

\hspace{0.5 cm}
\begin{minipage}{15.0 cm}
\textbf{Remark:} The contact moments occurring in \eqref{ABC_potentialbased_weakform} are not mechanically motivated. They are rather necessary from a mathematical point of view in order to 
enable exact energy conservation within the shifting interval. However, in Section \ref{sec:examples_example1}, it will be shown that in general the pure line-to-line contact formulation already generates 
(mechanically motivated) accumulated contact moments with respect to the closest-point normal vector $\mb{n}$. There, we will see that for a sensibly chosen ratio of 
$\varepsilon_{\perp}$ and $\varepsilon_{\parallel}$ (see Section \ref{sec:ABC_penaltyadjustment}), the model error between the \textit{ABC} formulation and the standard line-to-line contact formulation 
will not be increased in a noticeable manner by these algorithmic contact moments. Furthermore, as already mentioned above, the ratio of these contact moment contributions to moment contributions 
stemming from contact forces or external forces decreases with increasing penalty parameter. Nevertheless, an exact energy conservation can only be guaranteed if these terms are considered.\\
\end{minipage}

The final discrete version of \eqref{ABC_potentialbased_weakform} (or~\ref{ABC_forcebased_weakform}) basically consists of the standard contributions of point-contact (see Section \ref{sec:point}) and line-contact (see Section \ref{sec:line}) as well as the transition factor defined in \eqref{ABC_forcebased_transition} and discretized by \eqref{interpolation}.
The last missing term that has to be formulated in case of a potential-based transition is the variation $\delta k=(\partial k(z)/\partial \alpha) \cdot \delta  \alpha$: In \eqref{ABC_potentialbased_weakform}, the terms on the right have been formulated as variations with respect to $\alpha$ in order to illustrate the moment-character of these contributions. However, it is sensible to slightly reformulate these terms, since 
the transition factor has been formulated as function of $z=\cos \alpha$ and the dependence on the contact angle $\alpha$ is only of implicit nature:
\begin{align}
\label{ABC_potentialbased_momentcontributions1}
  \delta k(z) = \frac{\partial k(z)}{\partial \alpha} \delta  \alpha = \frac{\partial k(z)}{\partial z} \delta  z 
   \quad \text{with} \quad
     \frac{\partial k(z)}{\partial z}= \left\{\begin{array}{lll}
                                     0, & \alpha < \alpha_1 \\
                                     \frac{\pi}{2(z_1-z_2)} \sin \left(\pi\frac{z-z_2}{z_1-z_2}\right), & \alpha_2 \geq \alpha \geq \alpha_1 \\
                                     0, & \alpha > \alpha_2.
         \end{array}\right. .
\end{align}
Variation of the term $z(\mb{r}_1(\xi),\mb{r}_2(\eta))$ according to \eqref{ABC_forcebased_transition} leads, after some reformulations, to the following expression: 
\begin{align}
\begin{split}
\label{ABC_potentialbased_momentcontributions2}
   \delta  z & = \left[ \mb{v}_1^T \left(\frac{d \mb{r}_1^{\shortmid}}{d \mb{d}_{12}} \right) + \mb{v}_2^T \left(\frac{d \mb{r}_2^{\shortmid}}{d \mb{d}_{12}}\right) \right] \delta \mb{d}_{12} \\
   \text{with} \quad \mb{v}_1^T & = \frac{\bar{\mb{r}}_{2}^{\shortmid T}}{||\mb{r}_{1}^{\shortmid}||}\left[ \mb{I}_{3 \times 3} - \bar{\mb{r}}_{1}^{\shortmid} \otimes \bar{\mb{r}}_{1}^{\shortmid T} \right],
   \quad \mb{v}_2^T = \frac{\bar{\mb{r}}_{1}^{\shortmid T}}{||\mb{r}_{2}^{\shortmid}||}\left[ \mb{I}_{3 \times 3} - \bar{\mb{r}}_{2}^{\shortmid} \otimes \bar{\mb{r}}_{2}^{\shortmid T} \right],
   \quad \bar{\mb{r}}_{1}^{\shortmid} = \frac{\mb{r}_{1}^{\shortmid}}{||\mb{r}_{1}^{\shortmid}||},
   \quad \bar{\mb{r}}_{2}^{\shortmid} = \frac{\mb{r}_{2}^{\shortmid}}{||\mb{r}_{2}^{\shortmid}||} \\
    \text{and} \quad \frac{d \mb{r}_1^{\shortmid}}{d \mb{d}_{12}} &= \frac{\partial \mb{r}_1^{\shortmid}}{\partial \mb{d}_{12}} + \frac{\partial \mb{r}_1^{\shortmid}}{\partial \xi}\frac{d \xi}{d \mb{d}_{12}}
    =\left[ \left(\mb{N}_{1}^{\shortmid},\mb{0} \right) + \mb{r}_{1}^{\shortmid \shortmid} \frac{d \xi}{d \mb{d}_{12}} \right],
    \quad \frac{d \mb{r}_2^{\shortmid}}{d \mb{d}_{12}}= \frac{\partial \mb{r}_1^{\shortmid}}{\partial \mb{d}_{12}} + \frac{\partial \mb{r}_2^{\shortmid}}{\partial \eta}\frac{d \eta}{d \mb{d}_{12}}
    =\left[ \left(\mb{0},\mb{N}_{2}^{\shortmid} \right) + \mb{r}_{2}^{\shortmid \shortmid} \frac{d \eta}{d \mb{d}_{12}} \right],
\end{split}
\end{align}
where $\mb{I}_{3 \times 3}$ denotes the $3 \times 3$ unity matrix. In the derivation of \eqref{ABC_potentialbased_momentcontributions2}, we have already inserted the spatial discretization according to~\eqref{interpolation} and we have again solely considered the contribution of two beam elements with nodal degrees of freedom $\mb{d}_{12}=(\mb{d}_{1}^T,\mb{d}_{2}^T)^T$. The definition of the transition factor according to \eqref{ABC_forcebased_transition} and its variation according to ~\eqref{ABC_potentialbased_momentcontributions1} and \eqref{ABC_potentialbased_momentcontributions2} is valid for the point-contact and for the line-contact contribution. However, in the point-contact contribution all terms have to be evaluated at the closest point pair $\xi_c$ and $\eta_c$, while in the line-contact contribution all terms have to be evaluated at the Gauss point coordinates $\xi_{ij}$ (see also Section \ref{sec:line}) and the corresponding closest master points $\eta_c(\xi_{ij})$. In contrary to the gap function variation $\delta g$, the derivation of $\delta  z$ also requires the variations of the contact point coordinates $\xi$ and $\eta$ for a variationally consistent formulation of the weak form. The corresponding derivatives $d \xi / d \mb{d}_{12}$ and $d \eta / d \mb{d}_{12}$ of these coordinates for the cases of point-to-point and line-to-line contact are summarized in \ref{anhang:linearizationendpoint} and \ref{anhang:linearizationlinecontact}. Furthermore, the basic steps in order to derive the final residual and linearization contributions of the force-based and potential-based \textit{ABC} formulation are summarized in \ref{anhang:linearizationABC}.

%
\subsection{Choice of shifting angles}
\label{sec:ABC_choiceshiftingangles}
%

A sensible choice of the shifting angles $\alpha_1$ and $\alpha_2$ is crucial for a robust and efficient contact algorithm based on the presented \textit{ABC} formulation. For a given 
upper bound $\mu_{max}$ describing the maximal curvature admissible or expected for the considered example, \eqref{limitationspoint_requirementsecondderiv8} yields a lower bound 
for the shifting angle $\alpha_1$ above which a unique bilateral closest point solution necessary for the point-to-point contact formulation can be guaranteed. Thus, we choose $\alpha_1$ as
\begin{align}
\label{ABC_choiceshiftingangles}
     \alpha_1 = k_{\alpha_1} \arccos \left(1 - 2 \mu_{max} \right).
\end{align}
Here, $k_{\alpha_1}>1$ represents an additional safety factor. Next, the second shifting angle $\alpha_2$ has to be chosen. For efficiency reasons, this angle should be as small as possible.
However, the shifting interval should be large enough in order to ensure a model transition that is sufficiently smooth from a mechanical as well as from a numerical point of view.
Since the point-contact and the line-contact model approach each other with increasing penalty parameters, also the shifting interval can be chosen tighter with increasing penalty parameters. 
In our simulations, we have typically applied shifting intervals in the range of $\alpha_2-\alpha_1 \approx 1^{\circ} ... \, 5^{\circ}$. Having $\alpha_2$ prescribed, we can determine a 
lower bound for the required number of Gauss points of the line contact formulation by inserting $\alpha_{max}=\alpha_2$ into \eqref{limitationsline_GPdistance2}.

%
\subsection{Adjustment of point-to-point and line-to-line penalty parameters}
\label{sec:ABC_penaltyadjustment}
%
So far, $\varepsilon_{\perp}$ and $\varepsilon_{\parallel}$ represent two independent system parameters. In this section, we want to derive an optimal ratio of these two parameters, such that 
only one penalty parameter has to be user-defined while the second one can be determined automatically. Our criterion for this optimal choice is the minimization of the work contribution of the 
algorithmic contact moment contributions $m_{c \varepsilon \perp}$ and $m_{c \varepsilon \parallel}$. This approach is not only advantageous for the potential-based \textit{ABC} formulation: As a consequence 
of this choice, the difference between the work contributions of the force-based and the potential-based formulation, which represents the non-conservative work contributions of the force-based 
formulation, will be minimized. In the following, we do not search for the exact solution of the corresponding minimization problem, but rather for an approximate solution based on some 
simplifying assumptions. Thereto, we express the curves $\mb{r}_1(\xi)$ and $\mb{r}_2(\eta)$ as linear Taylor expansions with respect to the closest point $(\xi_c, \eta_c)$, thus leading 
to a constant contact angle within the line-to-line contact region. In this case, we can simplify the $m_{c \varepsilon \parallel}$-term:
\begin{align}
\varepsilon_{\parallel} \! \int \limits_0^{l_1} \! \langle g(\xi)\rangle^2 k(z(\xi)) \dfrac{\partial k(z(\xi))}{\partial \alpha} \delta \alpha(\xi) ds_1
\!=\varepsilon_{\parallel} \!\!\! \int \limits_{s_{1,c} - \Delta s_{\parallel}}^{s_{1,c} + \Delta s_{\parallel}} \!\!\! g(\xi)^2 k(z(\xi)) \dfrac{\partial k(z(\xi))}{\partial \alpha} \delta \alpha(\xi) ds_1 
\! \approx \varepsilon_{\parallel} \!\!\! \int \limits_{s_{1,c} - \Delta s_{\parallel}}^{s_{1,c} + \Delta s_{\parallel}} \!\!\! g_l(\xi)^2 ds_1 \cdot k(z_c) \dfrac{\partial k(z_c)}{\partial \alpha} \delta \alpha_c.
\end{align}
Here, $g_l(\xi)$ denotes the gap function based on linearly approximated beam geometries. This approximation is valid, since for a sensible choice of the penalty parameter and the shifting interval 
(see Section \ref{sec:ABC_choiceshiftingangles}) the length~$2 \Delta s_{\parallel}$ of the domain with active contact forces ($g < 0$) is small compared to the beam length. 
Furthermore, this length decreases with increasing penalty parameter. Next, we require that the contact moment work contributions balance each other:
\begin{align}
\label{ABC_penaltyadjustment1}
\varepsilon_{\parallel} \!\!\! \int \limits_{s_{1,c} - \Delta s_{\parallel}}^{s_{1,c} + \Delta s_{\parallel}} \!\!\! g_l(\xi)^2 ds_1 \cdot k(z_c) \dfrac{\partial k(z_c)}{\partial \alpha} \delta \alpha_c 
= \varepsilon_{\perp} g^2 k(z_c) \dfrac{\partial k(z_c)}{\partial \alpha} \delta  \alpha_c
\quad \rightarrow \quad
\varepsilon_{\perp} = \frac{\varepsilon_{\parallel}}{g^2} \int \limits_{s_{1,c} - \Delta s_{\parallel}}^{s_{1,c} + \Delta s_{\parallel}} \!\!\! g_l(\xi)^2 ds_1.
\end{align}
In general, equation \eqref{ABC_penaltyadjustment1} based on a linearly approximated geometry can not be fulfilled for arbitrary contact angles and gaps by one constant penalty parameter 
$\varepsilon_{\perp}$. Therefore, we only require that \eqref{ABC_penaltyadjustment1} is exactly fulfilled for the minimal admissible gap $g_{min}$ (going along with maximal contact work) 
and at the mean shifting angle $\bar{\alpha}_{12}:=(\alpha_1+\alpha_2)/2$:
\begin{align}
\label{ABC_penaltyadjustment2}
\frac{\varepsilon_{\perp}}{\varepsilon_{\parallel}} = \frac{1}{ g_{min}^2} \int \limits_{s_{1,c} - \Delta s_{\parallel}}^{s_{1,c} + \Delta s_{\parallel}} \!\!\! 
g_l(g_{min}, \bar{\alpha}_{12}, \xi)^2 ds_1 = \frac{\tilde{\Pi}_{c\varepsilon \parallel}(g_{min}, \bar{\alpha}_{12})}{\tilde{\Pi}_{c\varepsilon \perp}(g_{min})}.
\end{align}
Since we are not interested in an exact solution of \eqref{ABC_penaltyadjustment1}, but rather in an approximation providing at least a sensible order of magnitude, a comparatively rough bound for 
the maximal gap $g_{min}$ is sufficient. If the specific application prescribes some tolerable bound $g_{min}$, this value can be taken. With an implementation of the proposed \textit{ABC} formulation at hand, 
it is a simple task to perform a pre-processing step, where the normalized penalty potential $\tilde{\Pi}_{c \varepsilon \perp}(g_{min})$ of the point-contact and 
$\tilde{\Pi}_{c \varepsilon \parallel}(g_{min}, \bar{\alpha}_{12})$ 
of the line-contact (see Section~\ref{sec:algorithmicaspects_penaltylaws} for the definition of normalized penalty potentials and different penalty force laws) resulting from the contact interaction of two straight 
beams characterized by an enclosed angle $\bar{\alpha}_{12}$ and a gap $g_{min}$ are calculated. 
With these energies and a given penalty parameter $\varepsilon_{\parallel}$, the corresponding value of $\varepsilon_{\perp}$ can be calculated according to \eqref{ABC_penaltyadjustment2}. 
This procedure is valid for arbitrary penalty force/potential laws that depend linearly on the penalty parameter, such as for example the laws \eqref{penlaw_lin} and \eqref{penlaw_linposquad} (see Section~\ref{sec:algorithmicaspects_penaltylaws}). In case of a linear penalty law according to \eqref{penlaw_lin}, the integral appearing in \eqref{ABC_penaltyadjustment2} can also be approximated analytically. Assuming the most conservative estimation for the minimal gap, i.e. $g_{min}=-2R$ (intersection of the beam centerlines), and approximating the evolution of $g_l(\xi)$ as piecewise linear function in $\xi$ yields the following relation:
\begin{align}
\label{ABC_penaltyadjustment3}
\frac{\varepsilon_{\perp}}{\varepsilon_{\parallel}} \approx \frac{2}{ g_{min}^2} \! \int \limits_{s_{1,c}}^{s_{1,c} + \Delta s_{\parallel}} \!\!\! 
\left[-g_{min}\left(1-\frac{s_1-s_{1,c}}{\Delta s_{\parallel}}\right)\right]^2 ds_1 = \frac{2\Delta s_{\parallel}}{3}.
\end{align}
Additionally using the relation \eqref{limitationsline_GPdistance1b} in case of intersecting beam centerlines ($g_{min}=-2R \rightarrow g_{n,min}=-2$) in order to determine the integration length 
as $2\Delta s_{\parallel}=4R/ \sin{(\bar{\alpha}_{12}})$, we end up with the following simple expression:
\begin{align}
\label{ABC_penaltyadjustment4}
\frac{\varepsilon_{\perp}}{\varepsilon_{\parallel}} \approx \frac{4R}{3\sin{( \bar{\alpha}_{12})}}.
\end{align}
A similar analytical approximation could also be derived for the quadratically regularized penalty force law introduced in Section~\ref{sec:algorithmicaspects_penaltylaws}. However, since in practical simulations 
the relation $\bar{g} \ll R$ often holds (see again Section~\ref{sec:algorithmicaspects_penaltylaws} for the definition of $\bar{g}$), \eqref{ABC_penaltyadjustment4} can also be applied as approximation for this force law. In Section \ref{sec:numerical_examples}, it will be shown that in many cases a choice of 
$\varepsilon_{\perp}$ according to \eqref{ABC_penaltyadjustment4} is sufficient in order to reduce the non-conservative work contribution of the force-based formulation down to a practically 
tolerable level. Nevertheless, this behavior can be further improved by determining $\varepsilon_{\perp}$ via numerical integration of \eqref{ABC_penaltyadjustment2} based on the actually applied 
penalty law and a better estimation for the minimal gap $g_{min}$. With this suggestion at hand, only the line-to-line penalty parameter $\varepsilon_{\parallel}$ has to be prescribed 
independently. For the determination of $\varepsilon_{\parallel}$, the value of the cross section stiffness can serve as a starting point.\\

\hspace{0.5 cm}
\begin{minipage}{15.0 cm}
\textbf{Remark:} In \eqref{ABC_forcebased_weakform}, it seemed to be natural to introduce a scaling factor $\left[1-k(z)\right]$ for the point-to-point penalty parameter that is complementary to 
the scaling factor $k(z)$ of the line-to-line penalty parameter. With the knowledge of the last subsections, one could imagine an alternative description of the angle-dependent evolution of 
the point-to-point penalty parameter: Thereto, one could apply the force-based \textit{ABC} formulation, yet, not based on the applied $\left[1-k(z)\right]$-transition law in combination with a constant 
penalty parameter, but rather based on an individual point-to-point penalty parameter $\varepsilon_{\perp,i}$ at every contact point as additional unknown. In order to determine this additional unknown, 
one could formulate conditions similar to \eqref{ABC_penaltyadjustment1} for each individual contact point, such that the resulting evolution of $\varepsilon_{\perp,i}$ allows for exact energy 
conservation without the need for algorithmic contact moments as appearing in the potential-based \textit{ABC} formulation. However, in Section~\ref{sec:numerical_examples} it will be verified numerically that the error 
in the energy conservation of the simple force-based \textit{ABC} formulation in combination with a sensible, constant choice of $\varepsilon_{\perp}$ (see e.g. \eqref{ABC_penaltyadjustment2}) 
is in many cases negligible from a practical point of view. Thus, the numerical effort resulting from a formulation with additional unknowns $\varepsilon_{\perp,i}$ does not seem to be justified.
\end{minipage}

\subsection{Conservation properties}
\label{sec:ABC_conservation}
%

In \ref{anhang:conservation}, a rigorous analytical investigation of the spatially discretized variant of the proposed contact formulation concerning conservation of linear momentum, angular momentum 
and energy is performed. First, it is shown that these conservation properties are already fulfilled for the applied beam element formulation. 
Then, the additional contributions at the contact interfaces are considered. Concretely, the proposed \textit{ABC} formulation in combination with a
potential-based model transition according to \eqref{ABC_potentialbased_weakform} will be investigated, since it contains the variants ''force-based transition`` 
(neglect of the contact moment contributions and replacement of $k(z)$ by $k^2(z)$), ''pure point contact`` ($k(z)\equiv0$) and ''pure line contact`` ($k(z)\equiv1$) as special cases.
It will be proven that conservation of linear momentum, angular momentum and energy is preserved by the proposed finite element discretization. In Section~\ref{sec:examples_example2}, this statement 
will also be verified numerically. The question of preservation of these properties in the temporally discretized problem setting depends on the applied time integration scheme 
(see e.g. \cite{gonzalez2000}, \cite{kuhl1999} or \cite{simo1992}) and is not content of this work.

%
\section{Algorithmic aspects}
\label{sec:algorithmicaspects}
%
In the following Sections~\ref{sec:algorithmicaspects_search} and \ref{sec:algorithmicaspects_stepsizecontrol}, further information concerning the employed contact search algorithm and a step size control applied to the iterative displacement increments within the nonlinear solution scheme will be given. The latter method enables displacements per time step that are larger than the beam cross section radius, which is the typical time step size limitation of standard beam-to-beam contact algorithms. In Section~\ref{sec:algorithmicaspects_penaltylaws}, different penalty force laws are presented before the contact contributions of the beam endpoints are considered in Section~\ref{sec:algorithmicaspects_endpoint}.
%
\subsection{Contact Search Algorithm}
\label{sec:algorithmicaspects_search}
\begin{figure}[t]
 \centering
   \subfigure[Straight elements.]
   {
    \includegraphics[width=0.14\textwidth]{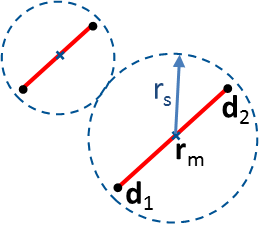}
    \label{fig:alorithmicaspects_searchbox1}
   }
   \hspace{0.04\textwidth}
   \subfigure[Curved elements.]
   {
    \includegraphics[width=0.14\textwidth]{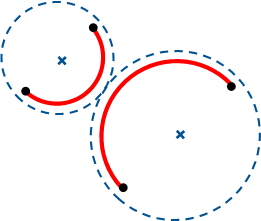}
    \label{fig:alorithmicaspects_searchbox2}
   }
   \hspace{0.02\textwidth}
   \subfigure[Subdivision of each finite element into $n_{seg}$ search segments until $\beta_{jl},\beta_{jr} < \beta_{max}$.]
   {
    \includegraphics[width=0.6\textwidth]{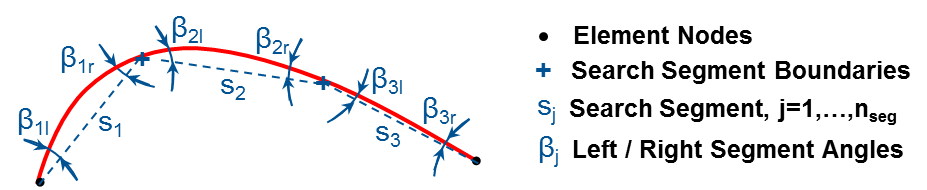}
    \label{fig:alorithmicaspects_secondsearchstep1}
   }
  \caption{Two-step search algorithm: 1) Octree search with spherical search boxes (\ref{fig:alorithmicaspects_searchbox1} and \ref{fig:alorithmicaspects_searchbox2}); 2) Subdivision in search segments (\ref{fig:alorithmicaspects_secondsearchstep1}).}
  \label{fig:alorithmicaspects_searchbox}
\end{figure}

%
The contact search algorithm combined with the proposed \textit{ABC} formulation consists of two search steps. The first step represents an element-based octree search and yields pairs of close-by 
finite elements (located on two different or in case of self-contact also on the same physical beam) that might potentially come into contact. The octree search is based on the assumption that 
(the 2D-projection of) the maximal deformation 
of an initially straight beam segment discretized by one third-order finite element according to Section~\ref{sec:beamformulation} does not exceed a half-circular shape. With this restriction in mind, 
we base the octree search on an intersection of spherical bounding boxes defined by
\begin{align}
\mb{r}_m=\frac{\mb{\hat{d}}^1+\mb{\hat{d}}^2}{2}, \quad r_s= (1+k_{r_s})\frac{||\mb{\hat{d}}^2-\mb{\hat{d}}^1||}{2}
\end{align}
and illustrated in $2$ dimensions in Figure \ref{fig:alorithmicaspects_searchbox}. Here, the introduced parameter $k_{r_s}$ represents an additional safety factor. While the spherical search box appears as a comparatively loose hull for straight beam elements (see Figure~\ref{fig:alorithmicaspects_searchbox1}) its application seems to be justified when considering strongly deformed beam elements as illustrated 
in Figure~\ref{fig:alorithmicaspects_searchbox2}. In order to further limit the size and number of beam segments where (a large number of) unilateral closest-point projections 
have to be evaluated, we apply a second search step. Thereto, we subdivide each beam element of the pairs found in the first step into $n_{seg}$ equidistant sub-segments as illustrated 
in Figure~\ref{fig:alorithmicaspects_secondsearchstep1}. Thereby, the number $n_{seg}$ of sub-segments is doubled until the angles $\beta_{jl}$ and $\beta_{jr}$ between the real geometry and the straight search segment at the left and right end of the 
search segments are smaller than a prescribed value $\beta_{max}$, i.e. $\beta_{jl},\beta_{jr} < \beta_{max} \quad \text{for} \quad j=1,...,n_{seg}$. To fulfill this criterion for arbitrary configurations, 
the number $n_{seg}$ is adapted dynamically in every Newton step.\\

\begin{figure}[t]
 \centering
   \subfigure[Creation of segment-wise conical search boxes.]
   {
    \includegraphics[width=0.42\textwidth]{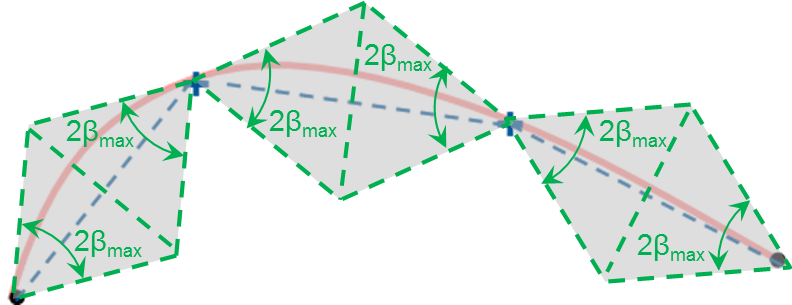}
    \label{fig:alorithmicaspects_secondsearchstep2}
   }
   \hspace{0.1\textwidth}
   \subfigure[Intersection of enveloping cylindrical boxes.]
   {
    \includegraphics[width=0.35\textwidth]{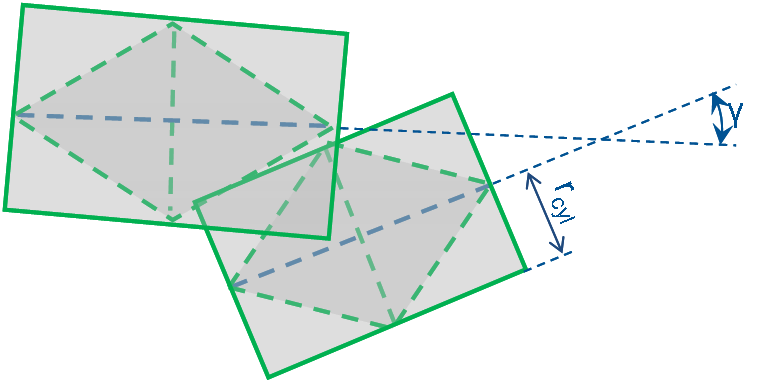}
    \label{fig:alorithmicaspects_secondsearchstep3}
   }
  \caption{Creation of segment-wise search boxes, intersection and determination of intersection angle $\gamma$.}
  \label{fig:alorithmicaspects_consearchbox}
\end{figure}

Next, we assume that for the search segments built by this procedure, the actual centerline geometry is completely enwrapped by a double cone with cone angle $2 \beta_{max}$ (see Figure~\ref{fig:alorithmicaspects_secondsearchstep2}). It seems to be rather intuitive when looking at Figure~\ref{fig:alorithmicaspects_secondsearchstep2} that for the applied third-order Hermite polynomials this assumption is justified. For the examples considered in Section~\ref{sec:numerical_examples}, this statement has also been verified numerically. Since the direct intersection of these double-cones is geometrically quite involved, we replace them by enwrapping cylinders (see Figure~\ref{fig:alorithmicaspects_secondsearchstep3}) with radius
\begin{align}
\label{search_cylinder_radius}
r_{cyl}= k_{cyl} \cdot \tan{(\beta_{max})} \cdot \frac{l_{seg}}{2} \quad  \text{with} \quad l_{seg}=\frac{l_{0}}{n_{seg}}.
\end{align}
Here, $l_{0}$ is the initial element length and $k_{cyl}$ a safety factor. After the geometrically close-by segment pairs have been determined via intersection of these cylindrical 
bounding boxes (which can be done very efficiently by an analytical CPP between the straight cylinder axes and a subsequent check of the segment endpoints), 
the intersection angle $\gamma$ can be determined, which enables an estimation of the potential contact angles possible for this segment pair, i.e. $\alpha \!\in\! [\gamma\!-\!2\beta_{max};\gamma\!+\!2\beta_{max}]$. 
Consequently, from the set of close segment pairs found in the second search step, we can create a subset of potential point contact segment pairs for all pairs satisfying 
$\gamma\!>\!\alpha_1\!-\!2\beta_{max}$ and a subset of potential line contact segments for all pairs satisfying $\gamma\!<\!\alpha_2\!+\!2\beta_{max}$. Considerable advantages result from this second search 
step: First, compared to the element-wise spherical bounding boxes, the tighter segment-wise cylindrical bounding boxes deliver a smaller set of potential contact pairs for which the closest point projections 
have to be solved iteratively. Secondly, the subdivision into potential point and line contact pairs again reduces the number of (expensive) unilateral closest point projections necessary at the slave beam Gauss points for the line contact formulation. This means that the computational savings of the \textit{ABC} formulation are twofold: On the one hand, the required Gauss point density can be reduced as compared to the pure line contact formulation, and on the other hand, the number of beam segments evaluated by the (unilateral CPP of the) line formulation is reduced to those enclosing small contact angles, while segments with larger contact angles are evaluated by the (bilateral CPP of the) cheaper point contact formulation. The third advantage resulting from the proposed two-stage contact search is related to the iterative solution of the nonlinear orthogonality conditions~\eqref{point_orthocond} 
for the closest points $\xi_c$ and $\eta_c$. Since the solution points $\xi_c$ and $\eta_c$ are necessary in order to determine the contact status (active or inactive) of a close-by segment pair, 
a robust and reliable solution scheme is mandatory, but not trivial to provide when considering beam elements of arbitrary interpolation order. The choice of the shifting angle $\alpha_1$ according to~\eqref{ABC_choiceshiftingangles} guarantees that a unique closest point solution exists for the segment pairs relevant for point-to-point contact, i.e. for pairs with $\alpha>\alpha_1$. 
On the one hand, the contact search also delivers a certain amount of potential point contact pairs for which the actual contact angle at the (a priori unknown) closest point is smaller than $\alpha_1$ and, consequently, 
for which no unique closest point solution can be guaranteed. On the other hand, the number of 
such segment pairs with $\alpha<\alpha_1$ is reduced drastically since the potential point contact segment pairs are already filtered on the basis of $\gamma>\alpha_1-2\beta_{max}$. By this means and as consequence of the good starting points (the search segment midpoints), 
the number of unconverged local Newton loops could be reduced drastically. Exemplarily, the example of Section~\ref{sec:examples_biopolymer}, i.e. a very complex contact scenario with more than $100$ independent point contact regions per time step, leads to less than one unconverged local Newton loop per $10$ time steps.\\


In a second step, it has to be checked that these potential contact pairs with unconverged bilateral closest point solution are indeed not relevant in terms of active contact force contributions. 
With the proposed \textit{ABC} formulation such a check basically comes at zero extra effort. Since the closest point projections of potential contact pairs lying in the transition range, i.e. $\gamma \in [\alpha_1-2\beta_{max}; \alpha_2+2\beta_{max}]$, 
are performed by both the point and the line contact formulation, one can use the results of the 
unilateral closest point projection associated with the line contact formulation in order to estimate the closest points $\xi_c$ and $\eta_c$ as well as the gap $g$ and 
the contact angle $\alpha_c$ at this location for pairs with unconverged bilateral closest point projection of the point contact formulation. In the cases $g>0$ or $\alpha<\alpha_1$, the corresponding unconverged pair is not relevant for point contact and the simulation can proceed. This procedure has been sufficient for all \textit{relevant} bilateral closest point projections of the examples in Section~\ref{sec:numerical_examples} 
to converge. However, in case a relevant bilateral closest point projection is not convergent, one could alternatively apply the estimation of the closest points $\xi_c$ and $\eta_c$ based on the unilateral closest point projection instead of the exact bilateral CPP solution. The strategies described above rely on convergent unilateral closest point projections. In \cite{meier2015b}, it has been shown that the solvability of this projection can be guaranteed for the entire range of possible contact angles. This prediction could be confirmed numerically: In combination with the applied two-stage contact search algorithm, all unilateral closest point projections carried out within the scope of examples considered in Section~\ref{sec:numerical_examples} have been convergent.\\

\begin{figure}[t]
  \centering
  \includegraphics[width=0.65\textwidth]{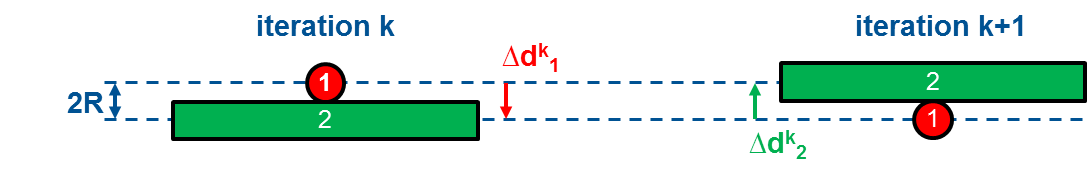}
  \caption{Undetected crossing of two beams as a consequence of displacement increments being too large.}
  \label{fig:stepsizecontrol}
\end{figure}

%
\subsection{Step Size Control}
\label{sec:algorithmicaspects_stepsizecontrol}
%

In Section \ref{sec:limitationsline}, it has already been shown that an increasing beam slenderness ratio requires an increasingly fine spatial ''contact discretization`` in the sense of a higher Gauss point density necessary for the line-to-line contact formulation. In this section, it will be shown that for standard beam-to-beam contact formulations also the maximal permissible time step size $\Delta t$ decreases with increasing beam slenderness ratio. Thereto, we consider two perpendicular beams as illustrated in Figure \ref{fig:stepsizecontrol}. If the norm of the iterative displacement increments $\Delta \mb{d}_1^k$ of beam 1 and $\Delta \mb{d}_2^k$ of beam 2 in the $k^{th}$ Newton iteration of a time step is larger than the
cross section diameter $2R$, the beams can cross completely without remaining penetration and therefore without contact being detected (see Figure~\ref{fig:stepsizecontrol}). Already for displacement norms in the range of the cross section radius $R$, the beam centerlines can cross, which results in a change of direction of the contact forces and, in turn, in an undetected crossing of the beams. To avoid such scenarios, we measure the inf-norm of the global iterative displacement increment vector $\Delta \mb{D}^k$ and scale it according to:
\begin{align}
\label{aspects_algorithm}
\text{while} \, (\,||\Delta \mb{D}^k||_{\infty}>R\,) \,\, \{\, \Delta \mb{D}^k= 0.5 \cdot \Delta \mb{D}^k \, \}.
\end{align}
The modified Newton scheme resulting from this algorithm does not only prevent undetected beam crossing, but it also enhances the robustness of the nonlinear solution process in general. This does 
especially apply to examples with strongly fluctuating external loads and high peak forces (see the example of Section \ref{sec:examples_biopolymer}) or abruptly opening contacts 
(e.g. when a beam slides across the end of a second beam; see also the example of Section \ref{sec:examples_rope}). In addition to algorithm \eqref{aspects_algorithm}, we check the following 
criterion in order to control the maximal penetration of the contacting beams:
\begin{align}
\label{aspects_criteria}
g > - k \cdot R \quad \text{with \quad} k \in [0;1]. 
\end{align}
In combination, criteria \eqref{aspects_algorithm} and \eqref{aspects_criteria} ensure that two beams can not cross each other without contact detection. The standard alternative to the procedure proposed in this section is to simply choose the time step size small enough, such that the inf-norm of the displacement increment per time step is smaller than the cross section radius, i.e.
\begin{align}
\label{aspects_alternativealgorithm}
||\mb{D}(t_{i})-\mb{D}(t_{i-1})||_{\infty}<R.
\end{align}
In general, such a procedure leads to a higher number of total Newton iterations, because convergence is required for every displacement step of size $R$ corresponding to one time step, while in case of algorithm \eqref{aspects_algorithm} some successive Newton iterations with (confined) displacement step size $R$ can take place before the converged solution of the considered time step is found. In Section~\ref{sec:examples_biopolymer}, this statement and the resulting efficiency gains will be confirmed.

%
\subsection{Penalty Laws}
\label{sec:algorithmicaspects_penaltylaws}
\begin{figure}[t]
 \centering
  \subfigure[Standard linear penalty law.]
   {
    \includegraphics[width=0.35\textwidth]{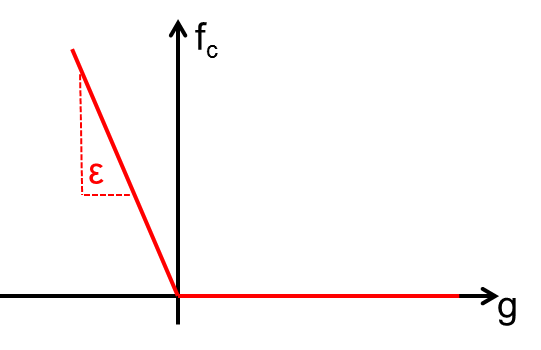}
    \label{fig:penlaw_lin}
   }
   \hspace{0.05 \textwidth}
   \subfigure[Linear penalty law with quadratic regularization.]
   {
    \includegraphics[width=0.35\textwidth]{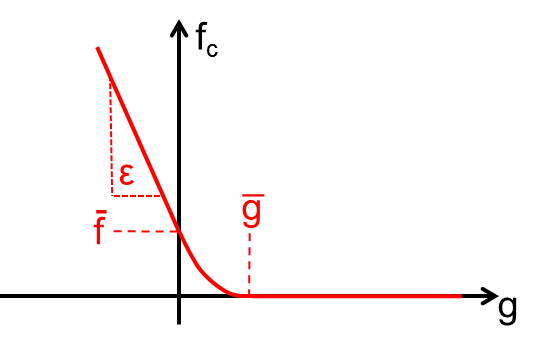}
    \label{fig:penlaw_linposquad}
   }
  \caption{Graphical visualization of standard and quadratically regularized penalty law.}
  \label{fig:penlaws}
\end{figure}

%
Up to now, we have considered the following linear penalty law as introduced in \eqref{line_pen_contactforce} and illustrated in Figure \ref{fig:penlaw_lin}:
\begin{align}
\label{penlaw_lin}
  f_{c\varepsilon}(g)=\left\{\begin{array}{ll}
			-\varepsilon g, \\
			0, \\
	  \end{array}\right. \quad \text{and} \quad
  \Pi_{c\varepsilon}(g)=\left\{\begin{array}{ll}
			\frac{\varepsilon}{2} g^2, & g \leq 0 \\
			0, & g > 0 \\
	  \end{array}\right..	  
\end{align}
In practical simulations, one often uses regularized penalty laws that allow for a smooth contact force 
transition (see Figure \ref{fig:penlaw_linposquad}). 
The quadratically regularized penalty law applied within this work has the following representation:
\begin{align}
\label{penlaw_linposquad}
  f_{c\varepsilon}(g)=\left\{\begin{array}{lll}
			\bar{f} - \varepsilon g,\\
			\frac{\varepsilon \bar{g} - \bar{f}}{\bar{g}^2} g^2 -\varepsilon g + \bar{f},\\
			0,\\
	  \end{array}\right. \,\,
	    \Pi_{c\varepsilon}(g)=\left\{\begin{array}{lll}
			\frac{\varepsilon}{2}g^2 - \bar{f} g + \frac{\varepsilon \bar{g}^2}{6.0}, & g \leq 0 \\
			-\frac{\varepsilon \bar{g} - \bar{f}}{3\bar{g}^2}g^3 + \frac{\varepsilon}{2}g^2 - \bar{f} g 
			+ \frac{\varepsilon \bar{g}^2}{6.0}, & 0 < g \leq \bar{g} \\
			0, & g > \bar{g} \\
	  \end{array}\right.
	  \quad \text{with} \quad \bar{f}=\frac{\varepsilon \bar{g}}{2}
\end{align}
For each force law an appropriate potential defined by $f_{c\varepsilon}= - \partial \Pi_{c\varepsilon} / \partial g$ and $\Pi_{c\varepsilon}(g=\bar{g})=0$ (with $\bar{g}=0$ for the linear force law~\eqref{penlaw_lin}) as well 
as a normalized potential $\tilde{\Pi}_{c\varepsilon}=\Pi_{c\varepsilon}/\varepsilon$ (see also Section~\ref{sec:ABC_penaltyadjustment}) can be derived. For simplicity, all derivations in the previous sections are based on a linear penalty law according to \eqref{penlaw_lin}. However, a more general form of these equations that is valid for arbitrary penalty laws can easily be derived by simply replacing all linear force-like expressions of the form $ -\varepsilon \langle g\rangle $ by $f_{c\varepsilon}(g)$ and all quadratic potential-like expressions of the form $ 0.5 \varepsilon \langle g \rangle^2 $ by $\Pi_{c\varepsilon}(g)$.\\

%
\subsection{Endpoint-to-line and endpoint-to-endpoint contact contributions}
\label{sec:algorithmicaspects_endpoint}
%
The contact formulations presented in the last sections have only considered solutions of the minimal distance problem within the element parameter domain $\xi,\eta \in [-1;1]$. However, a minimal distance solution can also occur in form of a boundary minimum at the physical endpoints of the contacting beams. Neglecting these boundary minima can lead to impermissibly large penetrations and even to an entirely undetected crossing of the beams. In \cite{meier2015b}, it has been shown that neglecting these contributions does not only lead to an inconsistency of the mechanical model itself, but also to a drastically reduced robustness of the nonlinear solution scheme, since initially undetected large penetrations can lead to considerable jumps in the contact forces during the iterations of a nonlinear solution scheme. In the numerical examples presented in Section~\ref{sec:numerical_examples}, these endpoint contact contributions will be considered in an identical manner as already derived in \cite{meier2015b}. The required residual and stiffness contributions are summarized in~\ref{anhang:linearizationendpoint}.

%
\section{Numerical examples}
\label{sec:numerical_examples}
%
The first two examples of this section aim at investigating the accuracy and consistency of the proposed \textit{ABC} formulation. The first example focuses on the contact force evolutions in the model transition range, while the second example verifies the conservation properties already shown theoretically in~\ref{anhang:conservation}. Finally, we want to verify the robustness and efficiency of the proposed contact algorithm when applied to practically relevant applications. Thereto, we employ the force-based \textit{ABC} formulation in combination with the quadratically regularized force law of Section~\ref{sec:algorithmicaspects_penaltylaws}, the endpoint contact according to Section~\ref{sec:algorithmicaspects_endpoint}, the search algorithm presented in Section~\ref{sec:algorithmicaspects_search} and the step size control as introduced in Section~\ref{sec:algorithmicaspects_stepsizecontrol}. Two applications are chosen in order to represent complex beam-to-beam contact interaction involving high slenderness ratios and arbitrary 
beam orientations. For all examples, a Newton-Raphson scheme is applied in order to solve the nonlinear system of equations $\mb{R}_{tot}$ as defined in~\eqref{global_system}. As convergence criteria, we check the Euclidean norms of the displacement increment vector $\Delta \mb{D}^k$ and of the residual vector $\mb{R}^k_{tot}$ at Newton iteration $k$. For convergence, these norms have to fall below prescribed tolerances $\delta_{\mb{R}}$ and $\delta_{\mb{D}}$, i.e. $||\mb{R}^k_{tot}||<\delta_{\mb{R}}$ and $||\Delta \mb{D}^k||<\delta_{\mb{D}}$. By default, these tolerances are chosen according to the following values: $\delta_{\mb{R}}=\delta_{\mb{D}}=1.0 \cdot 10^{-7}$.

%
\subsection{Example 1: Beam rotating on arc}
\label{sec:examples_example1}
%

\begin{figure}[t]
 \centering
 \includegraphics[width=0.4\textwidth]{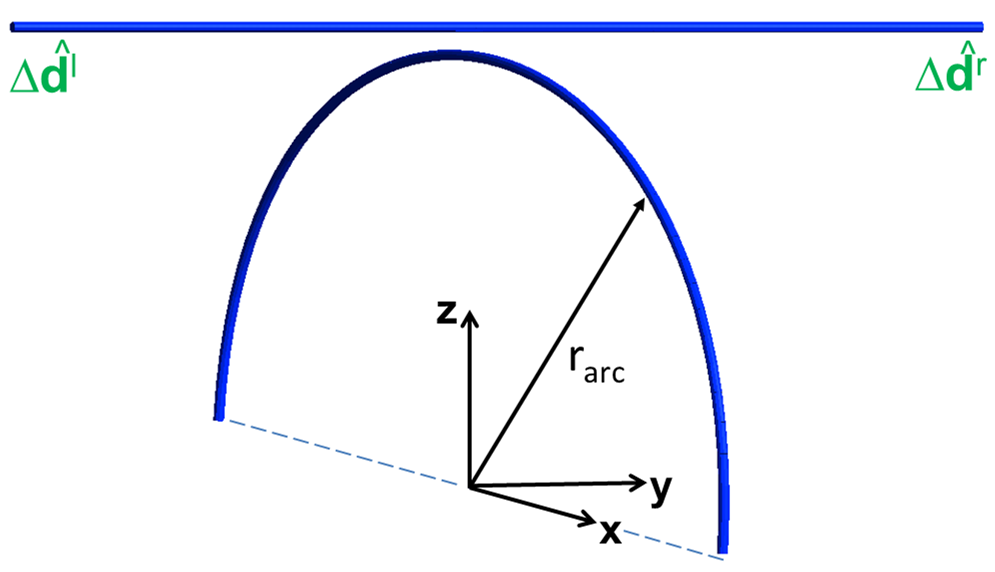}
 \caption{Straight flexible beam and rigid arc: $3D$-view of initial configuration.}
 \label{fig:ABC_example1_initialconfig}
\end{figure}
\begin{figure}[t]
 \centering
   \subfigure[$\theta_k=0^{\circ}$.]
   {
    \includegraphics[width=0.15\textwidth]{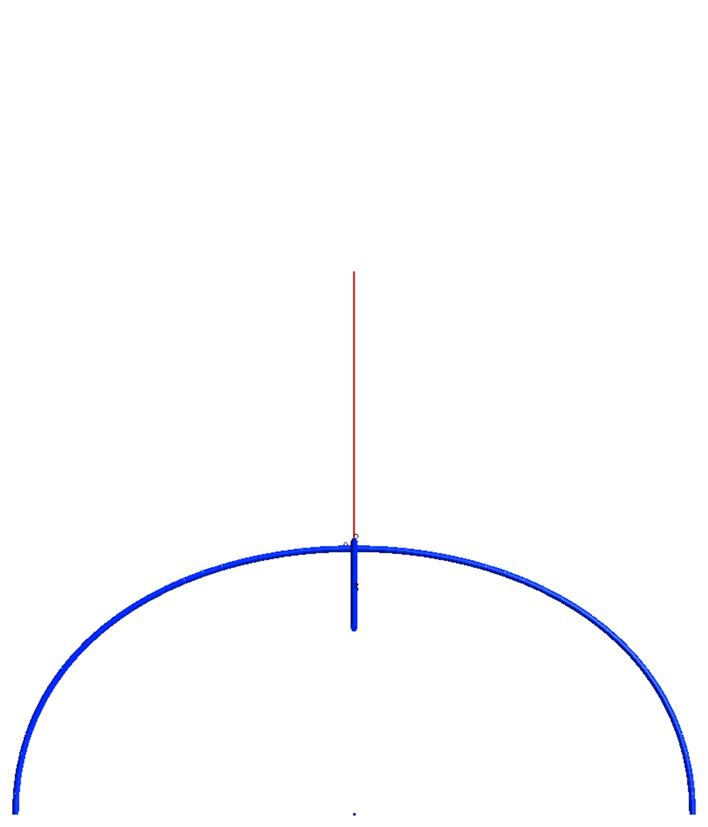}
    \label{fig:ABC_example1_step100}
   }
   \subfigure[$\theta_k=0.6 \cdot 90^{\circ}$.]
   {
    \includegraphics[width=0.15\textwidth]{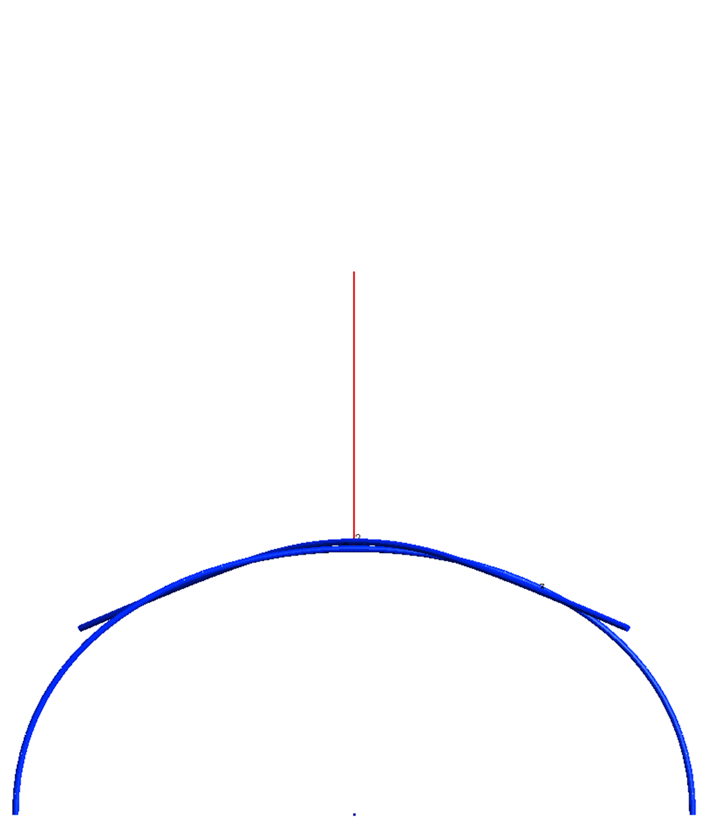}
    \label{fig:ABC_example1_step160}
   }
   \subfigure[$\theta_k=0.7 \cdot 90^{\circ}$.]
   {
    \includegraphics[width=0.15\textwidth]{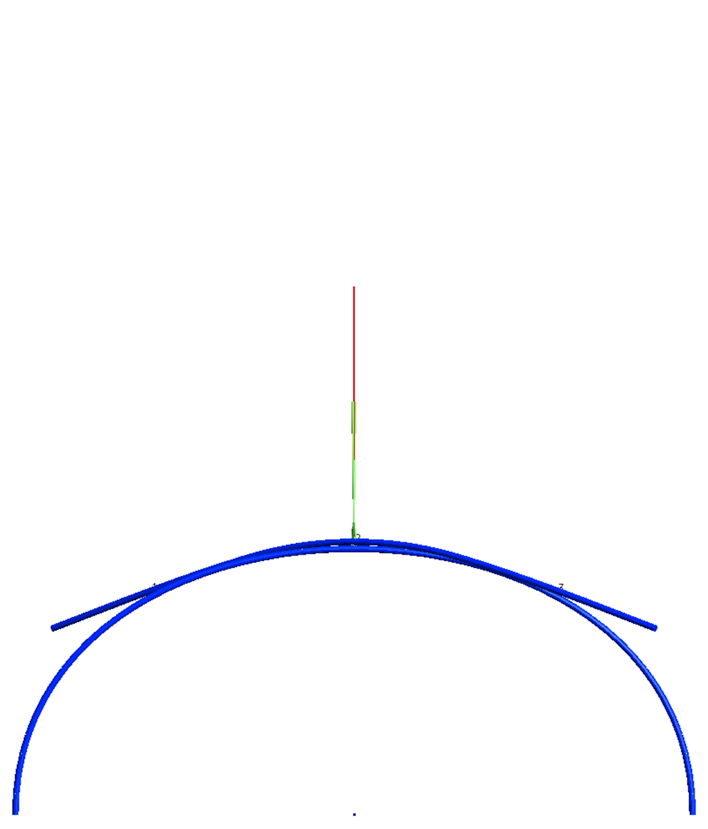}
    \label{fig:ABC_example1_step170}
   }
   \subfigure[$\theta_k=0.8 \cdot 90^{\circ}$.]
   {
    \includegraphics[width=0.15\textwidth]{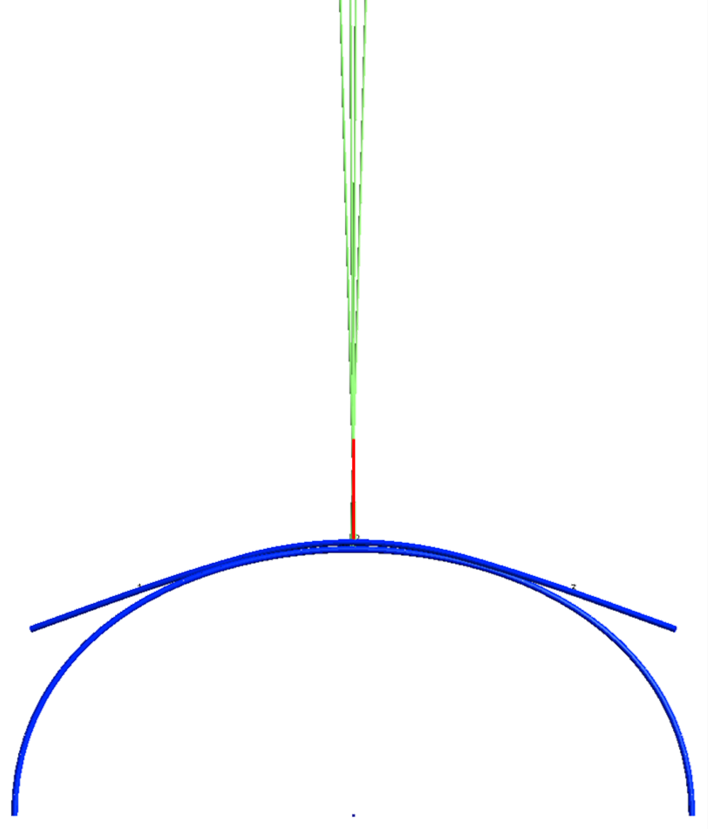}
    \label{fig:ABC_example1_step180}
   }
   \subfigure[$\theta_k=0.9 \cdot 90^{\circ}$.]
   {
    \includegraphics[width=0.15\textwidth]{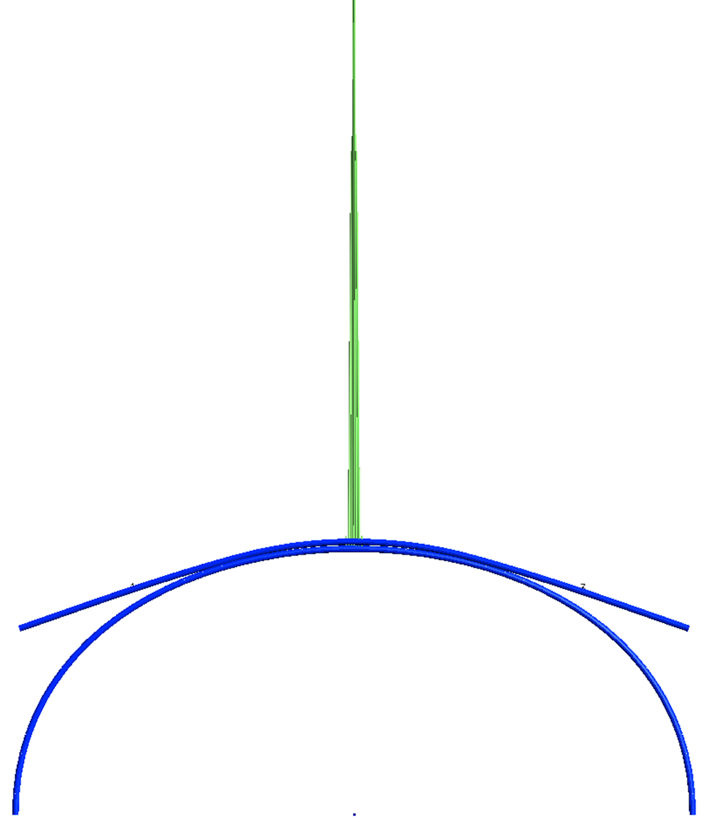}
    \label{fig:ABC_example1_step190}
   }
   \subfigure[$\theta_k=1.0 \cdot 90^{\circ}$.]
   {
    \includegraphics[width=0.15\textwidth]{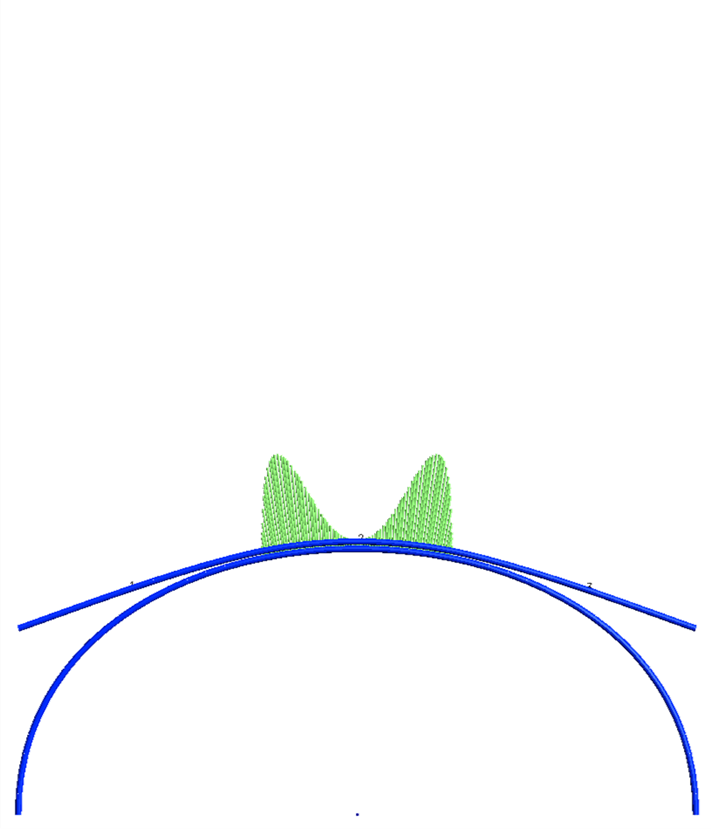}
    \label{fig:ABC_example1_step200}
   }
  \caption{Contact interaction of a straight flexible beam and a rigid arc: $xz$-view of deformed configuration at different rotation angles $\theta_k$.}
  \label{fig:ABC_example1_differentconfigs}
\end{figure}

The first static example consists of a completely fixed, rigid arc (beam 1 = slave) lying in the global $xz$-plane and being discretized by one beam element and a flexible straight beam (beam 2 = master)
that initially points into global $y$-direction and is discretized by three finite elements (see Figure \ref{fig:ABC_example1_initialconfig}). The following geometrical and material parameters have been chosen for this example: $E=1.0 \cdot 10^9$, $R=0.01$, $l_2=2$, $l_1=\pi r_{arc}$ with $r_{arc}=1.0$. Furthermore, a quadratically regularized 
penalty law with $\bar{g}=0.1R=0.001$ has been applied. In the following, the model transition between point- and line-contact will be investigated for different choices of the penalty 
parameters $\varepsilon_{\perp}$ and $\varepsilon_{\parallel}$. For the sake of better visualization, the comparatively large shifting interval limited by $\alpha_1=10^{\circ}$ and $\alpha_2=30^{\circ}$ has been chosen. Thereto, the endpoints of the master beam are first driven downwards (in negative $z$-direction) in a displacement controlled manner within $1000$ load steps until contact occurs. Then, with contact being active, 
the two endpoints of the master beam are moved on a circular path within further $n_l=4000$ load steps, such that the beam performs a full rotation with respect to the global $z$-axis, thus covering the whole range of 
possible contact angles. The only reason for the high number of load steps is a sufficiently high resolution required for the plots presented later on. The following Dirichlet conditions have been 
applied in the second stage of the deformation process:
\begin{align}
\label{ABC_example1_circularpath}
\begin{split}
  \Delta \hat{d}_{2,x}^l\!& =\!\frac{l_2}{2} \sin\left(\theta_k \right), \,\, \Delta \hat{d}_{2,y}^l \!=\!\frac{l_2}{2} \left[1 \!-\! \cos \left(\theta_k \right) \right], \,\, 
  \Delta \hat{d}_{2,x}^r\!=\!-\frac{l_2}{2} \sin\left(\theta_k \right), \,\, \Delta \hat{d}_{2,y}^r \!=\!-\frac{l_2}{2} \left[1 \!-\! \cos \left(\theta_k \right) \right], \\
  \Delta \hat{d}_{2,z}^l & =\Delta \hat{d}_{2,z}^r=-0.3
  \quad \text{with} \quad \theta_k=\dfrac{k \cdot 2\pi}{n_l} \,\, \text{for} \,\, k=1,...,n_l.
\end{split}
\end{align}
Since an axial displacement of the master beam is precluded by the applied Dirichlet fixation, we reduce the value of the beam cross section occurring in the axial stiffness by a factor of 
$100$, i.e. $A=0.01R^2\pi$, in order to end up with a deformation that is not completely dominated by the axial stiffness. Different states of deformation during the first quarter of the 
rotation process are illustrated in Figure~\ref{fig:ABC_example1_differentconfigs}. In the range of large contact angles, 
we observe a pure point-contact force (see Figures \ref{fig:ABC_example1_step100} and \ref{fig:ABC_example1_step160}) whose magnitude is illustrated by a red line.
Figures~\ref{fig:ABC_example1_step170} and \ref{fig:ABC_example1_step180} represent the realm of model transition: With decreasing contact angle, the magnitude of the point-contact force decreases, while the magnitude of the discrete Gauss point contributions to the line-contact force (illustrated by green lines) increases until a contact angle range of $\alpha<\alpha_1=10^{\circ}$, i.e. a pure line-contact state is reached (see Figure \ref{fig:ABC_example1_step190}). Finally, in Figure~\ref{fig:ABC_example1_step200} both beams lie within one plane ($\alpha=0^{\circ}$), thus leading to a state of pure line-contact.\\

\begin{figure}[t]
 \centering
   \subfigure[Low penalties: $\varepsilon_{\perp}\!=\!2 \cdot 10^4$, $\varepsilon_{\parallel}\!=\!5\! \cdot \!10^5$.]
   {
    \includegraphics[width=0.31\textwidth]{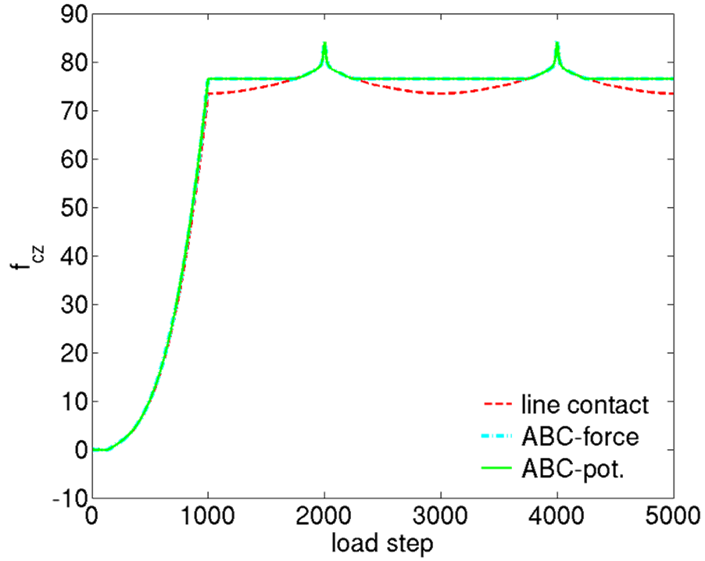}
    \label{fig:ABC_example1_force1}
   }
      \subfigure[Moderate penalties: $\varepsilon_{\perp}\!=\!2 \cdot 10^5$, $\varepsilon_{\parallel}\!=\!5 \!\cdot\! 10^6$.]
   {
    \includegraphics[width=0.31\textwidth]{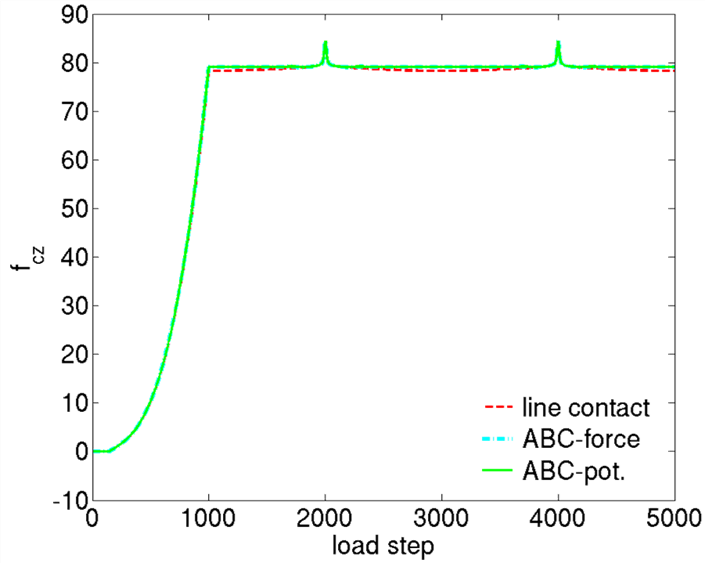}
    \label{fig:ABC_example1_force2}
   }
      \subfigure[High penalties: $\varepsilon_{\perp}\!=\!2 \cdot 10^6$, $\varepsilon_{\parallel}\!=\!5 \!\cdot \!10^7$.]
   {
    \includegraphics[width=0.31\textwidth]{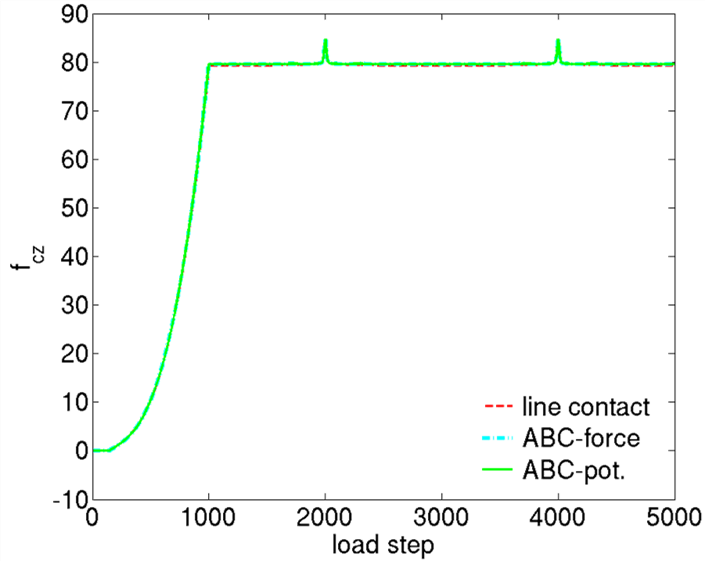}
    \label{fig:ABC_example1_force3}
   }
  \caption{Contact interaction of a straight flexible beam and a rigid arc: Evolution of accumulated contact force $f_{cz}$ over load steps.}
  \label{fig:ABC_example1_forces}
\end{figure}
\begin{figure}[t]
 \centering
   \subfigure[Low penalty: $\varepsilon_{\perp}\!=\!10^4$, $\varepsilon_{\parallel}\!=\!5\! \cdot \!10^5$.]
   {
    \includegraphics[width=0.31\textwidth]{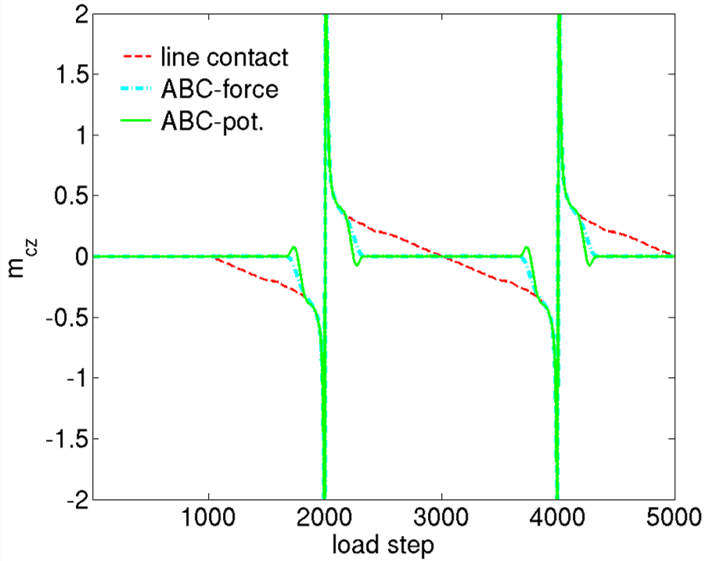}
    \label{fig:ABC_example1_moment1}
   }
      \subfigure[Moderate penalty: $\varepsilon_{\perp}\!=\!10^5$, $\varepsilon_{\parallel}\!=\!5 \!\cdot\! 10^6$.]
   {
    \includegraphics[width=0.31\textwidth]{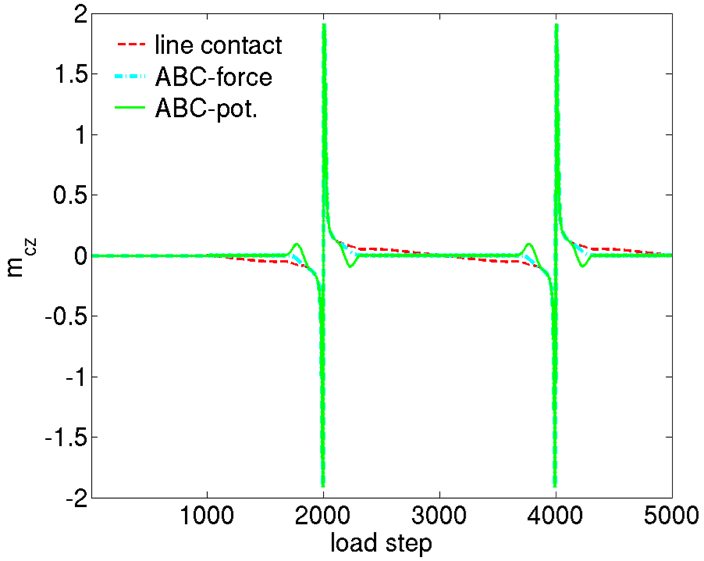}
    \label{fig:ABC_example1_moment2}
   }
      \subfigure[High penalty: $\varepsilon_{\perp}\!\!=\!\!10^6 \, / \,\, 5.4 \! \cdot \! 10^5$, $\varepsilon_{\parallel}\!\!=\!\!5 \!\cdot \!10^7$.]
   {
    \includegraphics[width=0.31\textwidth]{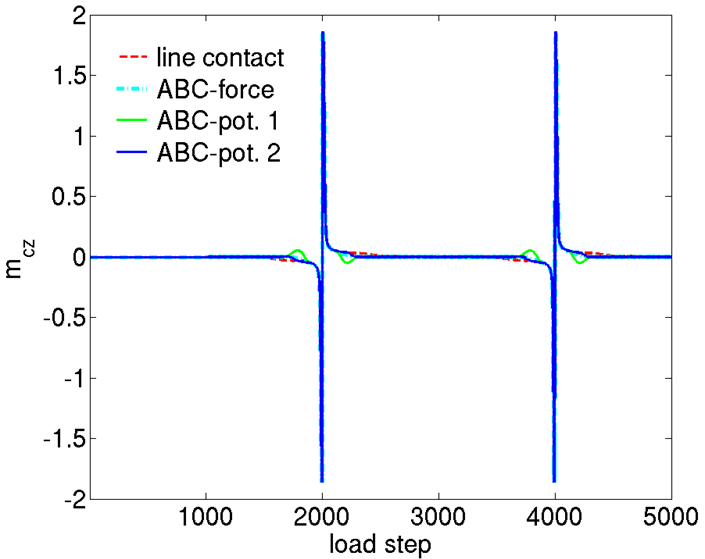}
    \label{fig:ABC_example1_moment3}
   }
  \caption{Contact interaction of a straight flexible beam and a rigid arc: Evolution of accumulated contact torque $m_{cz}$ over load steps.}
  \label{fig:ABC_example1_moments}
\end{figure}

In order to investigate the transition phase between the two contact formulations more closely, the accumulated contact force as well as the accumulated torque of the point and the line contact force 
with respect to the global $z$-axis are plotted in Figures \ref{fig:ABC_example1_forces} and \ref{fig:ABC_example1_moments}. The three individual plots in each case represent simulations with low 
($\varepsilon_{\parallel}=5 \cdot 10^5$, $\varepsilon_{\perp}=2 \cdot 10^4$), intermediate ($\varepsilon_{\parallel}=5 \cdot 10^6$, $\varepsilon_{\perp}=2 \cdot 10^5$) and high penalty 
($\varepsilon_{\parallel}=5 \cdot 10^7$, $\varepsilon_{\perp}=2 \cdot 10^6$) parameters. The three point-to-point penalty parameters $\varepsilon_{\perp}$ assigned to the three given 
line-to-line penalty parameters $\varepsilon_{\parallel}$ have been determined according to the approximation \eqref{ABC_penaltyadjustment4}, thus leading to $\varepsilon_{\perp} / \varepsilon_{\parallel} \approx 25$. Furthermore, in each plot the following three cases will be compared: a standard force-based \textit{ABC} formulation, a potential-based \textit{ABC} formulation  and finally a pure line-to-line contact formulation. For all cases, we have chosen $n_{II}=100$ integration intervals with $5$ Gauss points per interval. This high number has been chosen such that also the pure line-to-line contact formulation is able to properly resolve the range of large contact angles for the given, very rough spatial discretization. Let us first consider the accumulated contact forces resulting from a low penalty parameter as illustrated in Figure \ref{fig:ABC_example1_force1}. During the first $1000$ load steps, beam 2 is 
driven downwards. After approximately $200$ load steps, the beams come into contact and the contact forces rise. After $1000$ load steps the rotation starts. The two peaks occurring in all force 
plots at load step $2000$ and load step $4000$ represent configurations where both beams lie within one plane (see Figure \ref{fig:ABC_example1_step200}). These force peaks do not represent any numerical artifact or model error, but rather are expected from a mechanical point of view: beam 2 has to be deformed to a higher extent in order to pass this ''parallel`` configuration, which in turn leads to higher overall contact forces in this configuration. The pure line-to-line contact formulation (red dashed line) shows a smooth and steadily increasing contact force evolution in the range $\theta \in [0;90^{\circ}]$ (step $1000$ until step $2000$).\\

In contrary, the contact force evolutions of the force-based (blue dashed line) and potential-based (green solid line) \textit{ABC} formulation remain constant in the range of large contact angles. This is the 
expected evolution of the pure point-contact formulation (for the considered symmetrical problem), which is active in this angle range. After approximately $1700$ load steps ($\alpha=30^{\circ}$), the transition 
range begins, characterized by a visible increase of the contact force. Approximately at load step $1900$ ($\alpha=10^{\circ}$), the end of the transition interval is reached. From now on, the curves 
representing the \textit{ABC} formulations and the curve representing the pure line-to-line contact formulation are identical, since the \textit{ABC} formulation reduces to a pure line-to-line contact formulation for 
angles $\alpha<\alpha_1=10^{\circ}$. As expected, the difference between the pure line-to-line contact formulation and the more efficient \textit{ABC} formulations vanishes with increasing penalty factor (see Figures \ref{fig:ABC_example1_force2} and \ref{fig:ABC_example1_force3}). Furthermore, no distinctive difference between the contact force evolutions of the force-based and the potential-based \textit{ABC} formulation is visible. In Figure \ref{fig:ABC_example1_moments}, the accumulated torque of the contact forces with respect to the global $z$-axis is plotted. In this example, the resulting contact torque is a consequence 
of line contact force contributions that do not exactly point into global $z$-direction at all positions besides the rotation center at position $x=y=0$. Consequently, the contact torque contribution 
of the pure point-contact force vanishes. This is visible for the curves representing the \textit{ABC} formulation (see e.g. the blue chain line and the green solid line in Figure \ref{fig:ABC_example1_moment1}) 
in the range of large contact angles. The contact torque evolutions show a very steep gradient in the neighborhood of $\alpha=90^{\circ}$, which can mechanically be interpreted as a 
''snap-through`` behavior. Exactly at $\alpha=90^{\circ}$ (step 2000), the total torque vanishes as a consequence of the geometrical symmetry.\\ 

As expected, the difference between the force-based and the potential-based variant due to algorithmic contact moments appearing in the latter formulation is small compared to the model deviation between the pure line-to-line contact and the pure point-to-point contact (\textit{ABC} formulation in the range $\alpha<30^{\circ}$) and especially small compared to the total magnitude of the mechanically motivated contact torque peaks. Furthermore, this difference decreases with increasing penalty parameter. 
For the high-penalty case (see Figure \ref{fig:ABC_example1_moment3}), we have additionally plotted the variant based on a better penalty approximation $\varepsilon_{\perp}\!=\!5.4 \cdot 10^5$ (dark-blue solid line) determined via the numerical solution of \eqref{ABC_penaltyadjustment2} (with $g_{min}=0.0006$) 
instead of~\eqref{ABC_penaltyadjustment4}. In this case, almost no remaining difference between the force-based and the potential-based variant is visible. All other distinctions of the  different formulations, especially the mutual convergence of the three curves with increasing penalty parameter, are similar to the force 
evolutions above.\\

\begin{figure}[t]
 \centering
   \subfigure[Initial configuration.]
   {
    \includegraphics[width=0.235\textwidth]{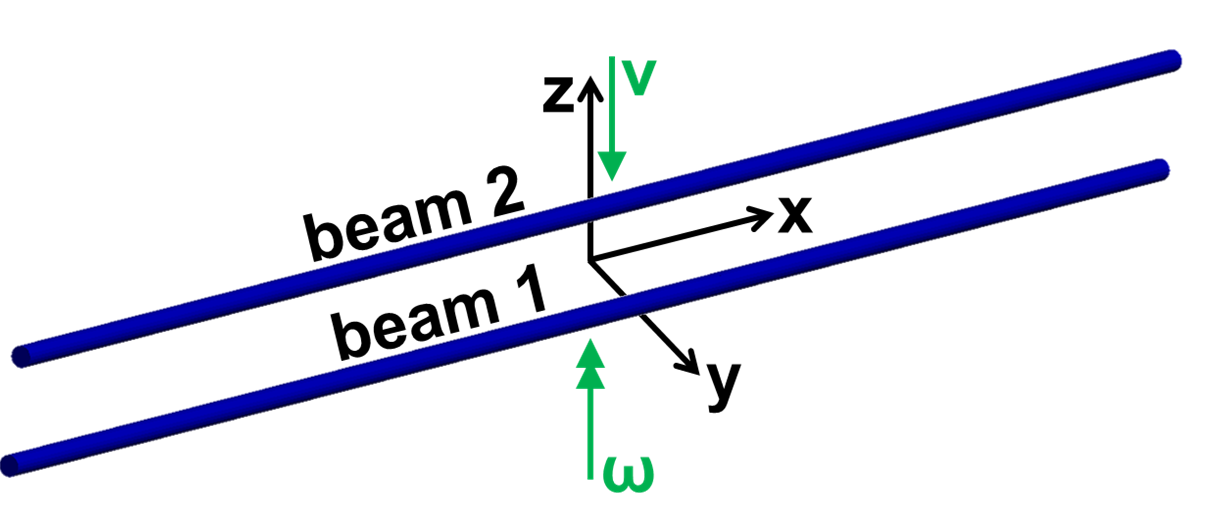}
    \label{fig:ABC_example2_step01}
   }
   \subfigure[Time step: 870.]
   {
    \includegraphics[width=0.235\textwidth]{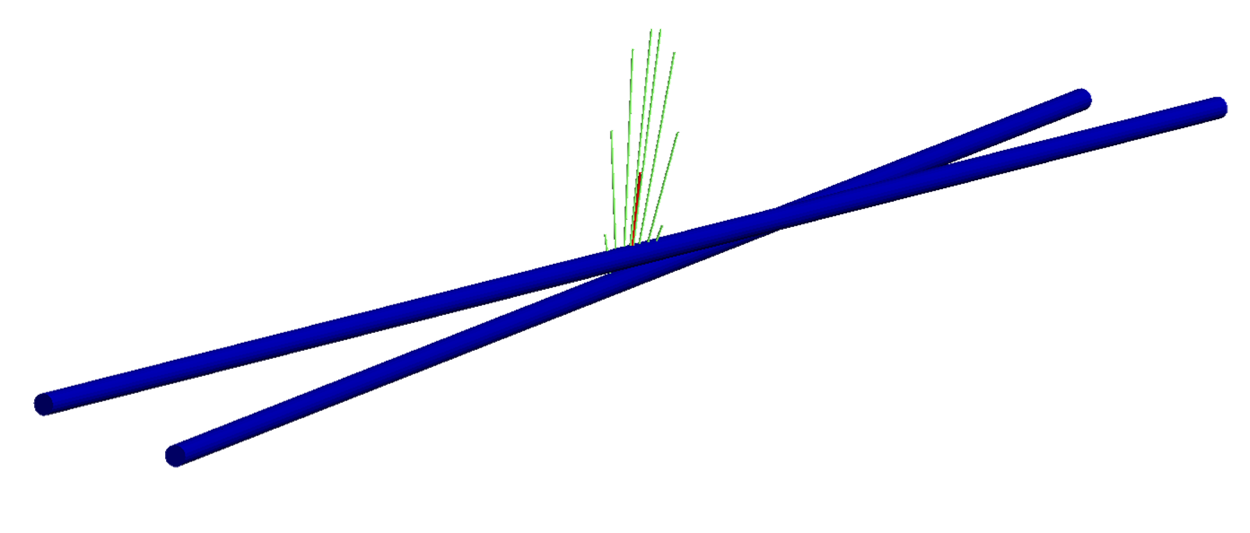}
    \label{fig:ABC_example2_step60}
   }
   \subfigure[Time step: 910.]
   {
    \includegraphics[width=0.235\textwidth]{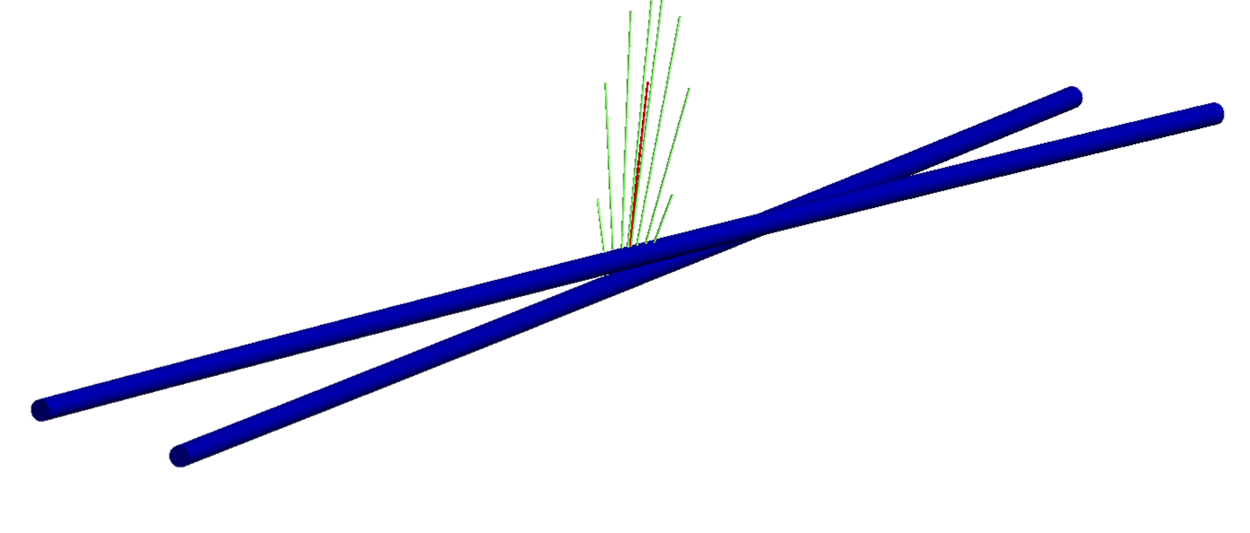}
    \label{fig:ABC_example2_step70}
   }
   \subfigure[Time step: 2000.]
   {
    \includegraphics[width=0.235\textwidth]{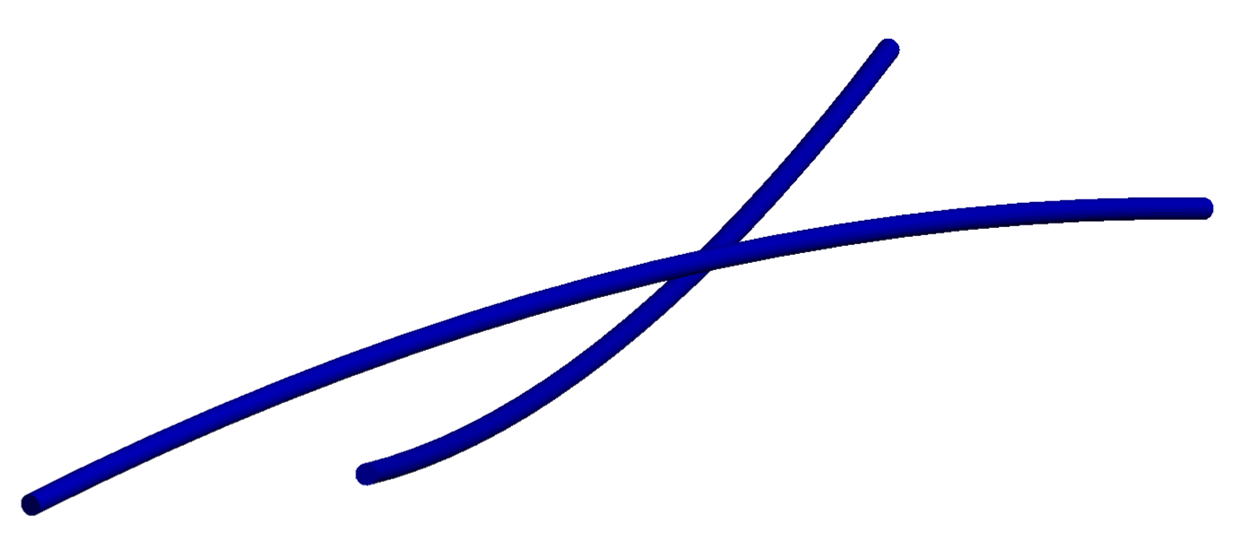}
    \label{fig:ABC_example2_step80}
   }
  \caption{Dynamic impact of rotating and translationally moving straight beam: Initial and deformed configurations at different time steps.}
  \label{fig:ABC_example2_differentsteps}
\end{figure}

%
\subsection{Example 2: Impact of free flying beams}
\label{sec:examples_example2}
%
The second example of this section aims at investigating the conservation properties (linear momentum, angular momentum and total energy) of the proposed \textit{ABC} formulation within 
a dynamic framework. Thereto, we consider two initially straight beams with $R=0.01$, $l=2$, $E=10^{-6}$ and densities $\rho_1=0.1$ and $\rho_2=0.05$. Initially, both beams are arranged in a parallel manner with a distance of $d=10R=0.1$ (see Figure \ref{fig:ABC_example2_step01}). Within the time interval $t \in [0;0.06]$, beam 2 (= master) is accelerated by a line load $\tilde{f}_z(\eta,t)=5\cdot10^{-7}\bar{f}_z(t)$ pointing in negative global $z$-direction and being constant along the beams length. The time scaling factor $\bar{f}_z(t)$ increases linearly from zero to one for $t \in [0;0.03]$ before it again decreases linearly from one to zero for $t \in [0.03;0.06]$. The slave beam (beam 1) is loaded by a line load $\tilde{f}_y(\xi,t)=2.5\cdot10^{-6}\cdot \bar{f}_y(t) \cdot \xi$ pointing in global $y$-direction and increasing linearly with $\xi \in [-1;1]$, which induces an angular momentum on the beam. The time scaling factor $\bar{f}_y(t)$ increases linearly from zero to one for $t \in [0;0.02]$ and decreases linearly from one to zero within the interval $t \in [0.02;0.04]$.\\

After this acceleration phase, the beams move freely until an impact of the two beams takes place. For time integration, we apply a generalized-$\alpha$ scheme without numerical dissipation ($\alpha_f=0.5$, $\alpha_m=0.5$, $\beta=0.25$ and $\gamma=0.5$) and a total simulation time $t_{end}=2.0s$. Since this time integrator can not guarantee exact energy conservation of the temporally discretized problem for arbitrary time step sizes (see e.g. \cite{simo1992}), we choose the time step size small enough (standard choice $\Delta t = 0.001s$), such that the impact of the spatial discretization on the conservation properties can be 
investigated with sufficient accuracy. The system parameters are chosen such that the impact takes place at a contact angle $\alpha \in [\alpha_1;\alpha_2]$ lying within the shifting interval. 
The configurations at the beginning of the simulation, at the beginning and the end of the impact and the end of the simulation are illustrated in Figures~\ref{fig:ABC_example2_step01} - 
\ref{fig:ABC_example2_step80}.\\


\begin{figure}[t]
 \centering
   \subfigure[Force-based: $\varepsilon_{\parallel}=3 \! \cdot \! 10^{-3}$,  $\varepsilon_{\perp}\!=3.1 \! \cdot \! 10^{-4}$, $\bar{g}=1 \! \cdot \! 10^{-3}$.]
   {
    \includegraphics[width=0.45\textwidth]{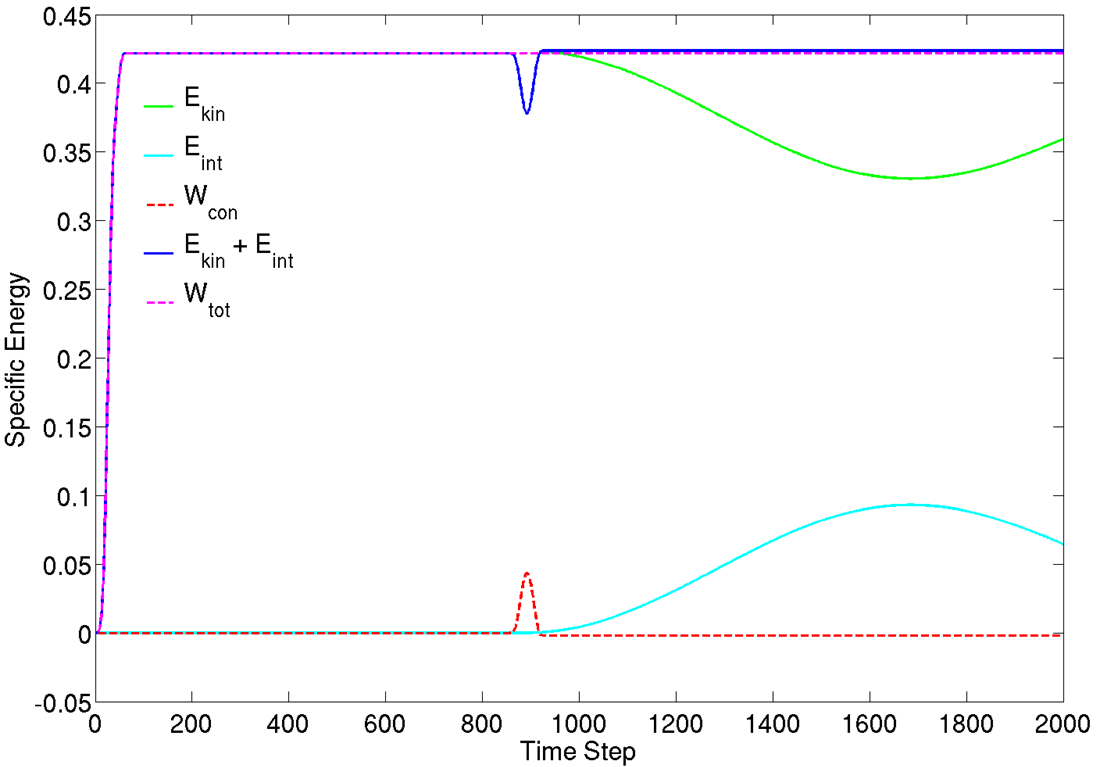}
    \label{fig:ABC_example2_energies1}
   }
   \subfigure[Potential-based: $\varepsilon_{\parallel}=3 \! \cdot \! 10^{-3}$,  $\varepsilon_{\perp}\!=3.1 \! \cdot \! 10^{-4}$, $\bar{g}=1 \! \cdot \! 10^{-3}$.]
   {
    \includegraphics[width=0.45\textwidth]{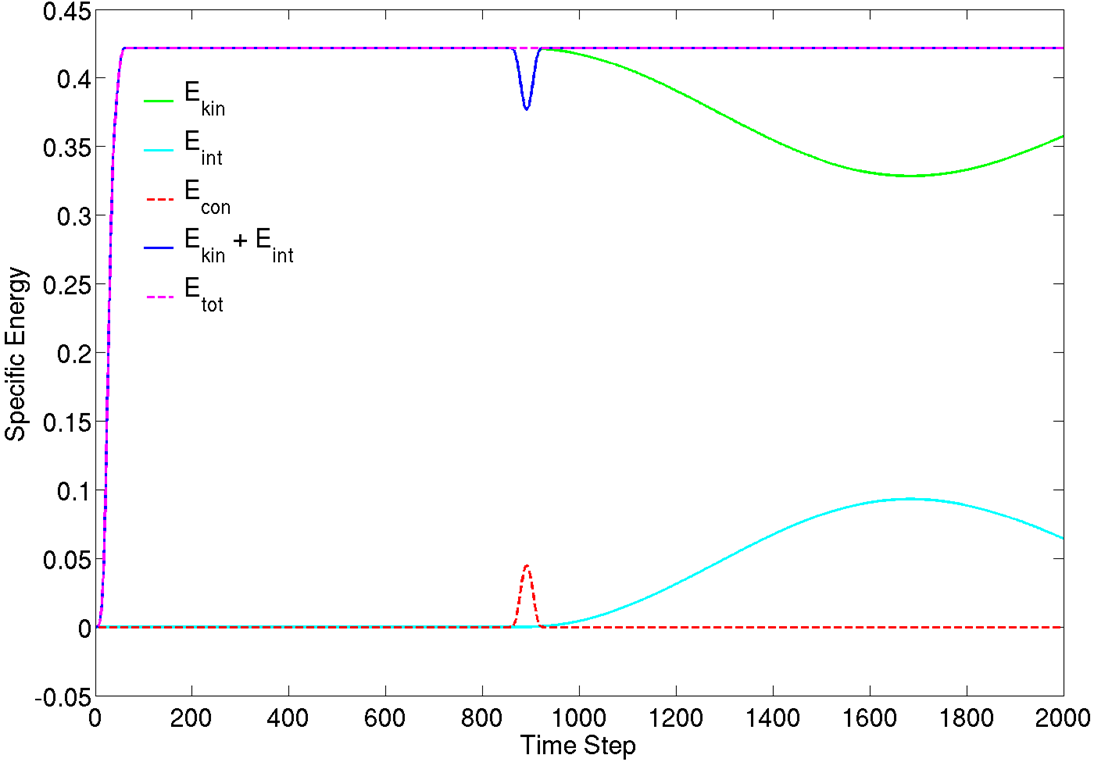}
    \label{fig:ABC_example2_energies2}
   }
   \subfigure[Force-based: $\varepsilon_{\parallel}=3 \! \cdot \! 10^{-3}$,  $\varepsilon_{\perp}\!=3.1 \! \cdot \! 10^{-6}$, $\bar{g}=1 \! \cdot \! 10^{-3}$.]
   {
    \includegraphics[width=0.45\textwidth]{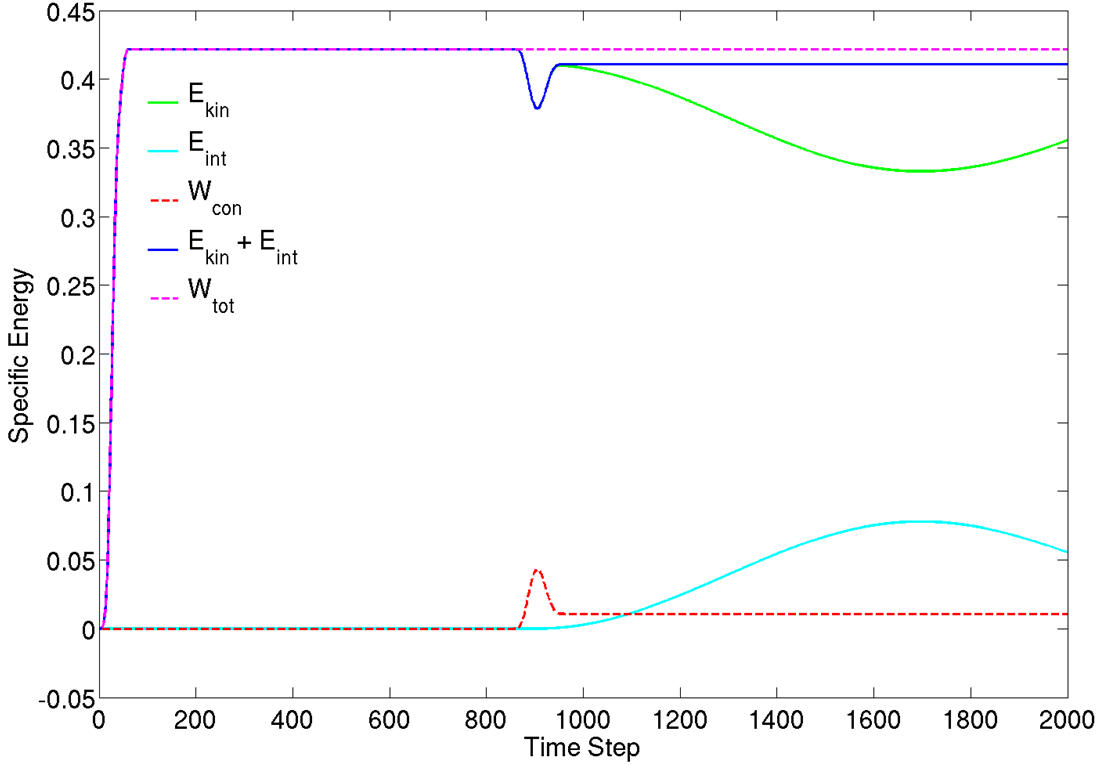}
    \label{fig:ABC_example2_energies3}
   }
   \subfigure[Force-based: $\varepsilon_{\parallel}=3 \! \cdot \! 10^{-1}$,  $\varepsilon_{\perp}\!=7.8 \! \cdot \! 10^{-5}$, $\bar{g}=1 \! \cdot \! 10^{-4}$.]
   {
    \includegraphics[width=0.45\textwidth]{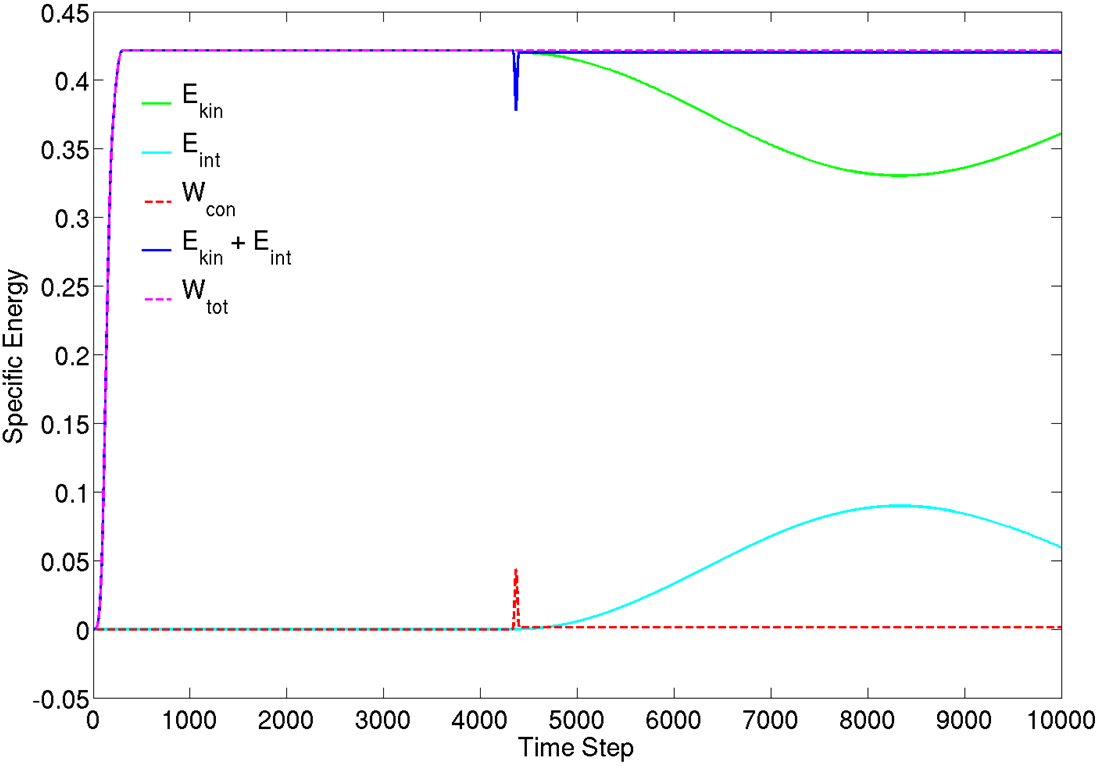}
    \label{fig:ABC_example2_energies4}
   }
  \caption{Dynamic impact of rotating and translationally moving straight beam: Conservation of energy.}
  \label{fig:ABC_example2_energies}
\end{figure}
\begin{figure}[t]
 \centering
   \subfigure[Conservation of linear momentum.]
   {
    \includegraphics[width=0.45\textwidth]{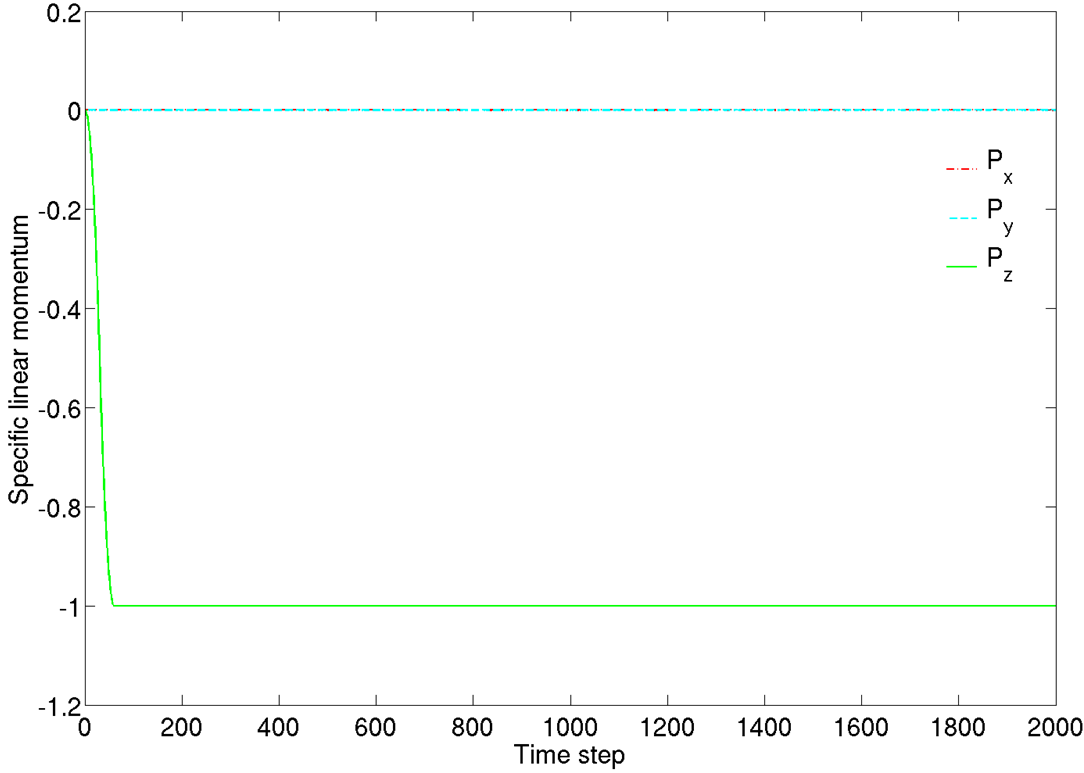}
    \label{fig:ABC_example2_linmomentum}
   }
   \subfigure[Conservation of angular momentum.]
   {
    \includegraphics[width=0.45\textwidth]{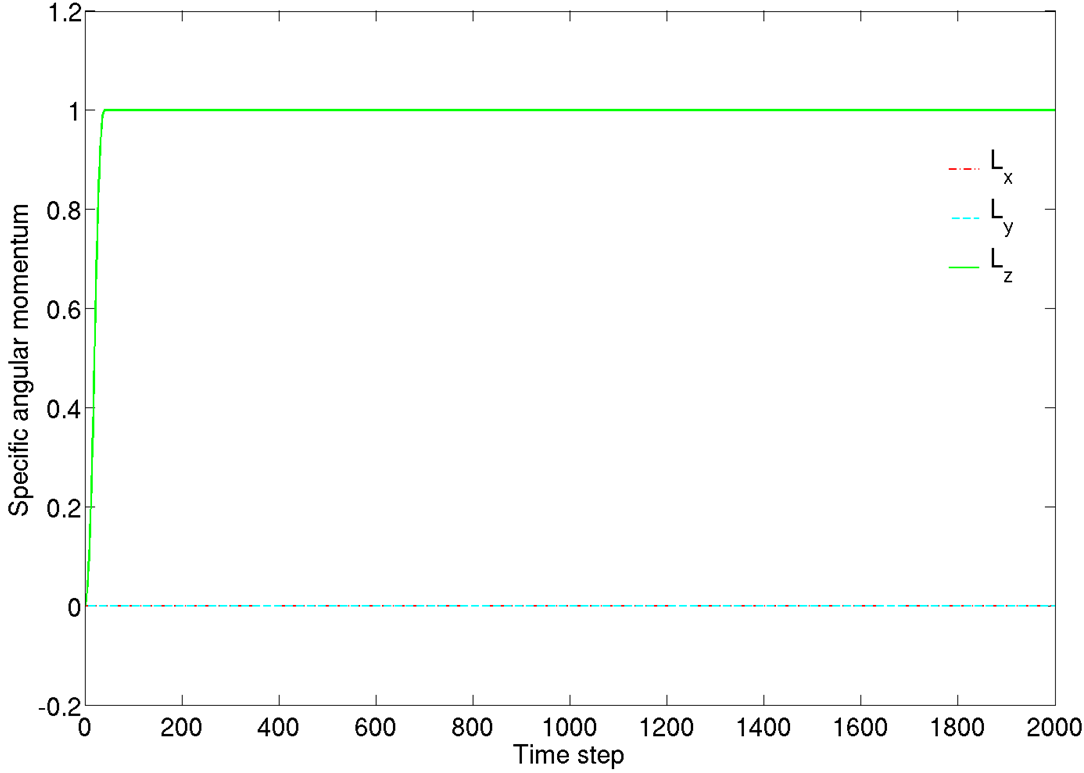}
    \label{fig:ABC_example2_angmomentum}
   }
  \caption{Dynamic impact of rotating and translationally moving straight beam: Conservation of linear and angular momentum.}
  \label{fig:ABC_example2_momentum}
\end{figure}

In Figure~\ref{fig:ABC_example2_energies}, the system energies are plotted for the force-based and potential-based \textit{ABC} formulation in combination with different penalty laws. 
Figure~\ref{fig:ABC_example2_energies1} represents the force-based variant in combination with a quadratically regularized penalty law with $\varepsilon_{\parallel}=3 \! \cdot \! 10^{-3}$,  
$\varepsilon_{\perp}=3.1 \! \cdot \! 10^{-4}$ and $\bar{g}=1 \! \cdot \! 10^{-3}$. Here, the point penalty parameter $\varepsilon_{\perp}$ has been determined on the basis of \eqref{ABC_penaltyadjustment4}. 
Besides the kinetic energy $E_{kin}$ and the internal elastic energy $E_{int}$ of the two beams, we have plotted the accumulated contact work determined by means of the following expression
(see also~\eqref{conservation_energy_eleandcontact})
\begin{align}
W_{con}(t)=\sum \limits_{t_i= \Delta t}^{t} \Delta \mb{D}(t_i)^T \mb{R}_{con}(t_i),
\end{align}
where $\mb{R}_{con}(t_i)$ denotes the total contact residual contribution and $\Delta \mb{D}(t_i)$ the increment of the total displacement vector at time step $t_i$. The notion of contact work is necessary since no potential is existent for the force-based formulation. Furthermore, we have plotted the sum of kinetic and internal energy $E_{kin}+E_{int}$ as well as 
the total work $W_{tot}=W_{con}+E_{kin}+E_{int}$ representing the sum of all three contributions. All mechanical energy and work contributions plotted in Figure~\ref{fig:ABC_example2_energies}
are normalized with the internal energy $E_0=EI \pi^2/(8l)$, which corresponds to a beam that has been elastically bent to a quarter-circle. Looking at Figure~\ref{fig:ABC_example2_energies1},
 one realizes that after the acceleration phase the total work $W_{tot}$ remains constant, which indicates that no relevant energy losses are caused by the applied time integrator. While 
 the total work basically consists of pure kinetic energy in the first half of the process, the dynamic impact (peak in the contact work) induces a deformation of the beams accompanied by an 
 increase in the internal elastic energy. However, after the contact has re-opened, we observe a remaining contact work in the range of $W_{con} \approx -0.002 \cdot E_0$, thus leading to an increase 
 in the energy $E_{kin}+E_{int}$ by the same amount. The fact that the contact work does not decrease to zero after the contact has re-opened again is a direct consequence of the 
 non-conservative nature of the force-based formulation. In contrary, the potential-based formulation (see Figure~\ref{fig:ABC_example2_energies2}) is able to represent exact conservation 
 (aside from possible losses caused by a non-conserving time integration scheme)
 of the total energy $E_{tot}=E_{con}+E_{kin}+E_{int}$, since a contact potential $E_{con}$ is existent that vanishes as soon as the contact re-opens. For comparison reasons, we have also plotted 
 a variant of the force-based formulation, where the penalty parameter $\varepsilon_{\perp}=3.1 \! \cdot \! 10^{-6}$ has been decreased by a factor of $100$ as compared to \eqref{ABC_penaltyadjustment4} (see Figure~\ref{fig:ABC_example2_energies3}). 
 In this case, the amount of accumulated non-conservative contact work that remains after the contact has re-opened increases to  $W_{con} \approx 0.011 \cdot E_0 $. On the contrary, when applying 
 a better approximation for the optimal penalty parameter $\varepsilon_{\perp}=2.4 \! \cdot \! 10^{-4}$ based on a numerical solution of \eqref{ABC_penaltyadjustment2} (with $g_{min}=-0.002$), 
 the remaining contact work drops to $W_{con} \approx -0.001 \cdot E_0$ (not illustrated in Figure~\ref{fig:ABC_example2_energies}). It has already been argued in Section~\ref{sec:ABC_potentialbased} that 
 the non-conservative work contributions of the force-based \textit{ABC} formulation decrease with increasing penalty parameter. Applying a penalty law with increased penalty parameters 
 $\varepsilon_{\parallel}=3 \! \cdot \! 10^{-1}$,  $\varepsilon_{\perp}=7.8 \! \cdot \! 10^{-3}$ ($\varepsilon_{\perp}$ determined with \eqref{ABC_penaltyadjustment2} and $g_{min}=-0.0002$) and 
 $\bar{g}=1 \! \cdot \! 10^{-4}$ results in a remaining contact work of~$W_{con} \approx - 0.000003 \cdot E_0$ (not illustrated in Figure~\ref{fig:ABC_example2_energies}). Even for a decrease of the 
 penalty parameter by a factor of $100$ to $\varepsilon_{\perp}=7.8 \! \cdot \! 10^{-5}$, the remaining non-conservative work does not exceed an amount of $W_{con} \approx 0.0013 \cdot E_0$ 
 in this case (see Figure~\ref{fig:ABC_example2_energies4}).
Finally, in Figure~\ref{fig:ABC_example2_momentum}, the linear and angular momentum normalized with the initial values introduced by the external forces are plotted for the force-based \textit{ABC} formulation 
and the quadratically regularized penalty law with $\varepsilon_{\parallel}=3 \! \cdot \! 10^{-3}$,  $\varepsilon_{\perp}=3.1 \! \cdot \! 10^{-4}$ and $\bar{g}=1 \! \cdot \! 10^{-3}$ 
(corresponding to Figure~\ref{fig:ABC_example2_energies1}). As already expected from analytical investigations (see~\ref{anhang:conservation}), the linear and angular momentum are exactly conserved. This also holds for the potential-based \textit{ABC} formulation and all investigated penalty laws.

%
\subsection{Example 3: Simulation of a biopolymer network}
\label{sec:examples_biopolymer}
%
In a first practically relevant example, we apply the presented contact algorithm in order to simulate the three-dimensional Brownian motion of filaments in biopolymer networks. Biopolymer networks are tight meshes of highly slender polymer filaments (e.g. Actin filaments) embedded in a liquid phase, often interconnected by means of a second molecule species (so-called cross-linkers). These networks can for example be found in biological cells. There, they crucially determine the mechanical properties of cells and highly relevant biological processes such as cell migration or cell division (see \cite{mueller2014}). In \cite{meier2015b}, an exemplary system of this type has already been analyzed by means of a pure line-to-line contact formulation. Here, we want to investigate the gain in computational efficiency when replacing a standard line contact formulation by the proposed \textit{ABC} formulation in combination with the contact search and the step size control introduced in Section~\ref{sec:algorithmicaspects}. Further information about the finite element model describing the Brownian motion of the considered filaments can for example be found in \cite{cyron2012} and is additionally summarized in \cite{meier2015b}.\\


\begin{figure}[t]
 \centering
   \subfigure[Undeformed initial configuration.]
   {
    \includegraphics[width=0.31\textwidth]{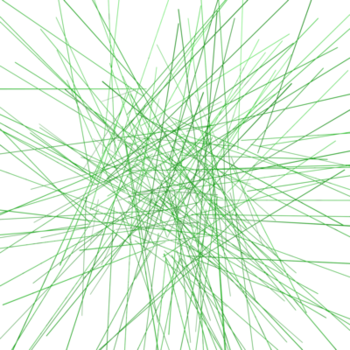}
    \label{fig:statmech_208filaments_step000}
   }
   \subfigure[Deformed configuration at step 190.]
   {
    \includegraphics[width=0.31\textwidth]{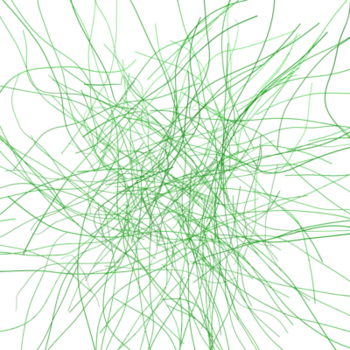}
    \label{fig:statmech_208filaments_step190}
   }
   \subfigure[Deformed configuration at step 500.]
   {
    \includegraphics[width=0.31\textwidth]{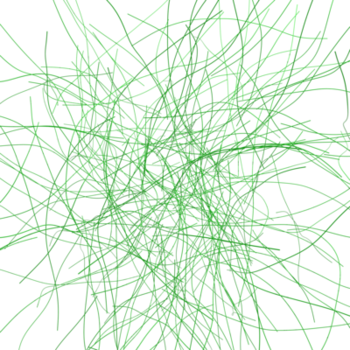}
    \label{fig:statmech_208filaments_step500}
   }
  \caption{Brownian dynamics simulation of the free diffusion of Actin filaments: Deformed configurations at different time steps.}
  \label{fig:statmech_208filaments_configs}
\end{figure}

\begin{figure}[h!]
 \centering
   \subfigure[Step 190: zoom factor 4.]
   {
    \includegraphics[width=0.48\textwidth]{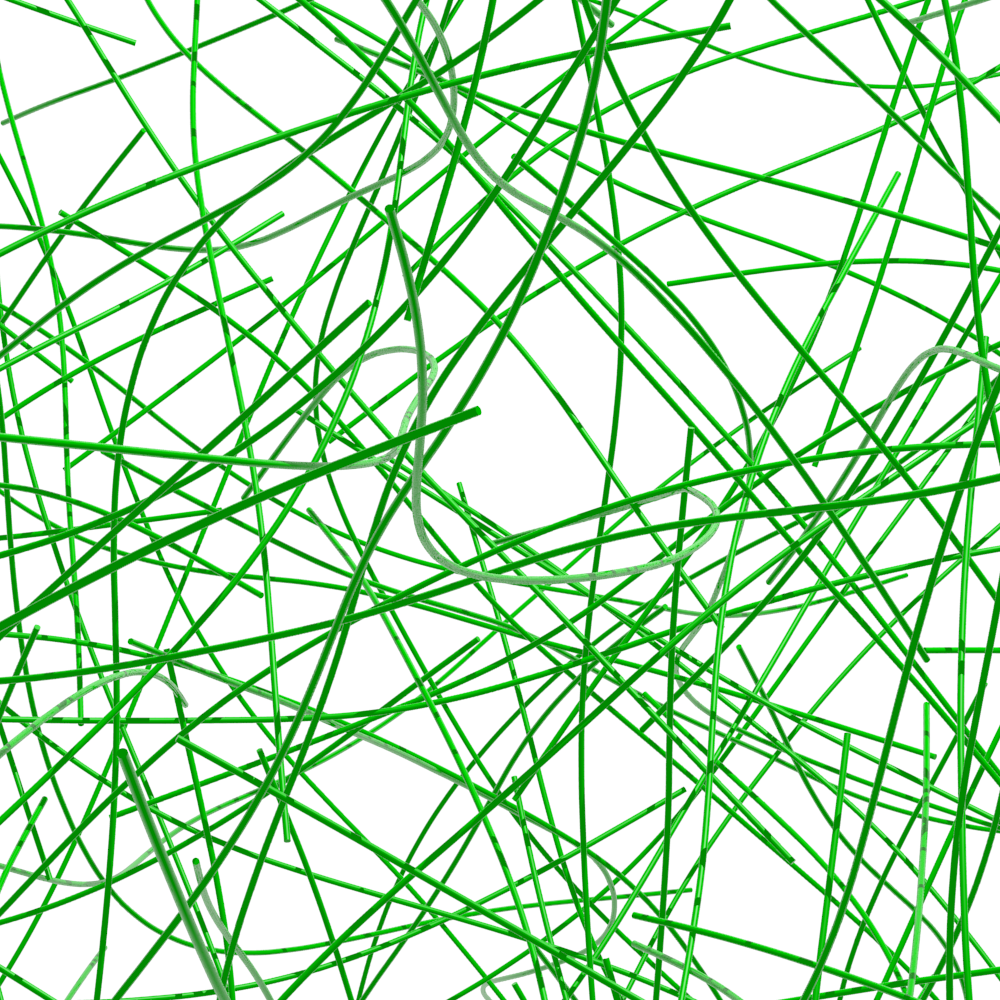}
    \label{fig:statmech_208filaments_step190_zoom4}
   }
      \subfigure[Step 190: zoom factor 8.]
   {
    \includegraphics[width=0.48\textwidth]{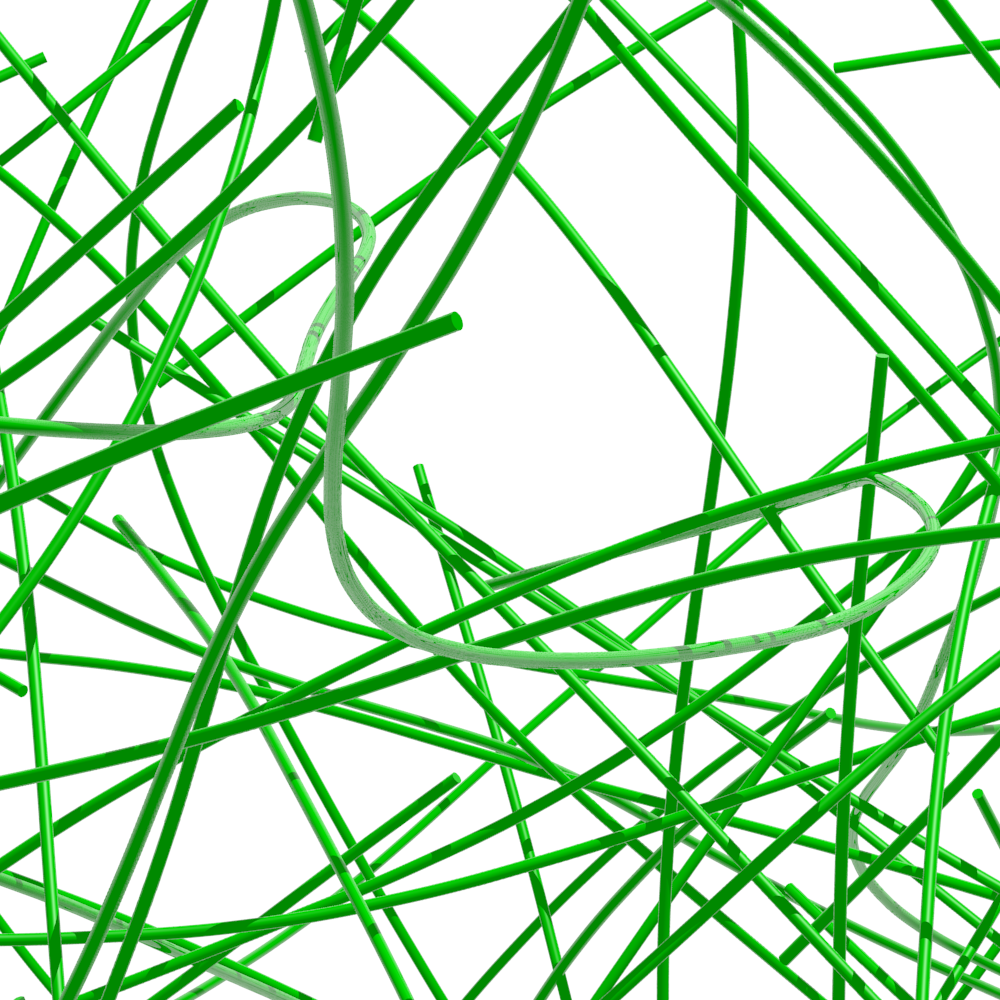}
    \label{fig:statmech_208filaments_step190_zoom8}
   }
  \caption{Brownian dynamics simulation of the free diffusion of Actin filaments: Final configuration with different zoom factors.}
  \label{fig:statmech_208filaments_configs}
\end{figure}

In the following, we consider a system of $208$ initially straight and randomly distributed Actin filaments with circular cross section of radius $R=2.45 \cdot 10^{-3}$, length $l=4$ and Young's modulus $E=1.3 \cdot 10^{9}$ (all quantities given in the units $mg$, $\mu m$, and $s$) as illustrated in Figure~\ref{fig:statmech_208filaments_step000}. All further physical system parameters describing the external forces acting on the filaments are identical to those applied in \cite{mueller2015} and can be found therein. The simulation has been performed by applying a spatial discretization with $32$ beam elements per filament, a time step size of $\Delta t=1.0 \cdot 10^{-4}$ and a total simulation time 
of $t_{end}=5.0 \cdot 10^{-1}$. Furthermore, the contact parameters have been chosen as $\varepsilon_{\perp}=1.0 \cdot 10^3, \varepsilon_{\parallel}=5.0 \cdot 10^4, \bar{g}=2.0 \cdot 10^{-3}, 
\alpha_1=9^{\circ}$ and $\alpha_2=11^{\circ}$ in combination with $20$ five-point integration intervals per element. The ratio $\varepsilon_{\parallel}/\varepsilon_{\perp} \approx 50$ 
results from~\eqref{ABC_penaltyadjustment4}. This Gauss point density has been determined on the basis of 
equation \eqref{limitationsline_GPdistance2} with $g_{n,min}=0.1$ in combination with an adequate safety factor. The spatial configurations at times $t=0.0, t=0.19$ and $t=0.5$ and 
corresponding detail views are illustrated in Figure~\ref{fig:statmech_208filaments_configs} (where for reasons of better visualization, the cross section radius has been scaled by a factor of $2$.). 
As a consequence of the excitatory stochastic forces employed in the considered Brownian dynamics model, the velocity field of these filaments is strongly fluctuating in space and in time, thus leading to drastic and frequent changes in the active contact set. This property in combination with the very high filament slenderness ratio of $\rho \approx 1600$, comparatively large time step sizes (maximal displacement per time $\Delta D_{max}=\max \, (||\mb{D}(t_{i})-\mb{D}(t_{i-1})||_{\infty}) \approx 10 \cdot R$) and complex geometrical contact configurations spanning the whole range of possible contact angles (see e.g. Figure~\ref{fig:statmech_208filaments_step190_zoom8}) make this example very challenging concerning the robustness and efficiency of the proposed contact algorithm. In case of non-convergence of the global Newton scheme within $50$ iterations, the time step size is halved before it is doubled again after $4$ successful Newton loops on the smaller time step level. Considering standard state-of-the-art beam contact algorithms, one would have to apply a line-to-line contact type formulation in order to represent not only intermediate and large contact angles but also the range of small contact angles, which occur with significant frequency in the considered type of application and which can not be resolved by a beam contact formulation of point-to-point type. In the following, we want to compare the proposed \textit{ABC} formulation with such a pure line-to-line contact formulation. According to~\eqref{limitationsline_GPdistance2}, we have chosen the number of integration intervals of the pure line contact formulation by a factor of five ($\sin(90^{\circ})/\sin(11^{\circ})\!\approx \! 5$) higher than for the \textit{ABC} formulation in order to resolve the most critical case $\alpha=90^{\circ}$ of the line contact model equivalently to the most critical case $\alpha=11^{\circ}$ of the \textit{ABC} formulation.\\

\begin{figure}[t]
 \centering
   \subfigure[Active point contacts.]
   {
    \includegraphics[width=0.31\textwidth]{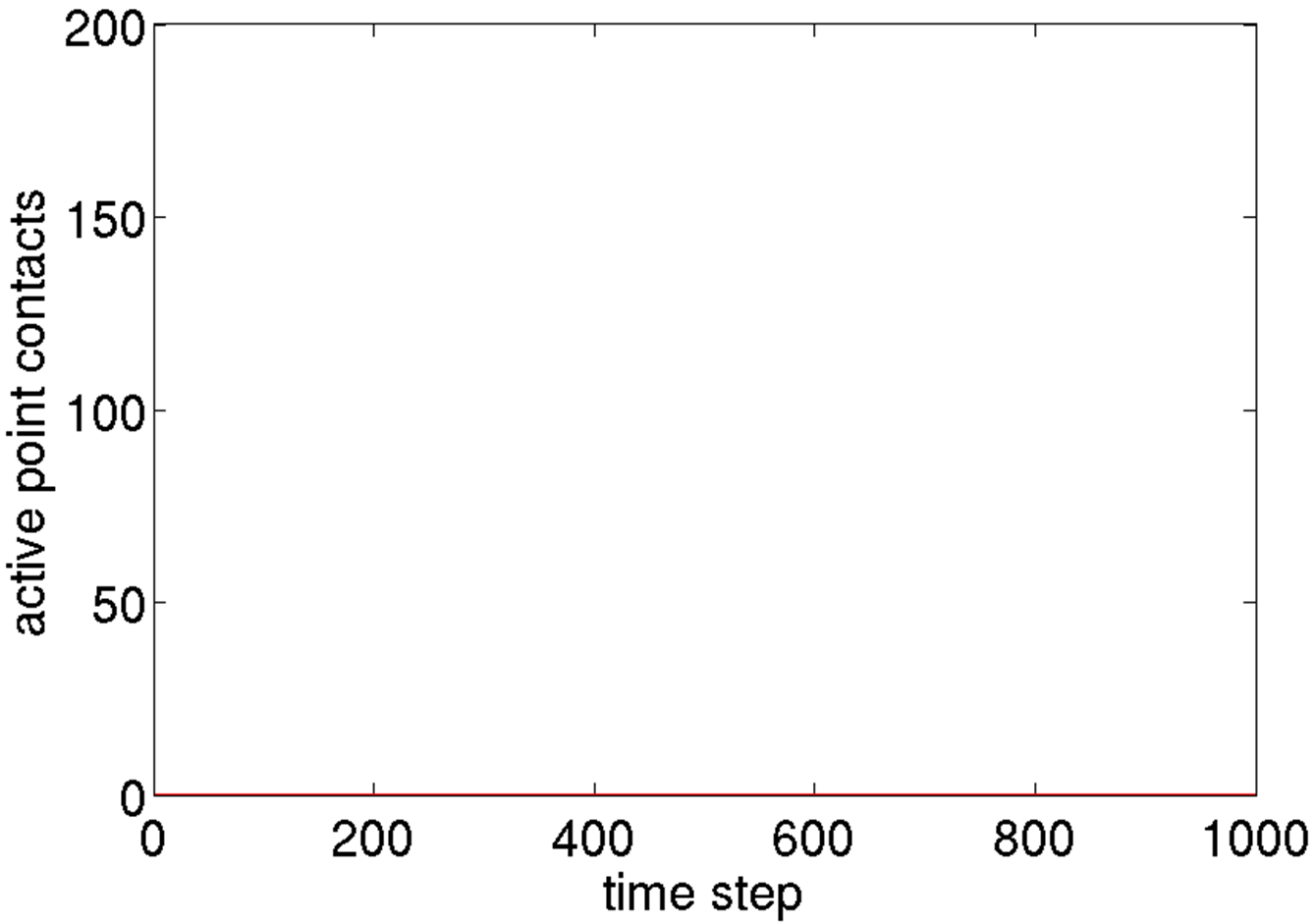}
    \label{fig:statmech_208filaments_pureline_activepoints}
   }
   \subfigure[Active line contact Gauss points.]
   {
    \includegraphics[width=0.315\textwidth]{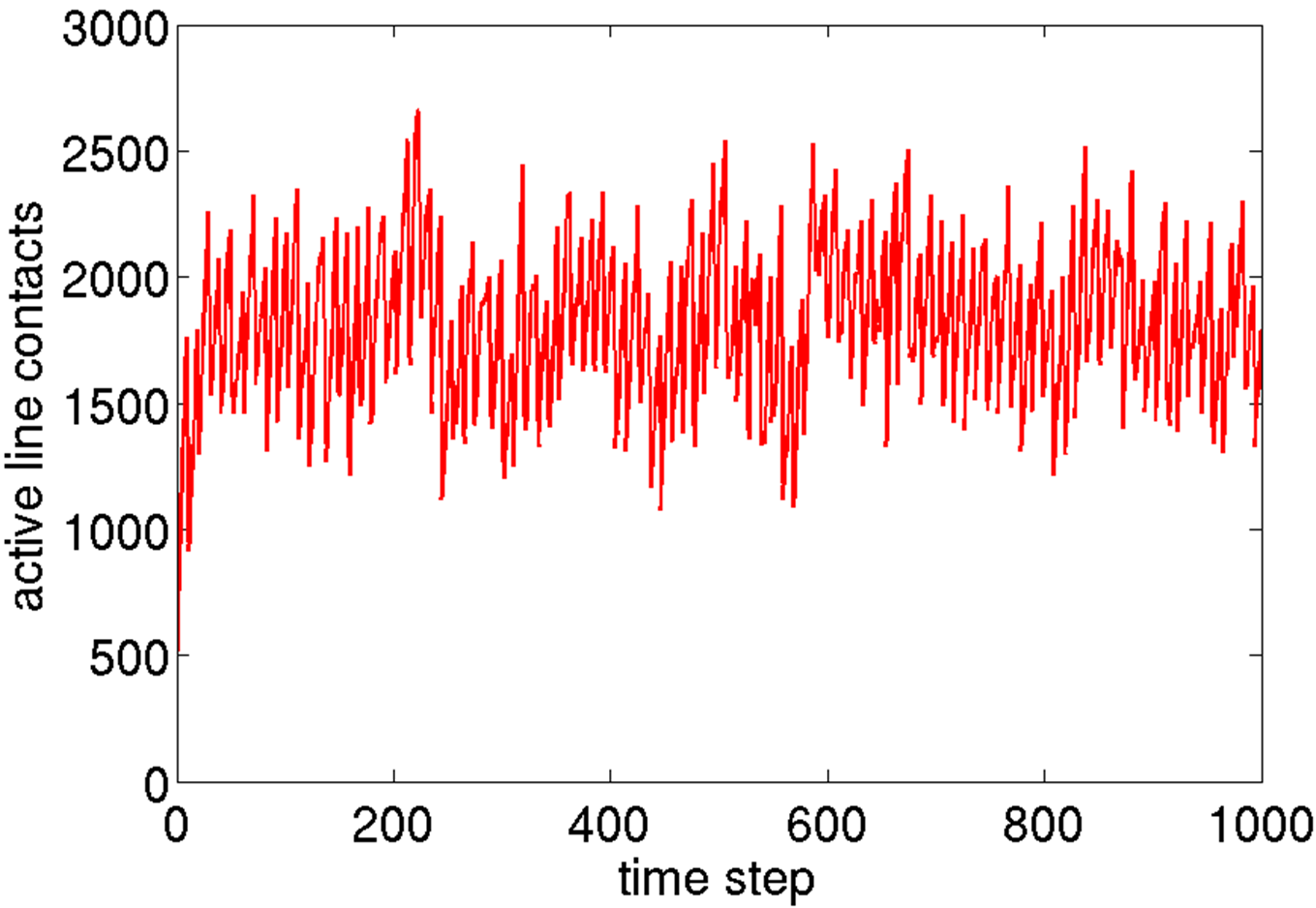}
    \label{fig:statmech_208filaments_pureline_activelines}
   }
   \subfigure[Active endpoint contacts.]
   {
    \includegraphics[width=0.305\textwidth]{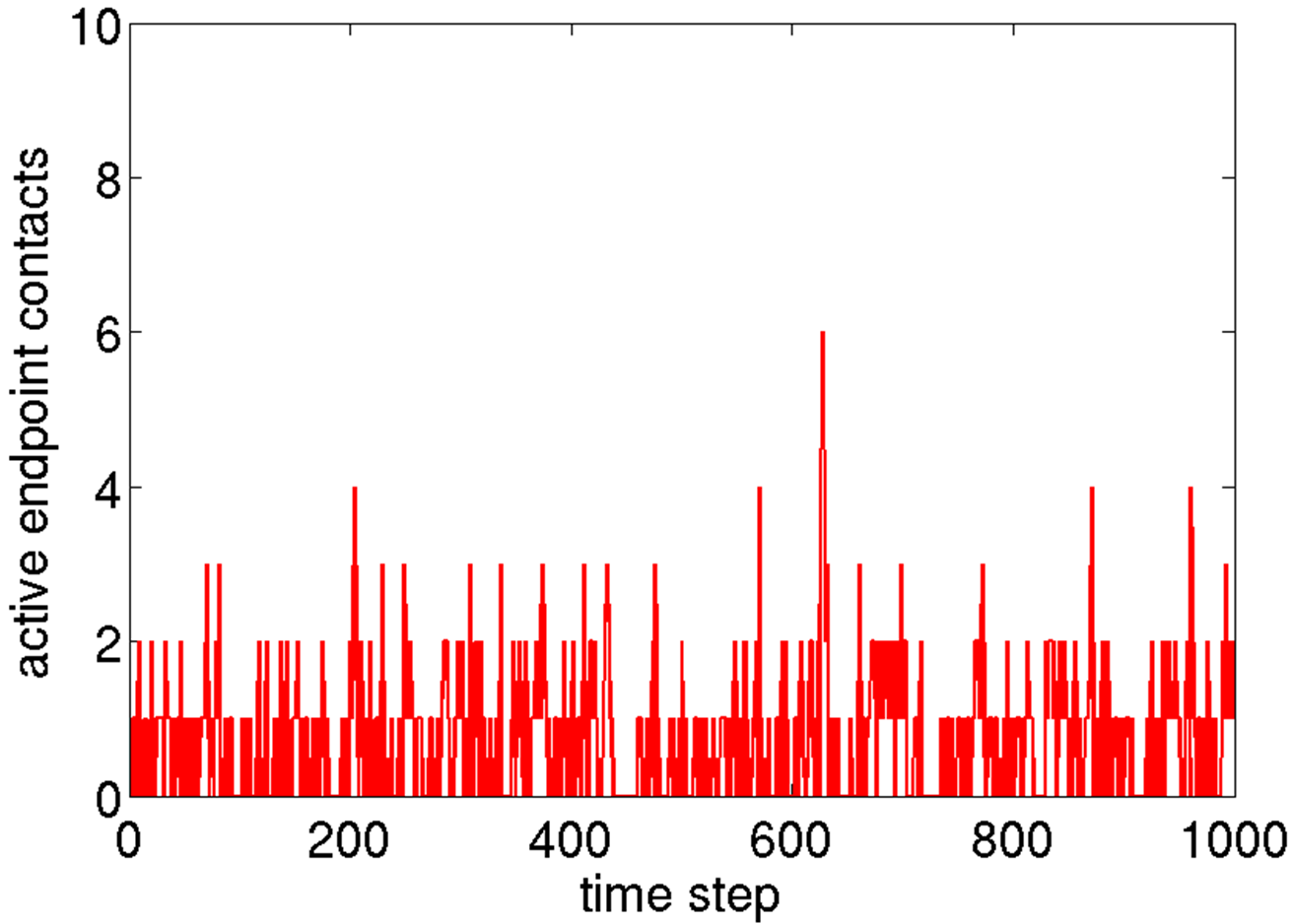}
    \label{fig:statmech_208filaments_pureline_activeendpoints}
   }
  \caption{Brownian dynamics simulation of the free diffusion of Actin filaments: Active contacts for pure line-to-line contact formulation.}
  \label{fig:statmech_208filaments_pureline_activecontacts}
\end{figure}
\begin{figure}[t]
 \centering
   \subfigure[Active point contacts.]
   {
    \includegraphics[width=0.31\textwidth]{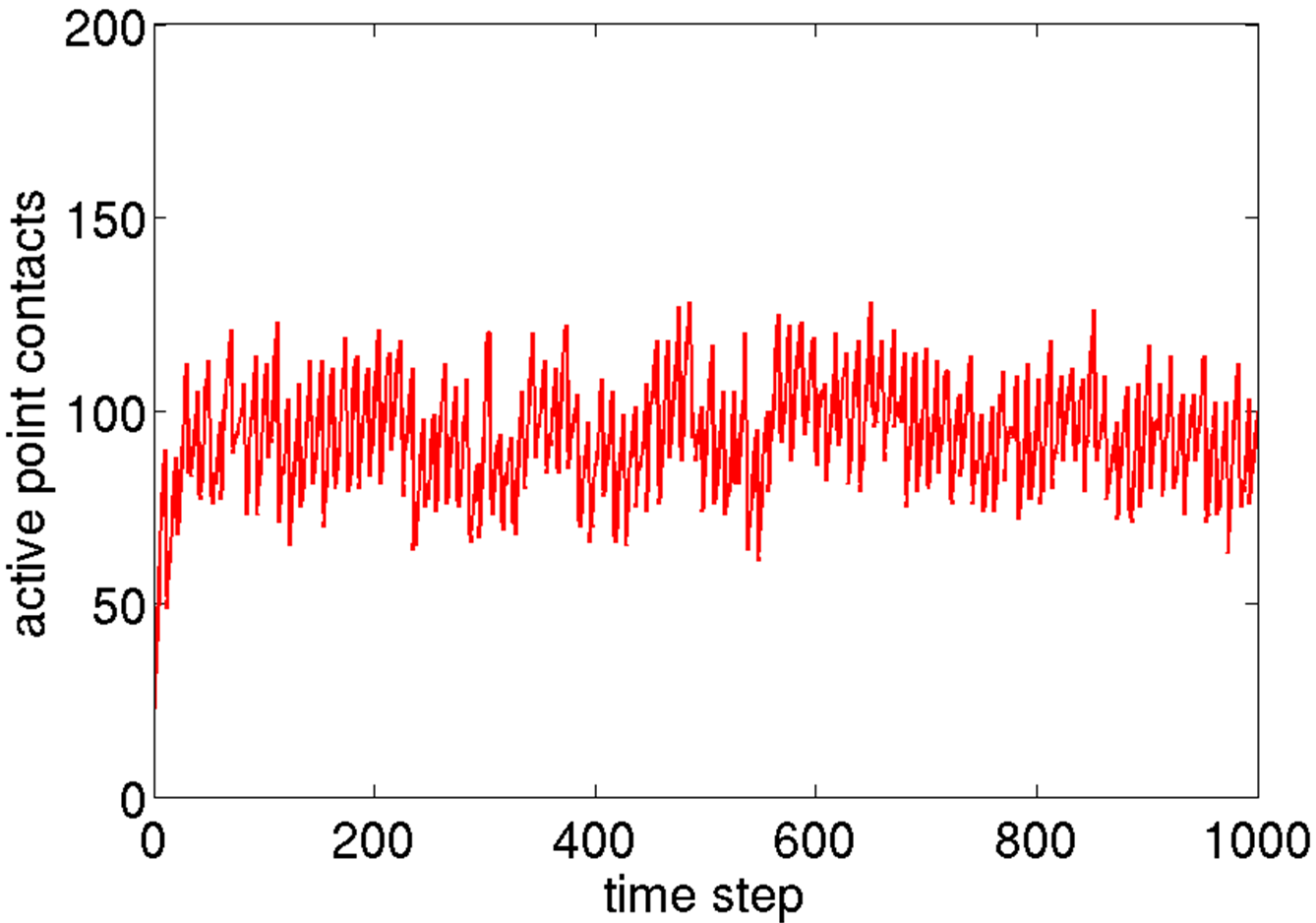}
    \label{fig:statmech_208filaments_abc_activepoints}
   }
   \subfigure[Active line contact Gauss points.]
   {
    \includegraphics[width=0.31\textwidth]{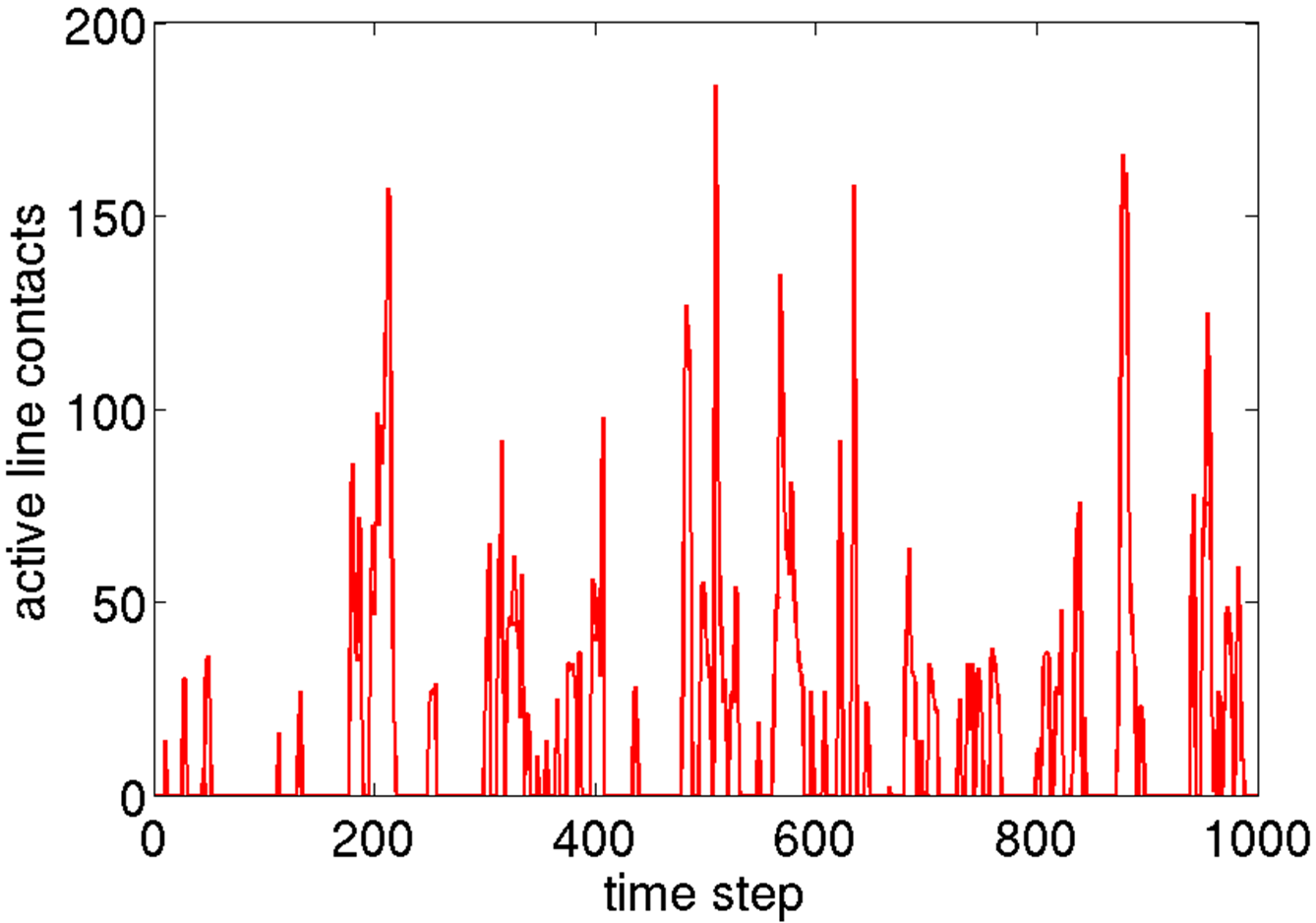}
    \label{fig:statmech_208filaments_abc_activelines}
   }
   \subfigure[Active endpoint contacts.]
   {
    \includegraphics[width=0.305\textwidth]{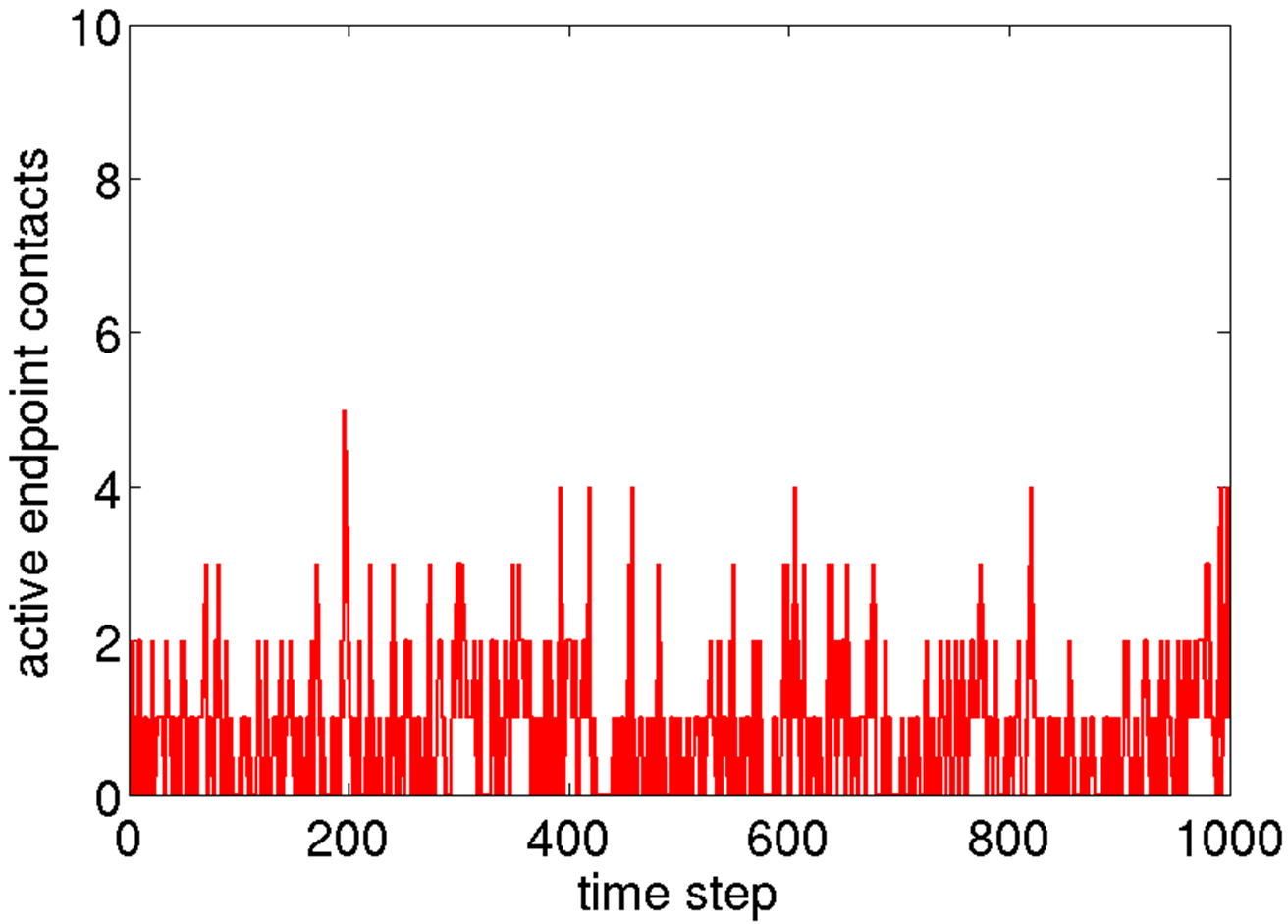}
    \label{fig:statmech_208filaments_abc_activeendpoints}
   }
  \caption{Brownian dynamics simulation of the free diffusion of Actin filaments: Active contacts for \textit{ABC} contact formulation.}
  \label{fig:statmech_208filaments_abc_activecontacts}
\end{figure}

In Figures~\ref{fig:statmech_208filaments_pureline_activecontacts} and \ref{fig:statmech_208filaments_abc_activecontacts}, the total number of active point contacts, active line contact Gauss points, and active beam endpoint contacts of the \textit{ABC} and the pure line-to-line contact formulation have been plotted over the simulation time for the time interval $t \in [0;0.1]$. While the number of active endpoint contacts is similar for both formulations and the number of active point contacts is, of course, zero for the pure line contact formulation, we observe that the new \textit{ABC} formulation could reduce the total number of active Gauss points in the line contact regime by approximately a factor of $10$. This reduction of computational effort by a factor of $10$ can be split into two individual contributions: First, the required Gauss point density could be reduced by a factor of $5$. Secondly, only a small proportion of the total set of active contacts, namely the subset with small contact angles, i.e. $\alpha<11^{\circ}$, had to be evaluated by the line-to-line contact formulation. In order to investigate also the influence of the applied two-stage contact search, we have additionally performed simulations of the \textit{ABC} formulation and the pure line contact formulation applying only the first search step, i.e. a standard octree-search with spherical bounding boxes. The resulting average contact evaluation time per Newton step (total CPU time for complete treatment of beam contact including contact search, closest-point projections, contact force and stiffness evaluations etc.) of the resulting four variants are plotted in Table~\ref{tab:contact_evaluation_times}. Accordingly, as compared to a standard beam contact algorithm consisting of a pure line contact formulation and a one-stage octree search (first line in Table~\ref{tab:contact_evaluation_times}), approximately a factor of $20$ could be saved by the proposed two-stage search.\\

These savings can be attributed to the tight cylindrical bounding boxes of the second search step yielding a very small set of potential contact segment pairs for which the active Gauss points have to be determined by means of an iterative solution of the unilateral closest point projection. For the \textit{ABC} formulation, the savings due to the second search stage (comparison of the third and fourth line in Table~\ref{tab:contact_evaluation_times}) are even more emphasized (approximately a factor of $50$), which can be explained as follows: The subdivision of the potential contact segment pairs into potential point-to-point and potential line-to-line contact segment pairs that is enabled by the two-stage search algorithm, leads to a considerable reduction of the number of unilateral closest point projections necessary in order to determine the active line contact Gauss points (which is typically much higher than the number of bilateral closest point projections in the point contact regime). Obviously, the computational savings resulting from this second search step overcompensate the required numerical effort. However, the efficiency of the second search stage depends on the number of search segments per finite element determined by the maximal segment angle $\beta_{max}$, in this example chosen as $\beta_{max}=1^{\circ}$. A comparison of the average contact evaluation times of the pure line-to-line and the \textit{ABC} formulation, both in combination with the two-stage contact search (second line and fourth line in Table~\ref{tab:contact_evaluation_times}) reveals another saving in computation time by a factor of $10$ that directly correlates with the reduced number of active Gauss points as already shown in Figures~\ref{fig:statmech_208filaments_pureline_activecontacts} and \ref{fig:statmech_208filaments_abc_activecontacts}. Thus, the new \textit{ABC} formulation in combination with the two-stage contact search leads to an overall saving by a factor of $200$ as compared to a standard line-to-line beam contact formulation with a one-stage octree search.\\

\begin{table}[t!]
\centering
\begin{tabular}{|p{3.0cm}|p{3.0cm}||p{3.0cm}|p{3.0cm}|} \hline
Formulation & Search Algorithm & $\max \, (n_{GP,tot})$ & $\bar{t}_c$ \rule{0pt}{2.6ex}\\ \hline 
Line-to-Line & 1-stage &$ \approx 2700 $&$ 1.7 \cdot 10^{1} \,\,\, sec. $ \rule{0pt}{2.6ex}\\ \hline
Line-to-Line & 2-stage &$ \approx 2700 $&$ 7.7 \cdot 10^{-1} sec.$ \rule{0pt}{2.6ex}\\ \hline
\textit{ABC} & 1-stage &$ \approx 180 $&$ 3.6 \cdot 10^{0} \,\,\, sec. $ \rule{0pt}{2.6ex}\\ \hline
\textit{ABC} & 2-stage &$ \approx 180 $&$ 7.6 \cdot 10^{-2} sec.$ \rule{0pt}{2.6ex}\\ \hline
\end{tabular}
\caption{Average contact evaluation times for different contact formulations and search algorithms.}
\label{tab:contact_evaluation_times}
\end{table}

Next, the influence of the step size control presented in Section~\ref{sec:algorithmicaspects_stepsizecontrol} will be investigated. In order to enable the corresponding investigations in an efficient manner, we want to investigate a second, smaller example of a biopolymer network consisting of only $37$ initially straight filaments as already considered in \cite{meier2015b}. The filaments of this second example are characterized by a reduced length $l=2.0$, are discretized by $8$ finite elements per filament and will be observed along a simulation time of $t \in [0.0;0.1]$. Furthermore, the line-to-line penalty parameter as well as the second shifting angle are slightly changed to $\varepsilon_{\parallel}=2.0 \cdot 10^4$ and $\alpha_2 = 15^{\circ}$. All other simulation parameters remain unchanged as compared to the first example.  In order to investigate the effectiveness of the algorithm proposed in Section~\ref{sec:algorithmicaspects_stepsizecontrol}, we have conducted one simulation with step size control of the iterative displacement increments per Newton step according 
to~\eqref{aspects_algorithm} based on a time step size of $\Delta t=1.0 \cdot 10^{-4}$ and one simulation without step size control. The standard procedure, and the simplest variant, of the latter case is based on a constant time step size that is small enough in order to avoid undetected crossing of beams. We have realized this by successively reducing the 
initial time step size $\Delta t=1.0 \cdot 10^{-4}$ by factors $0.5, 0.25, 0.1, 0.05, 0.025, 0.01$ etc. until the restriction of the displacement increment per time step~\eqref{aspects_alternativealgorithm} holds during the entire simulation, thus leading to a final time step size of $\Delta t=1.0 \cdot 10^{-7}$.\\
\begin{table}[t!]
\centering
\begin{tabular}{|p{1.9cm}|p{2.4cm}||p{2.4cm}|p{2.4cm}|p{2.4cm}|} \hline
SSC & Time Step Size & $\#$ Time Steps & $\#$ Total Iterations & $\#$ Iterations/Step\\ \hline 
No & $1.0 \cdot 10^{-4}$ &$ 1.0 \cdot 10^{3} $&$ \approx 2.0 \cdot 10^{4} $ & $\approx 20$ \rule{0pt}{2.6ex}\\ \hline
Yes & $1.0 \cdot 10^{-7}$ &$ 1.0 \cdot 10^{6} $&$ \approx 2.0 \cdot 10^{6} $ & $\approx 2$ \rule{0pt}{2.6ex}\\ \hline
\end{tabular}
\caption{Comparison of \textit{ABC} contact formulation with and without Step Size Control (SSC).}
\label{tab:number_newton_steps}
\end{table}

Table~\ref{tab:number_newton_steps} gives a comparison of the two variants ''with/without`` step size control (SSC). As a consequence of a considerably higher time step size 
(factor $1000$) and limited iterative displacement increments, the number of Newton iterations per time step is increased by a factor of $10$ for the variant with SSC, whereas 
the total number of Newton iterations during the entire simulation could be reduced approximately by a factor of $100$. The remarkable impact of these simple method can 
be explained by considering the following two aspects: First, similar to a pure time step size reduction, the step size control subdivides a given displacement into small 
sub-steps of size $R$. However, in contrary to a pure time step size reduction, the step size control does not require Newton convergence of the intermediate configurations generated 
by these sub-steps, a fact, that already saves a considerable number of overall Newton iterations. Secondly, the admissible constant time step size in case of a pure 
time step size reduction might be limited by a small number of individual time steps, whereas for the remaining time steps the displacement per time step might be much smaller than 
the beam cross section radius. The step size control on the other hand automatically adapts the number of sub-steps to the amount of total displacement within a time step, 
thus leading to the optimal number of sub-steps.\\


Of course, there exist applications where the maximal admissible time discretization error is the crucial limiting factor of the time step size. However, in many cases, especially when considering systems of highly slender filaments, the representation of the overall displacements on the length scales of the filament length are of practical interest, and not the resolution of the exact contact dynamics occurring on the length scale of the cross section radius and beyond. This applies in particular to non-deterministic systems such as the considered biopolymer networks, where averaged statistical statements efficiently generated out of a large number of individual stochastic realizations are relevant. There are many questions of interest in this field, e.g. the influence of mechanical contact interaction on filament diffusion or on the development of thermodynamically stable or unstable equilibrium phases \cite{mueller2015} in cross-linked biopolymer networks, where a robust contact simulation framework is required. In order to enable simulations along physically relevant time scales, computational efficiency is one of the key requirements for the employed algorithms.

%
\subsection{Example 4: Dynamic failure of a rope}
\label{sec:examples_rope}
\begin{figure}[t!]
 \centering
   \subfigure[Undeformed initial configuration.]
   {
    \includegraphics[width=0.31\textwidth]{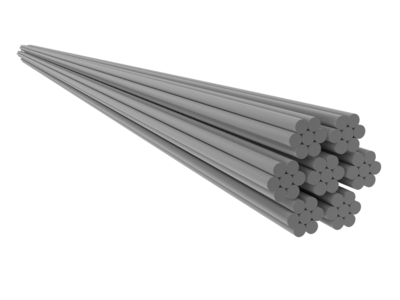}
    \label{fig:manybeams_twisting_step000}
   }
   \subfigure[Deformed configuration at load step 80.]
   {
    \includegraphics[width=0.31\textwidth]{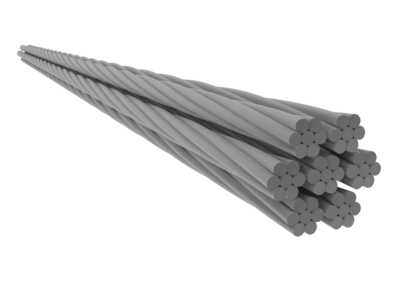}
    \label{fig:manybeams_twisting_step080}
   }
   \subfigure[Deformed configuration at load step 100.]
   {
    \includegraphics[width=0.31\textwidth]{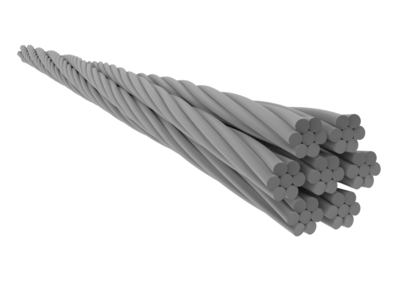}
    \label{fig:manybeams_twisting_step100}
   }
  \caption{Static simulation of the twisting process of a steel cable consisting of $7\times 7$ fibers: Deformed configurations at different load steps.}
  \label{fig:manybeams_twisting}
\end{figure}

%
In \cite{meier2015b}, the static twisting process of a rope has been investigated. The considered rope was built out of $7 \times 7$ individual fibers with length $l=5$, circular cross section of radius $R=0.01$, Young's Modulus $E=10^9$ and density $\rho=0.001$. The initial arrangement of the initially straight fibers in $7$ sub-bundles with $7$ fibers per sub-bundle as well as one intermediate and the final configuration of the twisted rope are illustrated in Figure~\ref{fig:manybeams_twisting}. For spatial discretization, $10$ beam elements per fiber were applied. The contact parameters were chosen to 
$\varepsilon_{\perp}=1.5 \cdot 10^4,\varepsilon_{\parallel}=5.0 \cdot 10^5, \bar{g}=0.1R=0.001, \alpha_1=23^{\circ}$ and $\alpha_2=25^{\circ}$ in combination with $7$ five-point integration intervals per element. The ratio $\varepsilon_{\parallel}/\varepsilon_{\perp} \approx 30$ resulted from equation~\eqref{ABC_penaltyadjustment4}. The Gauss point density was determined on the basis of equation \eqref{limitationsline_GPdistance2} with $g_{n,min}=0.1$ in combination with an adequate safety factor. The static equilibrium in the final configuration was enabled by applying proper Dirichlet conditions to all \textit{translational} degrees of freedom at both ends of the individual fibers besides the axial displacement components at one end of the rope. There, an axial tensile force $\bar{f}_{ax}=1000$ provided a certain degree of pre-stressing within the rope. The chosen shifting angles led to a pure line contact state in the final configuration.\\

\begin{figure}[t!]
 \centering
   \subfigure[Time: $0$.]
   {
    \includegraphics[width=0.31\textwidth]{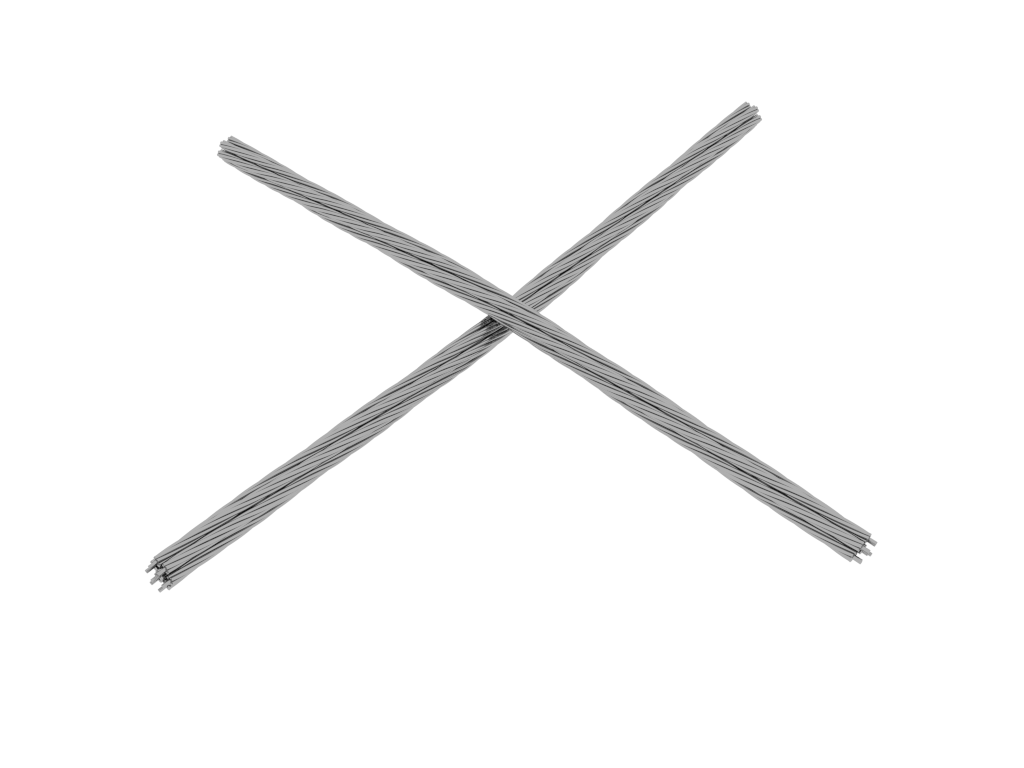}
    \label{fig:explosion_frame_000001}
   }
   \subfigure[Time: $t_{end}/8$.]
   {
    \includegraphics[width=0.31\textwidth]{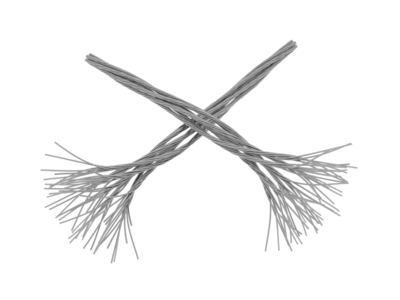}
    \label{fig:explosion_frame_000050}
   }   
   \subfigure[Time: $2t_{end}/8$.]
   {
    \includegraphics[width=0.31\textwidth]{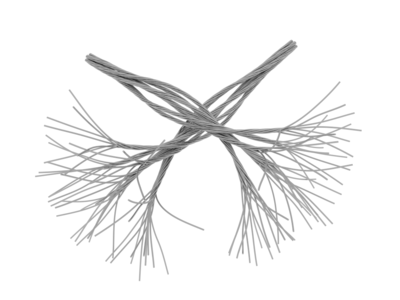}
    \label{fig:explosion_frame_000100}
   }     
   \subfigure[Time: $3t_{end}/8$.]
   {
    \includegraphics[width=0.31\textwidth]{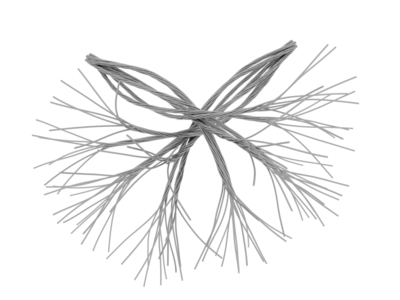}
    \label{fig:explosion_frame_000150}
   }  
   \subfigure[Time: $4t_{end}/8$.]
   {
    \includegraphics[width=0.31\textwidth]{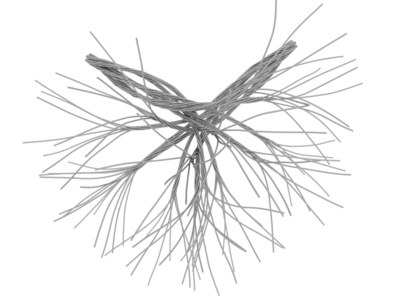}
    \label{fig:explosion_frame_000200}
   }     
   \subfigure[Time: $5t_{end}/8$.]
   {
    \includegraphics[width=0.31\textwidth]{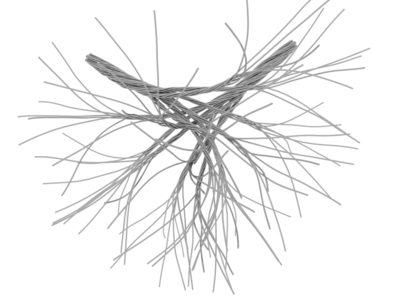}
    \label{fig:explosion_frame_000250}
   }     
   \subfigure[Time: $6t_{end}/8$.]
   {
    \includegraphics[width=0.31\textwidth]{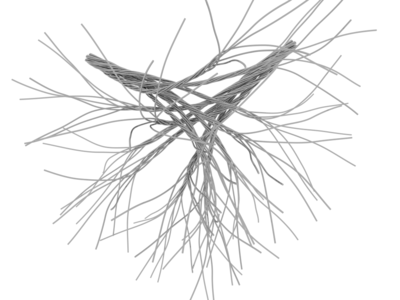}
    \label{fig:explosion_frame_000300}
   }     
   \subfigure[Time: $7t_{end}/8$.]
   {
    \includegraphics[width=0.31\textwidth]{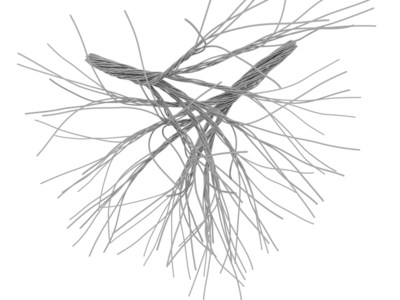}
    \label{fig:explosion_frame_000350}
   }     
   \subfigure[Time: $t_{end}$.]
   {
    \includegraphics[width=0.31\textwidth]{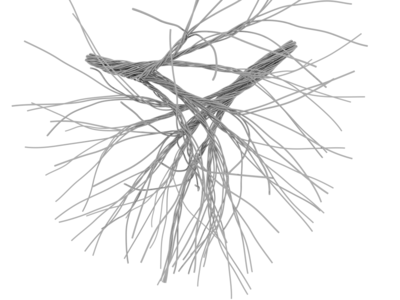}
    \label{fig:explosion_frame_000400}
   }     
  \caption{Simulation of dynamic failure of two steel cables in perpendicular contact: Deformed configurations at different time steps.}
  \label{fig:explosion_configs}
\end{figure}

\begin{figure}[t!]
 \centering  
   \subfigure[Time: $t_{end}$, small zoom factor.]
   {
    \includegraphics[width=0.48\textwidth]{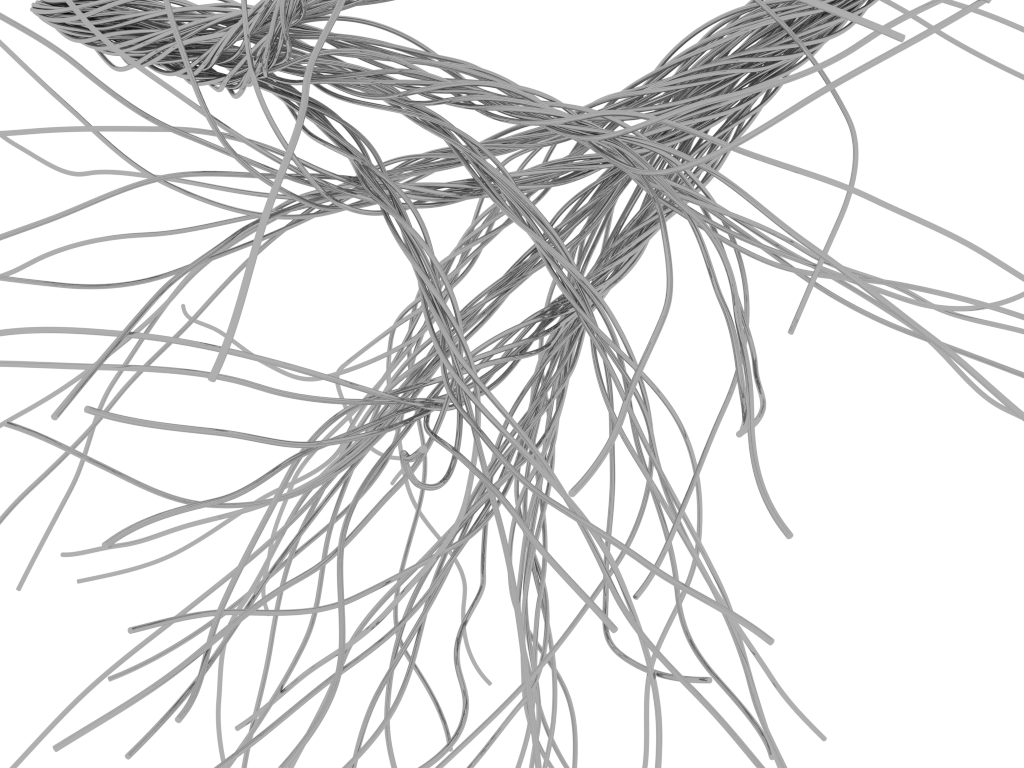}
    \label{fig:explosion_frame_000565}
   }         
   \subfigure[Time: $t_{end}$, intermediate zoom factor.]
   {
    \includegraphics[width=0.48\textwidth]{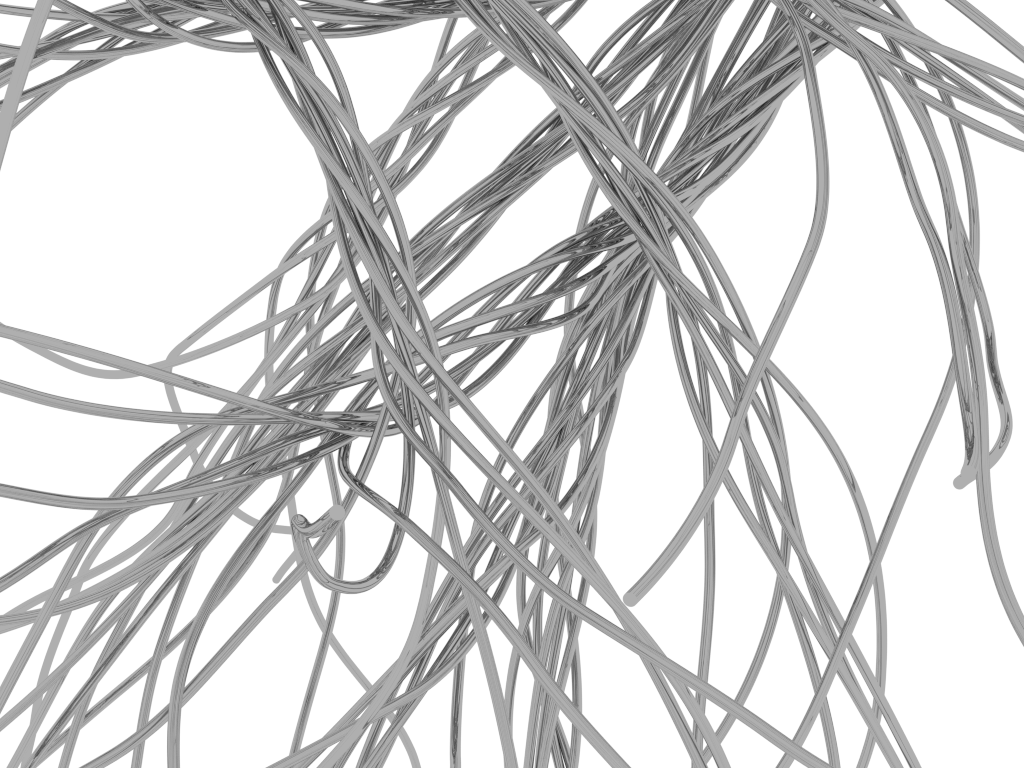}
    \label{fig:explosion_frame_000600}
   }
  \caption{Simulation of dynamic failure of two steel cables in perpendicular contact: Final configuration, small and intermediate zoom factor.}
  \label{fig:explosion_configs}
\end{figure}

\begin{figure}[t!]
 \centering         
   \subfigure[Time: $t_{end}$, high zoom factor.]
   {
    \includegraphics[width=0.97\textwidth]{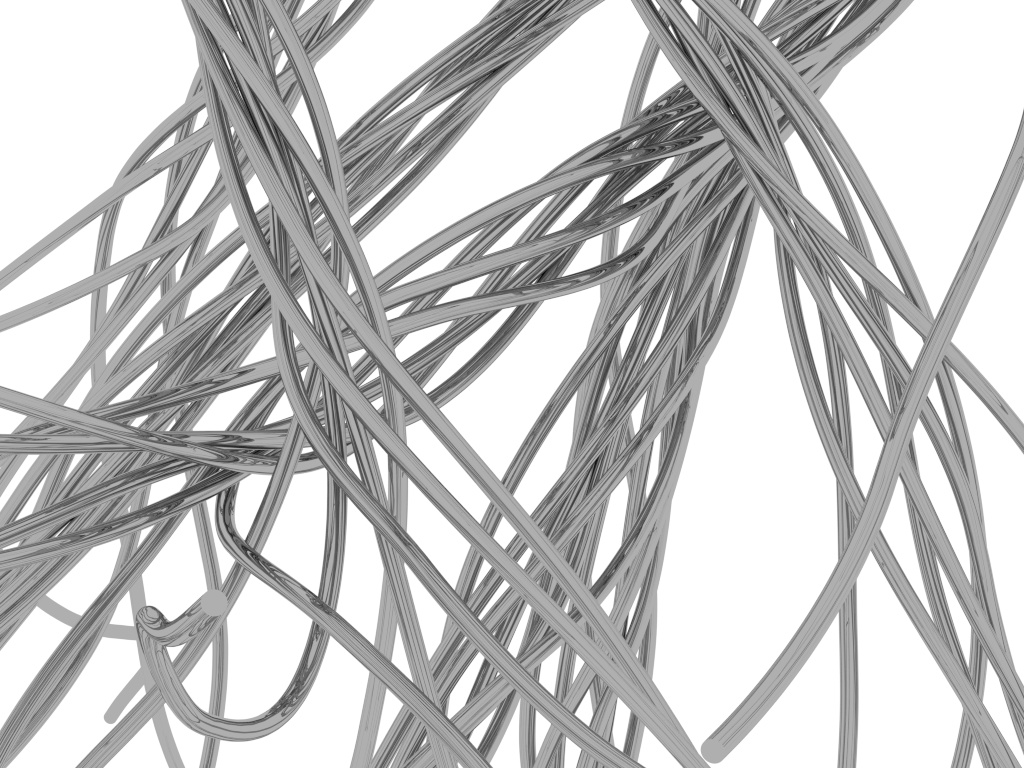}
    \label{fig:explosion_frame_000650}
   }
  \caption{Simulation of dynamic failure of two steel cables in perpendicular contact: Final configuration, high zoom factor.}
  \label{fig:explosion_configs}
\end{figure}

\begin{figure}[h]
 \centering
   \subfigure[Active point contacts.]
   {
    \includegraphics[height=0.25\textwidth]{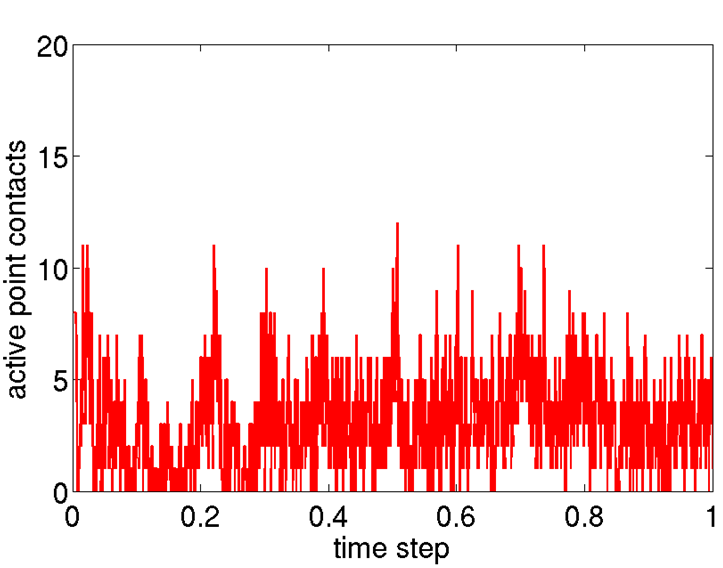}
    \label{fig:explosion_activepoints}
   }
   \subfigure[Active line contact Gauss points.]
   {
    \includegraphics[height=0.25\textwidth]{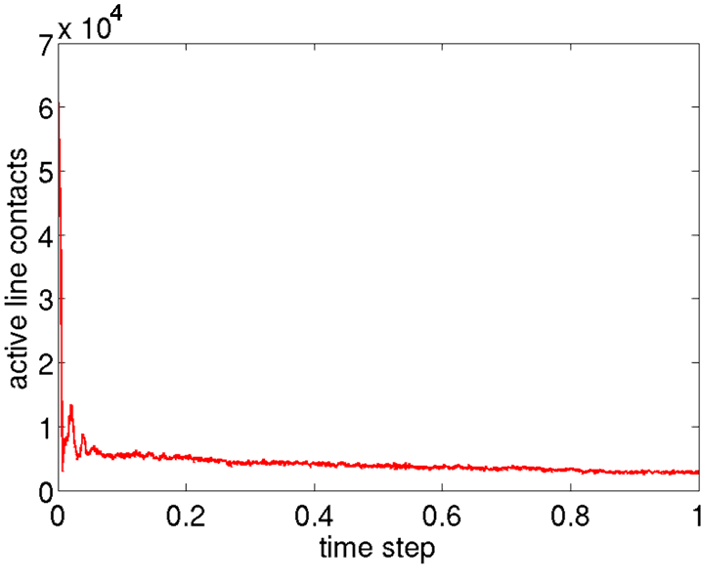}
    \label{fig:explosion_activelines}
   }
   \subfigure[Active endpoint contacts.]
   {
    \includegraphics[height=0.25\textwidth]{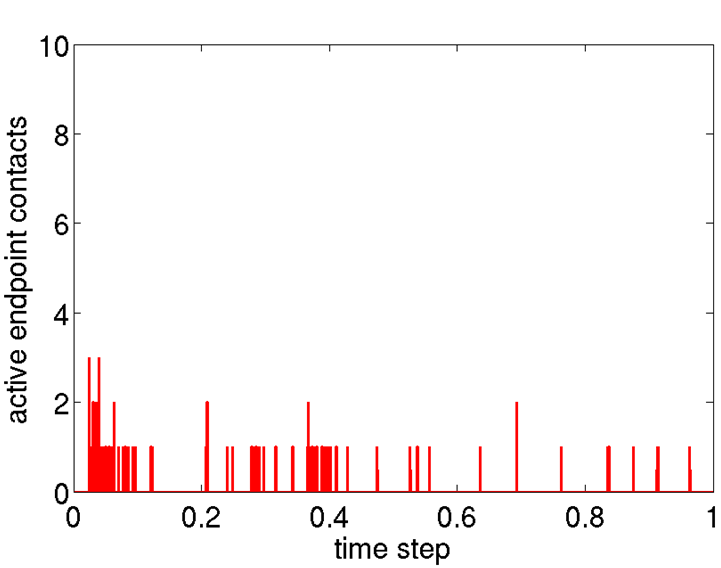}
    \label{fig:explosion_activeendpoints}
   }
  \caption{Simulation of dynamic failure of two steel cables in perpendicular contact: Active contacts.}
  \label{fig:explosion_activecontacts}
\end{figure}
\begin{figure}[t]
 \centering
   \subfigure[Minimal and maximal contact angle.]
   {
    \includegraphics[height=0.325\textwidth]{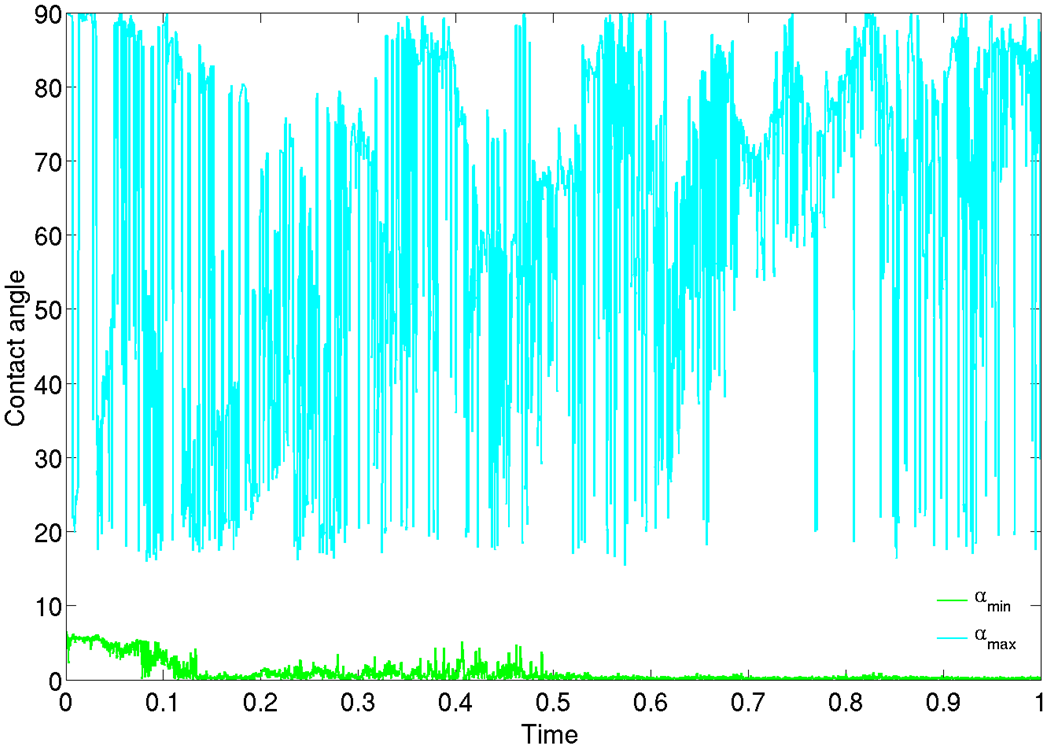}
    \label{fig:explosion_minmaxangle}
   }
    \subfigure[Evolution of total system energy.]
   {
    \includegraphics[height=0.325\textwidth]{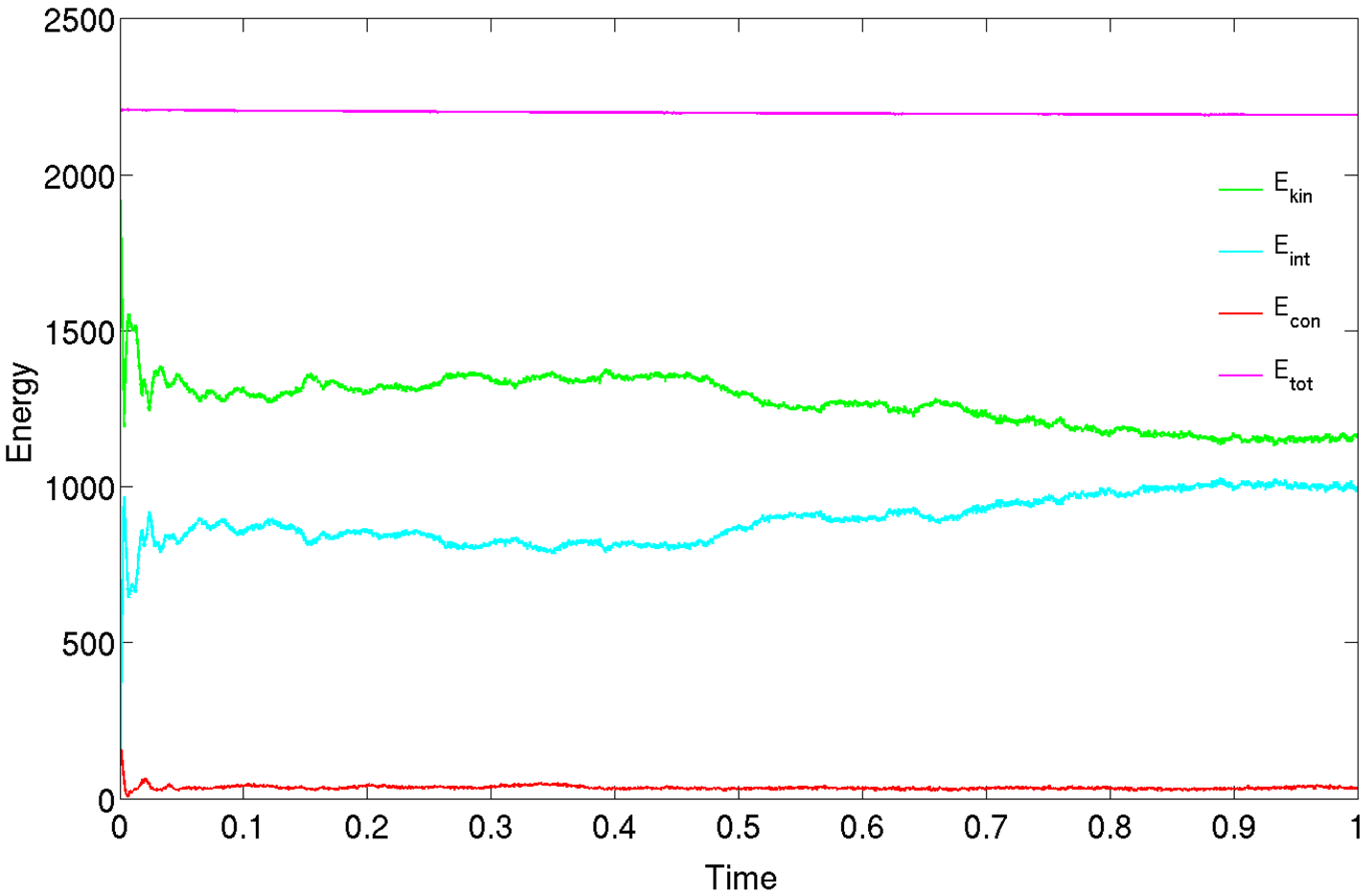}
    \label{fig:rope_dynamic_energy}
   }
  \caption{Simulation of dynamic failure of two steel cables in perpendicular contact: Contact angles and gaps.}
  \label{fig:explosion_minmaxangleandgap}
\end{figure}

Within this contribution, the dynamic failure of two ropes of this type shall be mimicked. In the initial, static equilibrium configuration, the two ropes are oriented in a perpendicular manner and contact each other (such that the imaginary undeformed rope centerlines would exactly cross each other) as illustrated in Figure~\ref{fig:explosion_frame_000001}. Additionally, the penalty parameters have been increased by a factor of ten, i.e. $\varepsilon_{\perp}=1.5 \cdot 10^5$ and $\varepsilon_{\parallel}=5.0 \cdot 10^6$, as compared to the static twisting process in~\cite{meier2015b}. Starting from this configuration, the Dirichlet fixations as well as the axial tensile forces at one of the two ends of each rope (bottom left and bottom right in Figure~\ref{fig:explosion_frame_000001}) are released, while the Dirichlet conditions at the other ends of the ropes (top left and top right in Figure~\ref{fig:explosion_frame_000001}) remain fixed. After having calculated the initial accelerations $\ddot{\mb{D}}_0$ of this non-equilibrium configuration, we start a dynamic simulation of the transient system evolution along a total simulation time of $t \in [0.0;1.5 \cdot 10^{-3}]$. For time discretization, we apply a generalized-$\alpha$ scheme in combination with a small 
amount of numerical dissipation provided by a spectral radius of $\rho_{\alpha}=0.95$ ($\alpha_f \!\approx\! 0.49$, $\alpha_m\!\approx\!0.46$, $\beta\!\approx\!0.26$ and $\gamma\!\approx\!0.53$) and a time step size of $\Delta t\!=\!1.0 \cdot 10^{-7}$. The deformed configurations at different time steps are illustrated in Figure~\ref{fig:explosion_configs}. Accordingly, the sudden release of the external (reaction-) forces leads to an initial wave propagation from the free end to the clamped end. During the entire process, a highly dynamic contact interaction between the two ropes and between the individual fibers within the ropes can be observed. This contact interaction again includes arbitrary three-dimensional contact configurations spanning the whole range of possible contact angles (see e.g. the detail views in Figures~\ref{fig:explosion_frame_000565}, \ref{fig:explosion_frame_000600} and \ref{fig:explosion_frame_000650}). This statement is confirmed by Figure~\ref{fig:explosion_minmaxangle}, where the minimal and maximal contact angle occurring in each time step is plotted over the simulation time. In Figures~\ref{fig:explosion_activecontacts}, the total number of active point-to-point contacts, active line-to-line contact Gauss points and active endpoint contacts is displayed. In this rather line-contact-dominated example, we observe a comparatively low number of point contacts in the range of $5-10$ active contacts per time step while the number of active line-to-line contact Gauss points decreases drastically from an initial value of $\approx 60000$ to $\approx 3000$ in the end of the simulation. The step size control according to Section~\ref{sec:algorithmicaspects_stepsizecontrol} allowed for comparatively large time steps, thus leading to displacements per time step in the range of four times the cross section radius, i.e. $\Delta D_{max}=\max \, (||\mb{D}(t_{i})-\mb{D}(t_{i-1})||_{\infty}) \approx 4 \cdot R$. According to Figure~\ref{fig:rope_dynamic_energy}, the total energy consisting of elastic, kinetic and contact contributions, is conserved very well despite the comparatively large time step size. The decline in total energy as a consequence of the numerical dissipation inherent to the generalized-$\alpha$ scheme with $\rho_{\alpha}=0.95$ is less than $1 \%$ during the total simulation time. Furthermore, due to the adjustment of point and line penalty parameter according to~\eqref{limitationspoint_requirementsecondderiv8}, no visible energy jumps of the force-based \textit{ABC} formulation could be observed when contact angles in the transition range~$\alpha \in [\alpha_1;\alpha_2]$ occurred. On the other hand, a simulation that has been performed without considering endpoint contacts (for comparison reasons) led to considerable jumps in the total energy by several percent - a result that underlines the importance of the endpoint contributions. 


\section{Conclusion}
\label{sec:conclusion}

The aim of this work was the development of an efficient and robust beam-to-beam contact formulation capable of modeling complex contact scenarios with arbitrary geometrical configurations in unstructured systems of highly slender fibers. It has been shown that line contact formulations represent very accurate and robust mechanical models in the range of small contact angles, whereas their computational efficiency considerably decreases with increasing contact angles. This fact can be attributed to the Gauss point densities required in the scope of high slenderness ratios. On the other hand, point contact formulations serve as sufficiently accurate and very efficient models in the regime of large contact angles, while they are inapplicable for small contact angles as a consequence of non-unique closest point projections. In order to combine the advantages of these basic formulations, a novel all-angle beam contact (\textit{ABC}) formulation has been developed that applies a point contact formulation in the range of large contact angles, a recently developed line contact formulation (see \cite{meier2015b}) in the range of small contact angles and a smooth model transition within a predefined contact angle interval. Concretely, two types of model transition have been investigated: a variationally consistent one formulated on penalty potential level as well as a simpler force-based model transition. It has been shown analytically that both variants exactly fulfill the essential conservation properties of linear and angular momentum. However, only the potential-based formulation fulfills exact conservation of energy. Nevertheless, for many fields of application the more efficient force-based model transition is recommended as method of choice, since the non-conservative work contributions of this variant can be minimized by choosing an optimal ratio of the point and line penalty parameters. Furthermore, optimal parameter choices concerning the required Gauss point densities and the model transition shifting angles have been derived. All deformation-dependent quantities have been consistently linearized, thus enabling the application within the framework of implicit time integration. Eventually, the proposed \textit{ABC} formulation has been supplemented by the contact contributions of the beam endpoints as introduced in \cite{meier2015b}.\\ 

Besides the requirement of highly resolved spatial contact discretizations, the modeling of thin fibers by means of standard beam contact formulations is also limited by small time step sizes. In order to address this limitation, we have proposed a step size control for the nonlinear solution scheme that enables displacements per time step far above the order of the cross section radius. Additionally, we have proposed a very efficient two-stage contact search consisting of an octree search with spherical bounding boxes in the first step and dynamically adapted search segments enwrapped by tight cylindrical search boxes in the second step. The second search step yields a very tight set of potential contact pairs and allows for subdividing this set into potential point-to-point and potential line-to-line contact pairs. This search strategy allows us to fully exploit the efficiency potential of the proposed all-angle beam contact formulation, with regard to two different aspects: On the one hand, a lower number of potential contact pairs has to be evaluated by the computationally more involved line-to-line contact formulation, viz. only the ones within the range of small contact angles. On the other hand, lower Gauss point densities are required within this range. Furthermore, the accuracy and consistency of the \textit{ABC} formulation in terms of conservation properties and contact force distributions has been verified numerically. Finally, two possible real-life applications for this formulation have been investigated in order to evaluate the robustness and efficiency of the overall contact algorithm. It could be verified that the proposed methods yield a beam-to-beam contact algorithm that combines a significant degree of robustness and universality in the implicit(!) numerical treatment of complex contact scenarios and arbitrary beam-to-beam orientations with a considerably increased computational efficiency compared to existing formulations, especially in the range of high slenderness ratios. For the investigated example, both the reduction in the total number of Newton iterations enabled by the step size control as well as the savings in contact evaluation time resulting from a combination of \textit{ABC} formulation and two-stage contact search are in the range of two orders of magnitude. Thus, the cumulative savings resulting from these two effects are in the range of four orders of magnitude. When considering examples with strong contact interaction, high beam slenderness ratios and comparatively rough spatial FEM discretizations enabled by powerful higher-order beam element formulations, it is precisely the contact evaluation time that may dominate the overall computational costs. Therefore, substantial savings in this scope are of highest practical relevance.

\appendix


%
\section{Residual contributions and linearization of the applied beam element formulation}
\label{anhang:reslin_beamelement}
%
The weak form of the balance equations of the considered torsion-free beam formulation (see also \cite{meier2015b}) reads
\begin{align}
\label{weakform}
\int \limits_0^l
  \Bigg[
  \delta \epsilon EA \epsilon 
 +\delta \boldsymbol{\kappa} EI \boldsymbol{\kappa} 
 +\delta \mb{r}^T \rho A \ddot{\boldsymbol{r}}
  \Bigg] ds
  -\int \limits_0^l
  \Bigg[ 
 \delta \mb{r}^T \mb{\tilde{f}}
 +\delta \boldsymbol{\theta^T_{\perp}} \mb{\tilde{m}_{\perp}}
  \Bigg] ds
 -\Bigg[\delta \mb{r}^T \bar{\mb{f}}+ \delta \boldsymbol{\theta^T_{\perp}} \bar{\mb{m}}_{\perp} \Bigg]_{\Gamma_{\sigma}} \hspace{-0.3cm}
 = 0. 
\end{align}
Here, $\rho$ is the mass density, $A$ the cross section area, $I$ the moment of inertia and $E$ the Young's modulus. Furthermore, $v=v(s)$ represents the material velocity field,
while $\epsilon=\epsilon(s)$ and $\kappa=\kappa(s)$ are the fields of axial tension and bending curvature. The quantities $\mb{\tilde{f}}$ and $\mb{\tilde{m}_{\perp}}$ denote distributed 
forces and moments, whereas $\bar{\mb{f}}$ and $\bar{\mb{m}}_{\perp}$ denote discrete point forces and 
moments on the Neumann boundary $\Gamma_{\sigma}$ of the beam. We have applied the following abbreviations:
\begin{align}
\label{abbriviations}
  \delta \epsilon = \frac{\delta \mb{r}^{\prime T} \mb{r}^{\prime}}{||\mb{r}^{\prime}||}, \quad
  \delta \boldsymbol{\kappa} = \frac{
                               ||\mb{r}^{\prime}||^2 \left(\delta \mb{r}^{\prime} \times \mb{r}^{\prime  \prime} + \mb{r}^{\prime} \times \delta \mb{r}^{\prime  \prime}\right)
                               -2 \left( \delta \mb{r}^{\prime T}\mb{r}^{\prime} \right) \left( \mb{r}^{\prime} \times \mb{r}^{\prime  \prime} \right)
                               }{||\mb{r}^{\prime}||^4} \quad \text{and} \quad
  \delta \boldsymbol{\theta_{\perp}}=\frac{\mb{r}^{\prime} \times \delta \mb{r}^{\prime}}{||\mb{r}^{\prime}||^2}.
\end{align}
The torsion-free beam theory is only applicable if the external moment vectors contain no components parallel to the centerline 
tangent vector, i.e. $\mb{r}^{\prime T}(s)\mb{\tilde{m}_{\perp}}(s) \equiv 0 \, \forall \, s \in [0,l]$ and $\mb{r}^{\prime T}\bar{\mb{m}}_{\perp} \equiv 0 \, \text{on} \,\Gamma_{\sigma}$.
Based on~\eqref{weakform}, the residual contributions $\mb{r}_{int}, \mb{r}_{kin}$ and $\mb{r}_{ext}$ of one beam element due to internal, inertia and external forces can be derived:
\begin{align}
\label{res_beamelement}
\begin{split}
   \mb{r}_{int}&=\int \limits_{-1}^{1} \left[ \mb{N}^{\prime T} \! \! \left(EA \mb{t_1}+EI\mb{t_2} \right)
   + \mb{N}^{\prime \prime T} \! EI\mb{t_3}  \right]\frac{l_{ele}}{2}d\xi,  \quad
  \mb{r}_{kin}= \int \limits_{-1}^{1} \! \mb{N}^T \! \rho A \ddot{\boldsymbol{r}} \frac{l_{ele}}{2}d\xi, \\
  \mb{r}_{ext}&= - \int \limits_{-1}^{1} \left[\mb{N}^T \mb{\tilde{f}} + \mb{N}^{\prime T} \! \left( \mb{\tilde{m} \times \mb{t_4}} \right) \right]\frac{l_{ele}}{2}d\xi
  -  \Bigg[\mb{N}^T \bar{\mb{f}} + \mb{N}^{\prime T} \! \left(\bar{\mb{m}} \times \mb{t_4} \right) \Bigg]_{\varGamma_{\sigma}} = 0.
\end{split}
\end{align}
Here, we have introduced the following additional abbreviations:
\begin{align}
\mb{t_1} := \frac{\mb{r}^{\prime}}{||\mb{r}^{\prime}||} \left( ||\mb{r}^{\prime}||-1 \right), \, \, \,
\mb{t_2} := \frac{2\mb{r}^{\prime}(\mb{r}^{\prime T} \mb{r}^{\prime \prime})^2}{||\mb{r}^{\prime}||^6}-\frac{\mb{r}^{\prime}(\mb{r}^{\prime \prime T}
\mb{r}^{\prime \prime })
+\mb{r}^{\prime \prime }(\mb{r}^{\prime T} \mb{r}^{\prime \prime})}{||\mb{r}^{\prime}||^4}, \, \, \,
\mb{t_3} := \frac{\mb{r}^{\prime \prime}}{||\mb{r}^{\prime}||^2}-\frac{\mb{r}^{\prime}(\mb{r}^{\prime T} \mb{r}^{\prime \prime})}{||\mb{r}'||^4}, \,\,\,
\mb{t_4} := \frac{\mb{r}^{\prime}}{||\mb{r}^{\prime}||^2}.
\end{align}
Consequently, we obtain the following expressions for the corresponding linearizations $\mb{k}_{int}, \mb{k}_{kin}$ and $\mb{k}_{ext}$:
\begin{align}
\label{linres_beamelement}
\begin{split}
   \mb{k}_{int}&=\frac{\partial \mb{r}_{int}}{\partial \mb{d}}=\int \limits_{-1}^{1} \left[ \mb{N}^{\prime T} \! \! \left(EA \frac{\partial \mb{t_1}}{\partial \mb{d}}
   +EI \frac{\partial \mb{t_2}}{\partial \mb{d}} \right)
   + \mb{N}^{\prime \prime T} \! EI \frac{\partial \mb{t_3}}{\partial \mb{d}}  \right]\frac{l_{ele}}{2}d\xi,  \quad
     \mb{k}_{kin}=\frac{\partial \mb{r}_{kin}}{\partial \mb{d}}= \rho A \ddot{d}_{,d} \int \limits_{-1}^{1} \! \mb{N}^T \mb{N} \frac{l_{ele}}{2}d\xi  = \text{const.}, \\
  \mb{k}_{ext}&=\frac{\partial \mb{r}_{ext}}{\partial \mb{d}}= - \int \limits_{-1}^{1} \left[\mb{N}^T \mb{\tilde{f}} + \mb{N}^{\prime T} \! \left( \mb{S}(\mb{\tilde{m}) \frac{\partial \mb{t_4}}{\partial \mb{d}}} \right) \right]\frac{l_{ele}}{2}d\xi
  -  \Bigg[\mb{N}^T \bar{\mb{f}} + \mb{N}^{\prime T} \! \left( \mb{S}(\bar{\mb{m}}) \frac{\partial \mb{t_4}}{\partial \mb{d}} \right) \Bigg]_{\varGamma_{\sigma}} = 0,
\end{split}
\end{align}
where $\ddot{d}_{,d}$ is typically a constant factor depending on the applied time integration scheme and $\mb{S}(.)$ is a skew-symmetric matrix that represents the 
cross-product, i.e. $\mb{S}(\mb{a})\mb{b}=\mb{a}\times \mb{b} \, \forall \, \mb{a}, \mb{b} \in \Re^3$. Additionally, we have:
\begin{align}
\begin{split}
 \frac{\partial \mb{t_1}}{\partial \mb{d}} & = \left[ \frac{\left(||\mb{r}^{\prime}||-1\right)}{||\mb{r}^{\prime}|| }  \mb{I}_3
 +\frac{1}{||\mb{r}^{\prime}||^3} \left(\mb{r}^{\prime} \otimes \mb{r}^{\prime T} \right)
 \right]\mb{N}^{\prime}, \\
 \frac{\partial \mb{t_2}}{\partial \mb{d}}
   & = \Bigg[ 
   \left\{ \frac{2(\mb{r}^{\prime T} \mb{r}^{\prime \prime})^2}{||\mb{r}^{\prime}||^6}-\frac{(\mb{r}^{\prime \prime T} \mb{r}^{\prime \prime})}{||\mb{r}^{\prime}||^4}  \right\} \mb{I}_3
  +\left\{ \frac{-12(\mb{r}^{\prime T} \mb{r}^{\prime \prime})^2}{||\mb{r}^{\prime}||^8}+\frac{4(\mb{r}^{\prime \prime T} \mb{r}^{\prime \prime})}{||\mb{r}^{\prime}||^6}  \right\} 
   \left(\mb{r}^{\prime} \otimes \mb{r}^{\prime T} \right)
  +\frac{4(\mb{r}^{\prime T} \mb{r}^{\prime \prime})}{||\mb{r}^{\prime}||^6} \left(\mb{r}^{\prime} \otimes \mb{r}^{\prime \prime T} \right) \\
  & \,\,\,\,\,\,\,\, +\frac{4(\mb{r}^{\prime T} \mb{r}^{\prime \prime})}{||\mb{r}^{\prime}||^6} \left(\mb{r}^{\prime \prime} \otimes \mb{r}^{\prime T} \right)
  -\frac{1}{||\mb{r}^{\prime}||^4} \left(\mb{r}^{\prime \prime} \otimes \mb{r}^{\prime \prime T} \right)
   \Bigg]\mb{N}^{\prime} \\
  & + \left[ -\frac{(\mb{r}^{\prime T} \mb{r}^{\prime \prime})}{||\mb{r}^{\prime}||^4}  \mb{I}_3
  +\frac{4(\mb{r}^{\prime T} \mb{r}^{\prime \prime})}{||\mb{r}^{\prime}||^6} \left(\mb{r}^{\prime} \otimes \mb{r}^{\prime T} \right)
  -\frac{2}{||\mb{r}^{\prime}||^4} \left(\mb{r}^{\prime} \otimes \mb{r}^{\prime \prime T} \right)
  -\frac{1}{||\mb{r}^{\prime}||^4} \left(\mb{r}^{\prime \prime} \otimes \mb{r}^{\prime T} \right)
  \right]\mb{N}^{\prime \prime}, \\
 \frac{\partial \mb{t_3}}{\partial \mb{d}} & = \!
  \left[ -\frac{(\mb{r}^{\prime T} \mb{r}^{\prime \prime})}{||\mb{r}^{\prime}||^4}  \mb{I}_3
 +\frac{4(\mb{r}^{\prime T} \mb{r}^{\prime \prime})}{||\mb{r}^{\prime}||^6} \left(\mb{r}^{\prime} \otimes \mb{r}^{\prime T} \right)
 -\frac{2}{||\mb{r}^{\prime}||^4} \left(\mb{r}^{\prime \prime} \otimes \mb{r}^{\prime T} \right)
 -\frac{1}{||\mb{r}^{\prime}||^4} \left(\mb{r}^{\prime} \otimes \mb{r}^{\prime \prime T} \right)
 \right]\mb{N}^{\prime} \\
  & + \left[ \frac{1}{||\mb{r}^{\prime}||^2}  \mb{I}_3
 -\frac{1}{||\mb{r}^{\prime}||^4} \left(\mb{r}^{\prime} \otimes \mb{r}^{\prime T} \right)
 \right]\mb{N}^{\prime \prime}, \\
  \frac{\partial \mb{t_4}}{\partial \mb{d}} & = \! \left[ \frac{1}{||\mb{r}^{\prime}||^2 }  \mb{I}_3
 -\frac{2}{||\mb{r}^{\prime}||^4}  \left(\mb{r}^{\prime} \otimes \mb{r}^{\prime T} \right)
 \right]\mb{N}^{\prime}
\end{split}
\end{align}

It can easily be shown that in the absence of external moments, i.e. $\mb{\tilde{m}}\!=\!\mb{\bar{m}}\!=\!\mb{0}$, the overall stiffness matrix is symmetric. In~\cite{meier2015}, the so-called \textit{MCS} method has been proposed in order to avoid membrane locking in the range of high beam slenderness ratios. If this method is applied, it is sensible to slightly reformulate the element residual contribution due to axial tension. 
Eventually, the original contribution $\mb{r}_{int,EA}$ and the alternative MCS contribution read:
\begin{align}
\label{res_beamelementmcs}
\begin{split}
   \mb{r}_{int,EA}&=EA \! \int \limits_{-1}^{1} \mb{N}^{\prime T}  \mb{t_1} \frac{l_{ele}}{2}d\xi = 
   EA \int \limits_{-1}^{1} \left(\frac{\partial \epsilon (\xi)}{\partial \mb{d}}\right)^T \! \! \epsilon(\xi) \, \frac{l_{ele}}{2}d\xi \quad \text{with} \quad
   \epsilon=||\mb{r}^{\prime}||-1, \quad 
   \frac{\partial \epsilon}{\partial \mb{d}}=\frac{\mb{r}^{\prime T}\mb{N}^{\prime}}{||\mb{r}^{\prime}||},\\
   \bar{\mb{r}}_{int,EA}&=
   EA \int \limits_{-1}^{1} \left(\frac{\partial \epsilon(\xi^i)}{\partial \mb{d}}\right)^T \! \!  L^i(\xi) \, L^j(\xi) \, \epsilon(\xi^j) \, \frac{l_{ele}}{2}d\xi 
   \quad \text{with} \quad i,j=1,2,3; \quad \xi^1\!\!=\!-1,\,\xi^2\!\!=\!0,\,\xi^3\!\!=\!1.
\end{split}
\end{align}
Accordingly, the corresponding contributions of the axial tension terms to the element stiffness matrix yield:
\begin{align}
\label{linres_beamelementmcs}
\begin{split}
   \mb{k}_{int,EA}&=
   EA \int \limits_{-1}^{1} \left[ \left(\frac{\partial^2 \epsilon (\xi)}{\partial \mb{d}^2}\right)^T \! \! \epsilon(\xi) 
    +\left(\frac{\partial \epsilon (\xi)}{\partial \mb{d}}\right)^T \left(\frac{\partial \epsilon (\xi)}{\partial \mb{d}}\right) 
\right] \frac{l_{ele}}{2} d\xi \quad \text{with} \quad
    \frac{\partial^2 \epsilon (\xi)}{\partial \mb{d}^2}=
    \frac{\mb{N}^{\prime T}}{||\mb{r}^{\prime}||}\left( \mb{I}_3 - \frac{\mb{r}^{\prime} \otimes \mb{r}^{\prime T}}{||\mb{r}^{\prime}||^2} \right)\mb{N}^{\prime},\\
       \bar{\mb{k}}_{int,EA}&=
   EA \int \limits_{-1}^{1} \left[ \left(\frac{\partial^2 \epsilon(\xi^i)}{\partial \mb{d}^2}\right)^T \! \!  L^i(\xi) \, L^j(\xi) \, \epsilon(\xi^j) 
   + \left(\frac{\partial \epsilon(\xi^i)}{\partial \mb{d}}\right)^T \! \!  L^i(\xi) \, L^j(\xi) \, \left(\frac{\partial \epsilon(\xi^i)}{\partial \mb{d}}\right) \right ]\frac{l_{ele}}{2}d\xi.
\end{split}
\end{align}
In equations \eqref{res_beamelementmcs} and \eqref{linres_beamelementmcs}, the summation convention over the repeated indices $i$ and $j$ applies.

%
\section{Conservation properties of the \textit{ABC} formulation and applied beam element}
\label{anhang:conservation}
%

The discretized weak form of the beam equilibrium equations~\eqref{weakform} (see \cite{meier2015b} for details) supplemented by the contact contributions \eqref{ABC_potentialbased_weakform} is satisfied for 
all test functions $\delta \mb{r}_h \in \mathcal{V}_h$. In the following, we choose the specific test function
\begin{align}
\label{rigid_translation}
\delta \mb{r}_{1h}\!=\! \delta \mb{r}_{2h}\!=\! \mb{u}_0 \in \mathcal{V}_h \quad \text{with} \quad \mb{u}_0^{\prime} = \mb{0}  
\quad \rightarrow \quad \delta \mb{d}_1(\mb{u}_0)=\delta \mb{d}_2(\mb{u}_0)=(\mb{u}_0^T,\mb{0}^T,\mb{u}_0^T,\mb{0}^T)^T,
\end{align}
representing a rigid body translation. Inserting this test function $\delta \mb{r}_h\!=\! \mb{u}_0$ into the weak form \eqref{weakform} leads to the following global force balance 
for the applied beam element in the absence of mechanical contact interaction:
\begin{align}
\label{conservation_linmomentum_ele}
\mb{\dot{L}}=\mb{F}_{ext}
\quad \text{with} \quad \mb{L}:= \int \limits_0^{l} \rho A \dot{\mb{r}} ds, \, \, \mb{F}_{ext}=\int \limits_0^{l} \mb{\tilde{f}} ds + \Big[\bar{\mb{f}}\Big]_{\Gamma_{\sigma}} \hspace{-0.3cm}
\end{align} 
and consequently to exact conservation of linear momentum $\mb{L}$ for the unloaded system, i.e. if $\mb{F}_{ext}=\mb{0}$. We will show below, that inserting \eqref{rigid_translation} 
into~\eqref{ABC_potentialbased_weakform} yields a vanishing overall contact contribution to the weak form:
\begin{align}
\label{conservation_linmomentum}
\delta \Pi_{c\varepsilon}(\delta \mb{r}_{1h}\!=\! \delta \mb{r}_{2h}\! =\! \mb{u}_0) = 0.
\end{align}
In other words, the discrete contact forces at the contact interface exactly balance each other and global conservation of linear momentum according to \eqref{conservation_linmomentum_ele} is 
preserved. In order to prove this statement, we first realize from \eqref{point_weakform} and \eqref{line_pen_weakform} that $\delta g(\delta \mb{r}_{1h}\!=\! \delta \mb{r}_{2h}\!=\!\mb{u}_0)=0$, which 
already yields vanishing contact force terms (terms on the left) in \eqref{ABC_potentialbased_weakform}. Furthermore, inserting \eqref{rigid_translation} into the expressions for 
$ d \xi / d \mb{d}_{12}$ and $ d \eta / d \mb{d}_{12}$ presented in \ref{anhang:linearizationendpoint} and \ref{anhang:linearizationlinecontact} delivers the trivial result that the closest 
point projections are not influenced by a rigid body translation of the entire system, i.e. 
$ d \xi (\delta \mb{r}_{1h}\!=\! \delta \mb{r}_{2h}\! =\! \mb{u}_0) / d \mb{d}_{12} = d \eta (\delta \mb{r}_{1h}\!=\! \delta \mb{r}_{2h}\!=\! \mb{u}_0) / d \mb{d}_{12} \!= \! \mb{0}$. 
Inserting this result together with $\mb{N}_1^{\prime} \delta \mb{d}_1(\mb{u}_0)=\mb{N}_2^{\prime} \delta \mb{d}_2(\mb{u}_0)=\mb{0}$ into \eqref{ABC_potentialbased_momentcontributions2} leads
to $\delta z=\delta \alpha =0$, and therefore also to vanishing contact moment contributions in \eqref{ABC_potentialbased_weakform}. This concludes the proof of 
conservation of linear momentum~\eqref{conservation_linmomentum}. In order to investigate conservation of angular momentum, we again choose a specific test function based on a 
spatially constant vector $\boldsymbol{\omega}_0$
\begin{align}
\label{rigid_rotation}
\delta \mb{r}_{ih}\!=\! \boldsymbol{\omega}_0 \times \mb{r}_{ih} \in \mathcal{V}_h \,\,\,\, \text{with} \,\,\,\, \boldsymbol{\omega}_0^{\prime} = \mb{0}  
\,\, \rightarrow \,\, \delta \mb{d}_i(\boldsymbol{\omega}_0)=((\boldsymbol{\omega}_0 \times \mb{\hat{d}}_i^{1})^T,(\boldsymbol{\omega}_0 \times \mb{\hat{t}}_i^{1})^T,
(\boldsymbol{\omega}_0 \times \mb{\hat{d}}_i^{2})^T,(\boldsymbol{\omega}_0 \times \mb{\hat{t}}_i^{2})^T)^T\!\!, \,\,\,\, i=1,2
\end{align}
representing a rigid body rotation. Inserting $\delta \mb{r}_h\!=\! \boldsymbol{\omega}_0 \times \mb{r}_{h}$ into the weak form \eqref{weakform} leads to the following global moment balance 
for the applied beam element in the absence of contact interaction
\begin{align}
\label{conservation_angmomentum_ele}
\mb{\dot{H}}=\mb{M}_{ext}
\quad \text{with} \quad \mb{H}:= \int \limits_0^l \mb{r} \times \rho A \dot{\mb{r}} ds, \, \, 
\mb{M}_{ext}=\int \limits_0^l \left(\mb{r} \times \mb{\tilde{f}} + \mb{\tilde{m}}\right) ds 
+ \Big[\mb{r} \times \bar{\mb{f}} + \bar{\mb{m}}\Big]_{\Gamma_{\sigma}} \hspace{-0.3cm}
\end{align} 
and consequently to exact conservation of angular momentum $\mb{H}$ for the unloaded system, i.e. if $\mb{M}_{ext}=\mb{0}$. We will show below that inserting \eqref{rigid_rotation} 
into~\eqref{ABC_potentialbased_weakform} yields a vanishing overall contact contribution to the weak form:
\begin{align}
\label{conservation_linmomentum}
\delta \Pi_{c\varepsilon}(\delta \mb{r}_{ih}\!=\! \boldsymbol{\omega}_0 \times \mb{r}_{ih}) = 0.
\end{align}
In other words, the contact moments at the contact interface exactly balance each other and conservation of angular momentum according to \eqref{conservation_angmomentum_ele} is 
preserved. In order to prove this statement, we first insert \eqref{rigid_rotation} into \eqref{point_weakform} and \eqref{line_pen_weakform}:
\begin{align}
\label{vanishing_gap}
\delta g(\delta \mb{r}_{ih}\!=\! \boldsymbol{\omega}_0 \times \mb{r}_{ih})=
\left(\boldsymbol{\omega}_0 \times \mb{r}_{1h} - \boldsymbol{\omega}_0 \times \mb{r}_{2h} \right)^T \mb{n} =  \boldsymbol{\omega}_0^T \left[ \left(\mb{r}_{1h}-\mb{r}_{2h} \right) \times \mb{n}\right]=0.
\end{align}
Again, inserting \eqref{rigid_rotation} into the expressions for $ d \xi / d  \mb{d}_{12}$ and $d \eta / d \mb{d}_{12}$ presented in \ref{anhang:linearizationendpoint} and \ref{anhang:linearizationlinecontact} yields 
the trivial result that the closest point projections are not influenced by a rigid body rotation, i.e. 
$d \xi (\delta \mb{r}_{ih}\!=\! \boldsymbol{\omega}_0 \times \mb{r}_{ih}) / d \mb{d}_{12}  = d \eta (\delta \mb{r}_{ih}\!=\! \boldsymbol{\omega}_0 \times \mb{r}_{ih}) / d \mb{d}_{12} \!= \! \mb{0}$. 
An evaluation of the remaining terms in \eqref{ABC_potentialbased_momentcontributions2} finally gives:
\begin{align}
\label{conservation_rigidrotation_deltaz}
   \delta  z  = \mb{v}_1^T \left(\boldsymbol{\omega}_0 \times \mb{r}_{1h}^{\prime} \right) + \mb{v}_2^T \left( \boldsymbol{\omega}_0 \times \mb{r}_{2h}^{\prime} \right) =
   \frac{\mb{r}_{2}^{\prime T}\left(\boldsymbol{\omega}_0 \times \mb{r}_{1h}^{\prime} \right)}{||\mb{r}_{1}^{\prime}||\,||\mb{r}_{2}^{\prime}||}+
   \frac{\mb{r}_{1}^{\prime T}\left(\boldsymbol{\omega}_0 \times \mb{r}_{2h}^{\prime} \right)}{||\mb{r}_{1}^{\prime}||\,||\mb{r}_{2}^{\prime}||}
   =0,
\end{align}
where we have used the relation $(\boldsymbol{\omega}_0 \times \mb{r}_{ih})^{\prime}=\boldsymbol{\omega}_0 \times \mb{r}_{ih}^{\prime}$. The results 
of \eqref{vanishing_gap} and \eqref{conservation_rigidrotation_deltaz} complete the proof of \eqref{conservation_linmomentum}.
Finally, we want to investigate the conservation of energy. Thereto, we choose the test functions according to
\begin{align}
\label{conservation_velocity}
\delta \mb{r}_{ih}\!=\! \dot{\mb{r}}_{ih} \in \mathcal{V}_h, \,\,\,\, i=1,2
\end{align}
representing the current velocity field. Inserting \eqref{conservation_velocity} into the weak form \eqref{weakform} leads to the following global mechanical power balance 
for the applied beam element in the absence of contact interaction
\begin{align}
\label{conservation_energy_ele}
\dot{E}_{kin}+\dot{E}_{int} = P_{ext} 
\quad \text{with} \quad 
P_{ext} = 
\int \limits_0^l
  \Bigg[ 
 \dot{\mb{r}}^T \mb{\tilde{f}}
 +\boldsymbol{\omega^T_{\perp}} \mb{\tilde{m}_{\perp}}
  \Bigg] ds
  +\Bigg[\dot{\mb{r}}^T \bar{\mb{f}} + \boldsymbol{\omega^T_{\perp}} \bar{\mb{m}}_{\perp} \Bigg]_{\Gamma_{\sigma}}
  \quad \text{and} \quad
\boldsymbol{\omega_{\perp}}=\frac{\mb{r}^{\prime} \times  \dot{\mb{r}}^{\prime}}{||\mb{r}^{\prime}||^2}
\end{align}
and consequently to exact energy conservation for the unloaded system, i.e. if $P_{ext}=0$. The contact contributions~\eqref{ABC_potentialbased_weakform} 
have been derived from the potential~\eqref{ABC_potentialbased_potential} under consistent consideration of the spatial discretization~\eqref{interpolation}, i.e.
\begin{align}
\begin{split}
\label{conservation_energy_contact}
\delta \Pi_{c\varepsilon }  = 
 \sum \limits_{i=1}^2 \left( \frac{\partial \Pi_{c\varepsilon}}{\partial \mb{r}_{i}} \frac{d \mb{r}_{i}}{d \mb{d}_{12}} \delta \mb{d}_{12}
+\frac{\partial \Pi_{c\varepsilon }}{\partial \mb{r}_{i}^{\prime}} \frac{d \mb{r}_{i}^{\prime}}{d \mb{d}_{12}} \delta \mb{d}_{12} \right), \quad
\frac{d \mb{r}_{1}}{d \mb{d}_{12}} & =\left[ \left(\mb{N}_{1},\mb{0} \right) + \mb{r}_{h}^{\prime} \frac{d \xi}{d \mb{d}_{12}} \right],
\frac{d \mb{r}_{1}^{\prime}}{d \mb{d}_{12}} =\left[ \left(\mb{N}_{1}^{\prime},\mb{0} \right) + \mb{r}_{1}^{\prime \prime} \frac{d \xi}{d \mb{d}_{12}} \right], \\
\frac{d \mb{r}_{2}}{d \mb{d}_{12}} & =\left[ \left(\mb{N}_{2},\mb{0} \right) + \mb{r}_{2}^{\prime} \frac{d \xi}{d \mb{d}_{12}} \right], \,\,
\frac{d \mb{r}_{2}^{\prime}}{d \mb{d}_{12}} =\left[ \left(\mb{N}_{2}^{\prime},\mb{0} \right) + \mb{r}_{2}^{\prime \prime} \frac{d \xi}{d \mb{d}_{12}} \right].
\end{split}
\end{align}
Therefore, by replacing the variations $\delta(.)$ with time derivatives $\dot{(.)}$, the correspondingly discretized contact contributions~\eqref{ABC_potentialbased_weakform} 
per definition represent the rate $\dot{\Pi}_{c \varepsilon }$ of the discrete penalty potential. After adding the contact terms to the contributions 
of internal, kinetic and external forces in \eqref{conservation_energy_ele}, we finally get
\begin{align}
\label{conservation_energy_eleandcontact}
\dot{E}_{kin}+\dot{E}_{int} +\dot{\Pi}_{c \varepsilon }= P_{ext}, 
\end{align}
which again implies conservation of the sum of kinetic, internal and penalty energy in the absence of external forces, i.e. if $P_{ext}=0$. In case of the non-conservative variant of a ''force-based transition'', no potential of the contact forces exists and the term $\dot{\Pi}_{\varepsilon }$ in 
\eqref{conservation_energy_eleandcontact} has to be replaced by the negative power of the contact forces $-P_{c \varepsilon }$, viz.
\begin{align}
\label{conservation_energy_eleandcontact}
P_{c \varepsilon} = 
\left(\dot{\mb{r}}_1-\dot{\mb{r}}_2 \right)^T \mb{f}_{c\varepsilon \perp}
+\int \limits_0^{l_1}
  \Bigg[ 
 \left(\dot{\mb{r}}_1-\dot{\mb{r}}_2\right)^T \mb{f}_{c\varepsilon \parallel}
  \Bigg] ds_1, \,\,\,\, 
\mb{f}_{c\varepsilon \perp}=f_{c\epsilon \perp}\mb{n}, \,\,\,\, 
\mb{f}_{c\varepsilon \parallel}=f_{c\varepsilon \parallel}\mb{n}.
\end{align}

%
\section{Linearization of point-to-point, endpoint-to-line and endpoint-to-endpoint contact contributions}
\label{anhang:linearizationendpoint}
%

Since the endpoint contact contributions can be regarded as a special case of the point contact formulation, we start with the linearization 
of this formulation. The linearization of~\eqref{point_discreteweakform} has the following general form: 
\begin{align}
\label{point_general_lin}
\mb{k}_{con,l}=\dfrac{d \mb{r}_{con,l}}{d \mb{d}_{12} }=
  \dfrac{\partial \mb{r}_{con,l}}{\partial \mb{d}_{12} } + \dfrac{\partial \mb{r}_{con,l}}{\partial \xi_c}\dfrac{d \xi_c}{d \mb{d}_{12} }
+ \dfrac{\partial \mb{r}_{con,l}}{\partial \eta_c}\dfrac{d \eta_c}{d \mb{d}_{12} }
  \quad \text{for} \quad l=1,2.
\end{align}
Here, the derivatives $d \xi_c/d \mb{d}_{12}$ and $d \eta_c/d \mb{d}_{12}$ stem from a linearization of the orthogonality conditions~\eqref{point_orthocond}:
\begin{align}
\begin{split}
\label{point_linorthogonality}
         \mb{A} (\xi_c,\eta_c) \cdot
         \left(
         \frac{d \xi_c}{d \mb{d}_{12}}^T,
         \frac{d \eta_c}{d \mb{d}_{12}}^T
         \right)^T
         & =  -\mb{B} (\xi_c,\eta_c), \\    
         \text{with} \quad \mb{A} & 
         =\left(
         \begin{matrix}
         p_{1,\xi} & p_{1,\eta} \\
         p_{2,\xi} & p_{2,\eta} \\     
         \end{matrix}
         \right)
         =\left(
         \begin{matrix}
         \mb{r}_{1,\xi}^T \mb{r}_{1,\xi} +(\mb{r}_1-\mb{r}_2)^T \mb{r}_{1,\xi \xi} & -\mb{r}_{1,\xi}^T \mb{r}_{2,\eta} \\
         \mb{r}_{1,\xi}^T \mb{r}_{2,\eta} & -\mb{r}_{2,\eta}^T \mb{r}_{2,\eta} +(\mb{r}_1-\mb{r}_2)^T \mb{r}_{2,\eta \eta}       
         \end{matrix}
         \right),\\
         \text{and} \quad \mb{B} & 
         =\left(
         \begin{matrix}
         p_{1,\mb{d}_{12}} \\
         p_{2,\mb{d}_{12}}      
         \end{matrix}
         \right)
         \hspace{0.058\textwidth}=\left(
         \begin{matrix}
         (\mb{r}_1-\mb{r}_2)^T \mb{N}_{1,\xi} + \mb{r}_{1,\xi}^T \mb{N}_{1} & -\mb{r}_{1,\xi}^T \mb{N}_{2} \\
         \mb{r}_{2,\eta}^T \mb{N}_{1} & (\mb{r}_1-\mb{r}_2)^T \mb{N}_{2,\eta} - \mb{r}_{2,\eta}^T \mb{N}_{2}       
         \end{matrix}
         \right) .
\end{split}
\end{align}
Here, the terms $p_{1,\xi}, p_{1,\eta}, p_{2,\xi}$ and $p_{2,\eta}$, which are collected in matrix $\mb{A}$, can be used for an iterative 
solution of the orthogonality conditions~\eqref{point_orthocond} for the unknown closest point coordinates $\xi_c$ and $\eta_c$ by means of 
a local Newton-Raphson scheme. The partial derivatives of the residual vectors with respect to $\mb{d}_{12}$
as occurring in~\eqref{point_general_lin} are given by:
\begin{align}
\label{point_discreteweakformapp2}
\begin{split}
  \frac{\partial \mb{r}_{con,1}}{\partial \mb{d}_{12}} & =  
  \varepsilon \left( 
  \mb{N}_{1}^T \mb{n} \frac{\partial g}{\partial \mb{d}_{12}}
  + g \mb{N}_{1}^T  \frac{\partial \mb{n}}{\partial \mb{d}_{12}}
  \right), \quad
    \frac{\partial \mb{r}_{con,2}}{\partial \mb{d}_{12}} =  
  \varepsilon \left( 
  \mb{N}_{2}^T \mb{n} \frac{\partial g}{\partial \mb{d}_{12}}
  + g \mb{N}_{2}^T  \frac{\partial \mb{n}}{\partial \mb{d}_{12}}
  \right), \\
  \frac{\partial g}{\partial \mb{d}_{12}} & = \mb{n}^T \left[ \mb{N}_{1}, -\mb{N}_{2} \right], \quad
  \frac{\partial \mb{n}}{\partial \mb{d}_{12}} = \frac{\mb{I}_3-\mb{n} \otimes \mb{n}^T}{||\mb{r}_1-\mb{r}_2||} \left[ \mb{N}_{1}, -\mb{N}_{2} \right].
\end{split}
\end{align}
Correspondingly, the partial derivatives with respect to the closest point coordinates $\xi_c$ and $\eta_c$ take the following form:
\begin{align}
\label{point_discreteweakformapp3}
\begin{split}
  \frac{\partial \mb{r}_{con,1}}{\partial \xi_c} & =  
  \varepsilon \left( 
  \mb{N}_{1}^T \mb{n} g_{,\xi}
  + g \mb{N}_{1,\xi}^T \mb{n}
  + g \mb{N}_{1}^T  \mb{n}_{,\xi}
  \right)\big|_{(\xi_c,\eta_c)}, \quad
  \frac{\partial \mb{r}_{con,2}}{\partial \xi_c}  =  
  \varepsilon \left( 
  \mb{N}_{2}^T \mb{n} g_{,\xi}
  + g \mb{N}_{2}^T  \mb{n}_{,\xi}
  \right)\big|_{(\xi_c,\eta_c)}, \\
   \frac{\partial \mb{r}_{con,1}}{\partial \eta_c} & =  
  \varepsilon \left( 
  \mb{N}_{1}^T \mb{n} g_{,\eta}
  + g \mb{N}_{1}^T  \mb{n}_{,\eta}
  \right)\big|_{(\xi_c,\eta_c)}, \quad
  \frac{\partial \mb{r}_{con,2}}{\partial \eta_c}  =  
  \varepsilon \left( 
  \mb{N}_{2}^T \mb{n} g_{,\eta}
  + g \mb{N}_{2,\eta}^T \mb{n}
  + g \mb{N}_{2}^T  \mb{n}_{,\eta}
  \right)\big|_{(\xi_c,\eta_c)}, \\
  g_{,\xi} &= \mb{n}^T \mb{r}_{1,\xi}, \quad
  g_{,\eta}= -\mb{n}^T \mb{r}_{2,\eta}, \quad
  \mb{n}_{,\xi}=\frac{\mb{I}_3-\mb{n} \otimes \mb{n}^T}{||\mb{r}_1-\mb{r}_2||}\mb{r}_{1,\xi}, \quad
  \mb{n}_{,\eta}=-\frac{\mb{I}_3-\mb{n} \otimes \mb{n}^T}{||\mb{r}_1-\mb{r}_2||}\mb{r}_{2,\eta}.
\end{split}
\end{align}
Depending on the case (point-, line- or endpoint-contact), \eqref{point_discreteweakformapp3} can be simplified due to $\mb{n}^T\mb{r}_{1,\xi}=0$ and/or $\mb{n}^T\mb{r}_{2,\eta}=0$.
In case of endpoint contact, only the partial derivatives $d \xi_c / d\mb{d}_{12}$ and $d \eta_c / d \mb{d}_{12}$ have to be adapted, while 
all other terms remain unchanged. In case of contact between an endpoint of beam 1, i.e. $\xi_c=-1$ or $\xi_c=1$, with a segment $\eta_c \in [-1;1]$ on beam 2, we consider the second 
line of~\eqref{point_linorthogonality} in order to determine $d \eta_c / d\mb{d}_{12}$, while $d \xi_c / d \mb{d}_{12}$ vanishes:
\begin{align}
\frac{d \xi_c}{d \mb{d}_{12}} = \mb{0} \quad \text{and} \quad
         \frac{d \eta_c}{d \mb{d}_{12}} = -\frac{p_{2,\mb{d}_{12}}}{p_{2,\eta}}.
\end{align}
Correspondingly, the condition $p_{2}(\eta_c)=0$ and the derivative $p_{2,\eta}$ can be used for an iterative determination of $\eta_c$. 
In case of contact between an endpoint of beam 2, i.e. $\eta_c=-1$ or $\eta_c=1$, with a curve segment $\xi_c \in [-1;1]$ on beam 1, we have to consider the first line of~\eqref{point_linorthogonality} 
in order to determine $d \xi_c / d \mb{d}_{12}$, while $d \eta_c / d \mb{d}_{12}$ vanishes:
\begin{align}
\frac{d \xi_c}{d \mb{d}_{12}} = -\frac{p_{1,\mb{d}_{12}}}{p_{1,\xi}} \quad \text{and} \quad
\frac{d \eta_c}{d \mb{d}_{12}} = \mb{0}.     
\end{align}
In this case, the condition $p_{1}(\xi_c)=0$ and the derivative $p_{1,\xi}$ can be used for an iterative determination of $\xi_c$.
When the contact between two endpoints is considered, i.e. $\xi_c=-1$ or $\xi_c=1$ and $\eta_c=-1$ or $\eta_c=1$, we have the condition:
\begin{align}
\frac{d \xi_c}{d \mb{d}_{12}}  = \frac{d \eta_c}{d \mb{d}_{12}} = \mb{0}.     
\end{align}

%
\section{Linearization of the line-to-line contact formulation}
\label{anhang:linearizationlinecontact}
%
The linearization of the contributions $\mb{r}_{con,1}^{ij}$ and $\mb{r}_{con,2}^{ij}$ of one individual Gauss point (see~\eqref{linediscreteweakform}) has the following form:
\begin{align}
\label{line_linearization_app}
\begin{split}
\mb{k}_{con,l}^{ij} =\dfrac{d \mb{r}_{con,l}^{ij}}{d \mb{d}_{12}} & =
  \dfrac{\partial \mb{r}_{con,l}^{ij}}{\partial \mb{d}_{12}} 
 +\dfrac{\partial \mb{r}_{con,l}^{ij}}{\partial \xi_{ij}}\dfrac{d \xi_{ij}}{d \mb{d}_{12}} 
 +\dfrac{\partial \mb{r}_{con,l}^{ij}}{\partial \eta_c}\dfrac{d \eta_c}{d \mb{d}_{12}}
 +\dfrac{\partial \mb{r}_{con,l}^{ij}}{\partial \xi_{1,i}}\dfrac{d \xi_{1,i}}{d \mb{d}_{12}}
 +\dfrac{\partial \mb{r}_{con,l}^{ij}}{\partial \xi_{2,i}}\dfrac{d \xi_{2,i}}{d \mb{d}_{12}}, \quad l=1,2, \\
  \text{with} \quad \dfrac{d \xi_{ij}}{d \mb{d}_{12}} & = \dfrac{\partial \xi_{ij}}{\partial \xi_{1,i}}\dfrac{d \xi_{1,i}}{d \mb{d}_{12}}+\dfrac{\partial \xi_{ij}}{\partial \xi_{2,i}}\dfrac{d \xi_{2,i}}{d \mb{d}_{12}}, \\
  \text{and} \quad \dfrac{d \eta_c}{d \mb{d}_{12}} & = \dfrac{\partial \eta_{c}}{\partial \xi_{ij}}\dfrac{d \xi_{ij}}{d \mb{d}_{12}}+\dfrac{\partial \eta_{c}}{\partial \mb{d}_{12}}.
\end{split}
\end{align}
We focus on the most general case with an integration interval segmentation being applied on both sides of the slave element.
In the line contact case, the orthogonality condition $p_2$ on beam 2 is relevant. Its linearization reads:
\begin{align}
\label{anhang_linf2}
 p_{2,\xi} \frac{d \xi}{d \mb{d}_{12}} + p_{2,\eta_c} \frac{d \eta_c}{d \mb{d}_{12}} = -p_{2,\mb{d}_{12}} \quad \rightarrow \quad
  \frac{d \eta_c}{d \mb{d}_{12}} = \Bigg( 
 \underbrace{\frac{-p_{2,\xi_{ij}}}{p_{2,\eta}}}_{=\frac{\partial \eta_c}{\partial \xi_{ij}}} 
 \cdot \frac{d \xi_{ij}}{d \mb{d}_{12}} 
 +\underbrace{\frac{-1}{p_{2,\eta}}p_{2,\mb{d}_{12}}}_{=\frac{\partial \eta_c}{\partial \mb{d}_{12}}} 
 \Bigg) \Bigg|_{\,(\xi_{ij},\eta_c(\xi_{ij}))}.
\end{align}
With the help of \eqref{line_integrationparamvalues}, the linearization $d \xi_{ij} / d \mb{d}_{12}$ of the evaluation points on the slave beam follows as
\begin{align}
\frac{d \xi_{ij}}{d \mb{d}_{12}} = \dfrac{\partial \xi_{ij}}{\partial \xi_{1,i}}\dfrac{d \xi_{1,i}}{d \mb{d}_{12}}+\dfrac{\partial \xi_{ij}}{\partial \xi_{2,i}}\dfrac{d \xi_{2,i}}{d \mb{d}_{12}}
\quad \text{with} \quad \dfrac{\partial \xi_{ij}}{\partial \xi_{1,i}} = \frac{1.0-\bar{\xi}_j}{2}
\quad \text{and} \quad \dfrac{\partial \xi_{ij}}{\partial \xi_{2,i}} = \frac{1.0+\bar{\xi}_j}{2},
\end{align}
where $\bar{\xi}_j$ are constant Gauss point coordinates. Since $\eta$ is fixed at the master beam endpoints, one obtains from \eqref{anhang_linf2}:
\begin{align}
 \frac{d \xi_{1,i}}{d \mb{d}_{12}} = \Bigg(\underbrace{\frac{-1}{p_{2,\xi}}p_{2,\mb{d}_{12}}}_{=\frac{\partial \xi_{B1}}{\partial \mb{d}_{12}}}\Bigg) \Bigg|_{\,(\xi_{B1}(\eta_{EP}),\eta_{EP})}
 \quad \text{and} \quad
 \frac{d \xi_{2,i}}{d \mb{d}_{12}} = \Bigg(\underbrace{\frac{-1}{p_{2,\xi}}p_{2,\mb{d}_{12}}}_{=\frac{\partial \xi_{B2}}{\partial \mb{d}_{12}}}\Bigg) \Bigg|_{\,(\xi_{B2}(\eta_{EP}),\eta_{EP})}.
\end{align}
Since the linearizations  $\partial \mb{r}_{con,l}^{ij} / \partial \xi_{1,i}$ and $\partial \mb{r}_{con,l}^{ij} / \partial \xi_{2,i}$ solely stem from the explicit dependence of the 
total Jacobian $J(\xi_{ij},\xi_{1,i},\xi_{2,i})$ on the boundary coordinates $\xi_{1,i}$ and $\xi_{2,i}$, these linearizations can be rewritten as follows:
\begin{align}
\label{anhang_linearizationxiB}
\dfrac{\partial \mb{r}_{con,l}^{ij}}{\partial \xi_{1,i}} = \dfrac{\mb{r}_{con,l}^{ij}}{J(\xi_{ij},\xi_{1,i},\xi_{2,i})}\cdot J_{,\xi_{1,i}}(\xi_{ij},\xi_{1,i},\xi_{2,i}), \quad
\dfrac{\partial \mb{r}_{con,l}^{ij}}{\partial \xi_{2,i}} = \dfrac{\mb{r}_{con,l}^{ij}}{J(\xi_{ij},\xi_{1,i},\xi_{2,i})}\cdot J_{,\xi_{2,i}}(\xi_{ij},\xi_{1,i},\xi_{2,i}) \quad \text{with} \quad l=1,2.
\end{align}
The linearizations of the Jacobian occurring in \eqref{anhang_linearizationxiB} follow directly from their definition in equation \eqref{line_totaljacobian}:
\begin{align}
J_{,\xi_{1,i}}(\xi_{ij},\xi_{1,i},\xi_{2,i}) = -\frac{J_{ele}(\xi(\bar{\xi}_i))}{2}, \quad
J_{,\xi_{2,i}}(\xi_{ij},\xi_{1,i},\xi_{2,i}) =  \frac{J_{ele}(\xi(\bar{\xi}_i))}{2}.
\end{align}
The derivative $\partial \mb{r}_{con,l}^{ij}/\partial \mb{d}_{12}$ with respect to $\mb{d}_{12}$ shows strong similarities to the corresponding terms in~\ref{anhang:linearizationendpoint}:
\begin{align}
\label{linediscreteweakform}
\begin{split}
  \dfrac{\partial \mb{r}_{con,1}^{ij}}{\partial \mb{d}_{12}}  \! &= \! w_j J(\xi_{ij},\xi_{1,i},\xi_{2,i}) \varepsilon \frac{\partial g(\xi_{ij})}{\partial \mb{d}_{12}}  \mb{N}_{1}^T(\xi_{ij}) \mb{n}(\xi_{ij})
  + w_j J(\xi_{ij},\xi_{1,i},\xi_{2,i}) \varepsilon  g(\xi_{ij}) \mb{N}_{1}^T(\xi_{ij}) \frac{\partial \mb{n}(\xi_{ij})}{\partial \mb{d}_{12}},\\
  \dfrac{\partial \mb{r}_{con,2}^{ij}}{\partial \mb{d}_{12}}  \! &= \!  -w_j J(\xi_{ij},\xi_{1,i},\xi_{2,i}) \varepsilon  \varepsilon \frac{\partial g(\xi_{ij})}{\partial \mb{d}_{12}}  \mb{N}_{2}^T(\eta_{c}(\xi_{ij})) \mb{n}(\xi_{ij})
  -w_j J(\xi_{ij},\xi_{1,i},\xi_{2,i}) \varepsilon  g(\xi_{ij})  \mb{N}_{2}^T(\eta_{c}(\xi_{ij})) \frac{\partial \mb{n}(\xi_{ij})}{\partial \mb{d}_{12}}.
\end{split}
\end{align}
The terms $\partial g /\partial \mb{d}_{12}$ and $\partial \mb{n}/\partial \mb{d}_{12}$ are identical to the ones presented in~\eqref{point_discreteweakformapp2}. The partial derivatives 
of the residual contributions $\mb{r}_{con,1}^{ij}$ and $\mb{r}_{con,2}^{ij}$ with respect to the evaluation points $\xi_{ij}$ and $\eta_c$ have the following form:
\begin{align}
\begin{split}
  \dfrac{\partial \mb{r}_{con,1}^{ij}}{\partial \xi_{ij}}  \! &= \! w_j J_{,\xi_{ij}}(\xi_{ij},\xi_{1,i},\xi_{2,i}) \varepsilon g(\xi_{ij})  \mb{N}_{1}^T(\xi_{ij}) \mb{n}(\xi_{ij})
  + w_j J(\xi_{ij},\xi_{1,i},\xi_{2,i}) \varepsilon g_{,\xi_{ij}}(\xi_{ij})  \mb{N}_{1}^T(\xi_{ij}) \mb{n}(\xi_{ij})\\
  & + w_j J(\xi_{ij},\xi_{1,i},\xi_{2,i}) \varepsilon g(\xi_{ij})  \mb{N}_{1,\xi_{ij}}^T(\xi_{ij}) \mb{n}(\xi_{ij})
  + w_j J(\xi_{ij},\xi_{1,i},\xi_{2,i}) \varepsilon g(\xi_{ij})  \mb{N}_{1}^T(\xi_{ij}) \mb{n}_{,\xi_{ij}}(\xi_{ij}),\\
  \dfrac{\partial \mb{r}_{con,2}^{ij}}{\partial \xi_{ij}}  \! &= - w_j J_{,\xi_{ij}}(\xi_{ij},\xi_{1,i},\xi_{2,i}) \varepsilon g(\xi_{ij})  \mb{N}_{2}^T(\xi_{ij}) \mb{n}(\xi_{ij})
  - w_j J(\xi_{ij},\xi_{1,i},\xi_{2,i}) \varepsilon g_{,\xi_{ij}}(\xi_{ij})  \mb{N}_{2}^T(\xi_{ij}) \mb{n}(\xi_{ij})\\
  & \,\, \, \,\,\, - w_j J(\xi_{ij},\xi_{1,i},\xi_{2,i}) \varepsilon g(\xi_{ij})  \mb{N}_{2}^T(\xi_{ij}) \mb{n}_{,\xi_{ij}}(\xi_{ij}),\\  
   \dfrac{\partial \mb{r}_{con,1}^{ij}}{\partial \eta_c}  \! &= \!
    w_j J(\xi_{ij},\xi_{1,i},\xi_{2,i}) \varepsilon g_{,\eta_c}(\xi_{ij})  \mb{N}_{1}^T(\xi_{ij}) \mb{n}(\xi_{ij})
   + w_j J(\xi_{ij},\xi_{1,i},\xi_{2,i}) \varepsilon g(\xi_{ij})  \mb{N}_{1}^T(\xi_{ij}) \mb{n}_{,\eta_c}(\xi_{ij}),\\
   \dfrac{\partial \mb{r}_{con,2}^{ij}}{\partial \eta_c}  \! &=
   - w_j J(\xi_{ij},\xi_{1,i},\xi_{2,i}) \varepsilon g_{,\eta_c}(\xi_{ij})  \mb{N}_{2}^T(\xi_{ij}) \mb{n}(\xi_{ij})
   - w_j J(\xi_{ij},\xi_{1,i},\xi_{2,i}) \varepsilon g(\xi_{ij})  \mb{N}_{2,\eta_c}^T(\xi_{ij}) \mb{n}(\xi_{ij})\\
   & \,\, \, \,\,\, - w_j J(\xi_{ij},\xi_{1,i},\xi_{2,i}) \varepsilon g(\xi_{ij})  \mb{N}_{2}^T(\xi_{ij}) \mb{n}_{,\eta_c}(\xi_{ij}).
\end{split}
\end{align}
The partial derivatives of $g$ and $\mb{n}$ are identical to the ones presented in~\eqref{point_discreteweakformapp3}. The partial derivative 
$J_{,\xi_{ij}}=J_{ele,\xi_{ij}}(\xi_{2,i}-\xi_{1,i})/2$ is only relevant in case of a non-constant element Jacobian $J_{ele}$.
It is emphasized that this most general linearization in~\eqref{line_linearization_app} is only necessary for slave elements with valid master beam 
endpoint projections. In practical simulations, for the vast majority of contact element pairs this is not the case, 
i.e. $d \xi_{1,i} / d \mb{d}_{12} = \mb{0}$ and $d \xi_{2,i} / d \mb{d}_{12} = \mb{0}$.

%
\section{Residual and linearization of the \textit{ABC} formulation}
\label{anhang:linearizationABC}
%

In a first step, the residual and linearization terms of the \textit{ABC} formulation with force-based model transition will be considered. The residual contributions directly follow from inserting the discretized weak forms~\eqref{point_discreteweakform} and \eqref{line_pen_discreteweakform_multipleIS} into~\eqref{ABC_forcebased_weakform}. Following the chain rule, the corresponding linearization consists of the basic linearizations of the point-to-point and line-to-line formulations according to~\ref{anhang:linearizationendpoint} and~\ref{anhang:linearizationlinecontact} scaled by the transition factor occurring in~\eqref{ABC_forcebased_weakform} and supplemented by additional terms containing the linearization of the transition factor itself. The linearization of the transition factor follows directly from~\eqref{ABC_potentialbased_momentcontributions1} and~\eqref{ABC_potentialbased_momentcontributions2} by replacing the variation $\delta \mb{d}_{12}$ with the increment $\Delta \mb{d}_{12}$.
According to~\eqref{ABC_potentialbased_weakform}, the residual of the \textit{ABC} formulation with potential-based model transition basically consists of the residual terms of the variant with force-based model transition (with squared transition factor $k(z)^2$ instead of $k(z)$; terms on the left-hand side) and additional contact moment contributions composed of energy-like scalar terms of the form $\varepsilon g^2$ multiplied with the transition factor and the variation of the transition factor according to~\eqref{ABC_potentialbased_momentcontributions1} and~\eqref{ABC_potentialbased_momentcontributions2}. The linearization of the potential-based variant is straight-forward, but more involved than for the force-based variant, since the linearization of the transition factor variation~\eqref{ABC_potentialbased_momentcontributions1} and~\eqref{ABC_potentialbased_momentcontributions2} is required. For that reason, we employed a convenient automatic differentiation tool instead of deriving this linearization analytically.



%
\bibliographystyle{plain}
\bibliography{allgemein.bib}
%
%
\end{document}